\documentclass[aps,prl,reprint,showpacs]{revtex4-1}
\usepackage{graphicx}
\usepackage{subfigure}
\usepackage{dcolumn}
\usepackage{bm}
\usepackage{hyperref}
\usepackage{indentfirst}
\usepackage{braket}
\usepackage{amsmath}
\usepackage{amssymb}
\usepackage[normalem]{ulem}
\hypersetup{colorlinks=true, citecolor=blue, urlcolor=blue, linkcolor=blue}
\begin{document}
 
\title{Quantum Phase Diagram and Spontaneously Emergent Topological Chiral Superconductivity in Doped Triangular-Lattice Mott Insulators}

\author{Yixuan Huang$^1$}

\author{Shou-Shu Gong$^2$}
\email{shoushu.gong$@$buaa.edu.cn}

\author{D. N. Sheng$^1$}
\email{donna.sheng1$@$csun.edu}

\affiliation{$^1$Department of Physics and Astronomy, California State University, Northridge, California 91330, USA \\
$^2$Department of Physics, Beihang University, Beijing 100191, China}

\date{\today}

\begin{abstract}
The topological superconducting state is a highly sought-after quantum state hosting topological order and Majorana excitations. In this work, we explore the mechanism to realize the topological superconductivity (TSC) in the doped Mott insulators with time-reversal symmetry (TRS). Through large-scale density matrix renormalization group study of an extended triangular-lattice $t$-$J$ model on the six- and eight-leg cylinders, we identify a $d+id$-wave chiral TSC with spontaneous TRS breaking, which is characterized by a Chern number $C=2$ and quasi-long-range superconducting order. We map out the quantum phase diagram by tuning the next-nearest-neighbor (NNN) electron hopping and spin interaction. In the weaker NNN-coupling regime, we identify a pseudogaplike phase with a charge stripe order coexisting with fluctuating superconductivity, which can be tuned into $d$-wave superconductivity by increasing the doping level and system width.
The TSC emerges in the intermediate-coupling regime, which has a transition to a $d$-wave superconducting phase with larger NNN couplings. The emergence of the TSC is driven by geometrical frustrations and hole dynamics which suppress spin correlation and charge order, leading to a topological quantum phase transition.
\end{abstract}

\maketitle

\textit{Introduction.}---The fractional quantum Hall states discovered in two-dimensional (2D) electron systems under external magnetic fields~\cite{tsui1982, laughlin1983} are remarkable states of matter demonstrating topological orders and fractionalized excitations~\cite{halperin1984, wen1990topological, wen1991mean}.
In 2D Mott insulators, geometrical frustration and quantum fluctuations can suppress magnetic order and lead to a topologically ordered quantum spin liquid (QSL)~\cite{balents2010spin,zhou2017quantum,broholm2020quantum}.
Tuning Mott insulators with doping, more exotic phases including unconventional superconductivity (SC) and non-Fermi liquid emerge~\cite{anderson1987resonating,lee2006doping,keimer2015quantum,proust2019remarkable,wen1996theory,fradkin2015colloquium,senthil2005cup,balents2007dual,sachdev2010},
which are central topics in condensed matter physics.
Interestingly, there is a class of time-reversal-symmetry (TRS) breaking QSL named the chiral spin liquid (CSL), which was first proposed by Kalmeyer and Laughlin (KL) as the analog of the fractional quantum Hall state~\cite{kalmeyer1987equivalence}.
Remarkably,  doping a CSL may lead to $d+id$-wave topological superconductivity (TSC) through the condensation of paired fractional quasiparticles~\cite{laughlin1988,wen1989chiral,lee1989}.

Recently, the KL-CSL has been theoretically discovered in the kagome spin systems with competing interactions~\cite{he2014,gong2014,bauer2014,gong2015}, and near the metal-insulator transition in the triangular Hubbard model~\cite{szasz2020chiral,chen2021quantum,wietek2021} through spontaneous TRS breaking.
Numerical studies on the doped CSL in these systems~\cite{gong2015, szasz2020chiral} have uncovered either a Wigner crystal solid or a nonsuperconducting chiral metal~\cite{jiang2017holon,peng2021doping,zhu2022doped}, which challenge the original proposal of realizing a TSC~\cite{laughlin1988,wen1989chiral,lee1989} and demonstrate the richness of doped frustrated systems~\cite{song2021doping,baskaran2003electronic,kumar2003superconductivity,wang2004doped,watanabe2004,braunecker2005edge,weber2006magnetism,gan2006superconducting,zhou2008nodal,chen2013unconventional,motrunich2004,kiesel2013,arovas2022hubbard,gannot2020SU,peng2021gapless,aghaei2020efficient,jiang2021possible}.
A  breakthrough comes from density matrix renormalization group (DMRG) studies, which have identified a $d+id$-wave TSC by doping either a CSL~\cite{jiang2020topological, huang2021topological} or a weak Mott insulator~\cite{huang2021topological} in the triangular-lattice $t$-$J$ model with three-spin chiral coupling $J_{\chi}$ breaking TRS explicitly.
Despite the exciting progress, the mechanism of realizing TSC in the systems with TRS remains an outstanding issue, which demands unbiased numerical study beyond mean-field and variational treatments~\cite{baskaran2003electronic,kumar2003superconductivity,wang2004doped,watanabe2004,braunecker2005edge,weber2006magnetism,gan2006superconducting,zhou2008nodal,Gu2013Time,xu2018topological,zhou2022chiral,belanger2022superconductivity}.
Focusing on TRS triangular systems, previous DMRG study of the doped $J_1$-$J_2$  QSL identified a $d$-wave SC~\cite{jiang2021superconductivity} while the rich interplay among conventional orders, hole dynamics and spin fluctuations  has not been extensively explored in such systems, which may provide a new mechanism to realize TSC through spontaneous TRS breaking.

\begin{figure}
\centering
\includegraphics[width=0.9\linewidth]{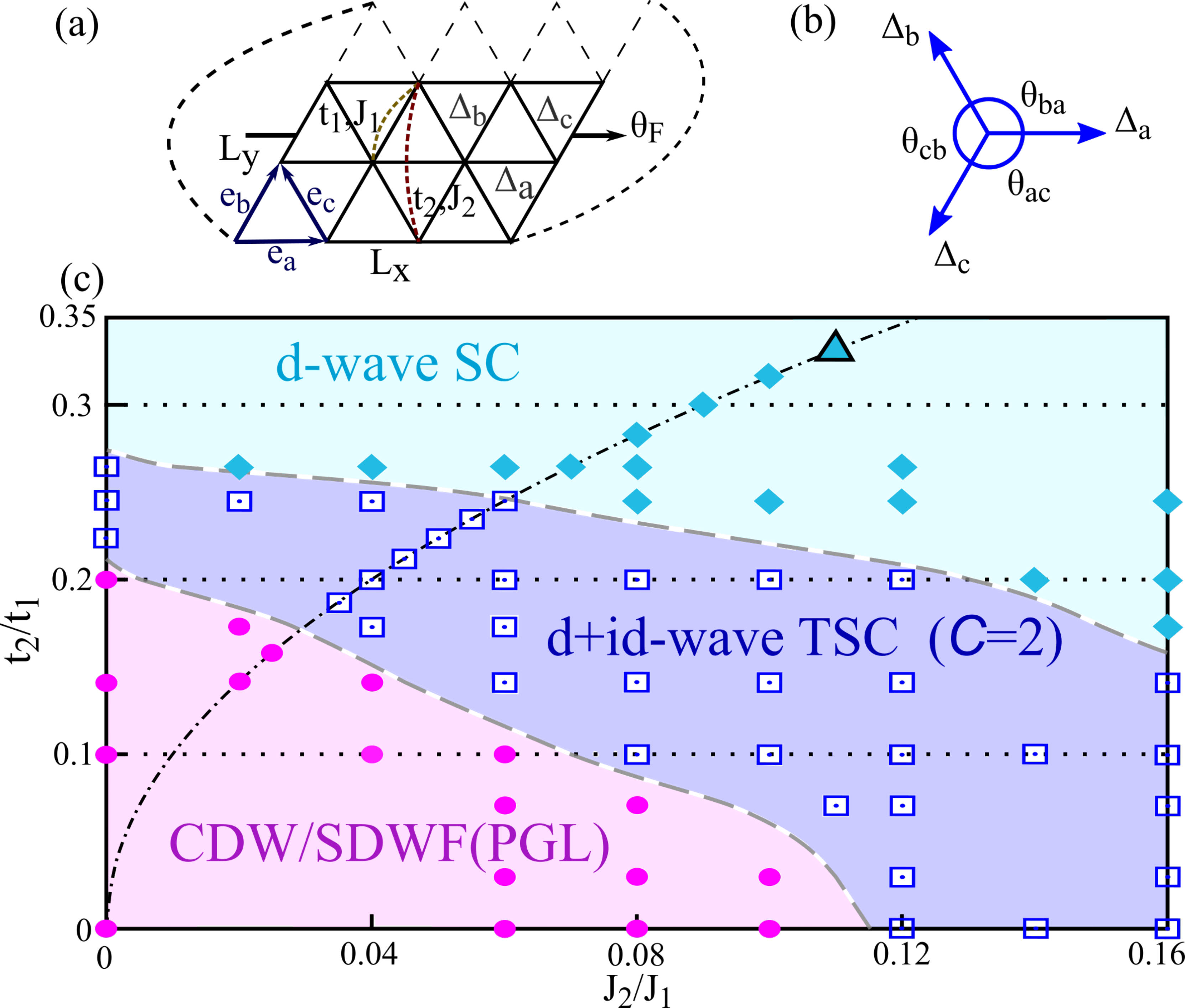}
\includegraphics[width=1\linewidth]{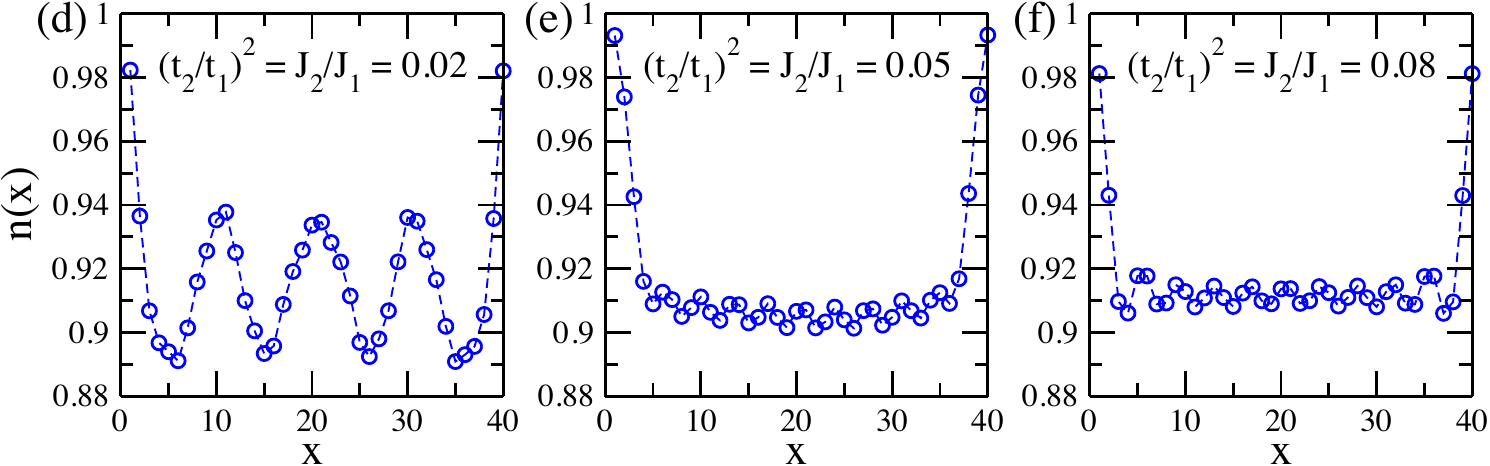}
\caption{Global quantum phase diagram. (a) Schematic figure of the triangular $t$-$J$ model with the NN and NNN hoppings $t_1, t_2$ and spin interactions $J_1, J_2$. 
$\theta_F$ is the magnetic flux threading in the cylinder. $\Delta_{a,b,c}$ define the pairing order parameters of the NN bonds along the $e_{a,b,c}$ directions. (b) The relative phases  between  $\Delta_{\alpha} = |\Delta_{\alpha}|e^{i \theta_{\alpha}}$ ($\alpha = a,b,c$), defined as $\theta_{\alpha\beta} = \theta_{\alpha} - \theta_{\beta}$. (c) The quantum phase diagram obtained on the $L_y = 6$ cylinder with doping level $\delta = 1/12$. We identify a pseudogaplike (PGL) phase with CDW + SDWF, a $d+id$-wave TSC phase, and a $d$-wave SC phase. The dotted dashed line denotes $J_2/J_1 = (t_2/t_1)^2$. The symbols mark the studied parameters, and the cyan triangle marks the studied parameter in Ref.~\cite{jiang2021superconductivity}. (d)-(f) The charge density profile in the three phases. $n(x)$ is the charge density per site in each column $x$, obtained on the $40 \times 6$ cylinder with M=12000.}
\label{Fig_1phase_diagram}
\end{figure}

Experimentally, triangular-lattice compounds are among the most promising candidates for hosting topological states, including the QSL candidates of weak Mott insulators~\cite{kurosaki2005mott, itou2007spin, yamashita2008thermodynamic}, 
the $d+id$-wave TSC candidates Na$_x$CoO$_2$·yH$_2$O~\cite{takada2003superconductivity,schaak2003,fujimoto2004} and Sn/Si(111) systems~\cite{ming2022evidence}, and the twisted transition metal dichalcogenides (TMD) moir\'e systems which can simulate the Hubbard and related $t$-$J$ model ~\cite{wu2018hubbard,Tang2020}.
The correlated insulators and possible SC states discovered in these systems~\cite{an2020interaction,schrade2021nematic,scherer2021mathcal} also call for theoretical understanding of the rich interplay among the experimentally tunable parameters such as electronic hopping and interaction.

In this Letter, we study the quantum phases in the extended triangular $t$-$J$ model using  
DMRG simulations.
By tuning the ratios of the next-nearest-neighbor (NNN) to nearest-neighbor (NN) hopping $t_{2}/t_1$ and spin interaction $J_{2}/J_1$, we find a pseudogaplike phase with  charge density wave (CDW) order  at small NNN couplings, which coexists with  both the strong spin density wave  fluctuation (SDWF) and 
fluctuating superconductivity (FSC) showing a tendency to develop into a $d$-wave SC on wider nine-leg cylinder.
With growing $t_2/t_1$ or (and) $J_2/J_1$, we identify a phase transition to an emergent $d+id$-wave TSC~\cite{laughlin1988,wen1989chiral,lee1989,read2000paired, senthil1999spin,zhou2008nodal} characterized by a topological Chern number $C=2$, through spontaneous TRS breaking.
The SC pairing correlations show algebraic decay with the power exponent $K_{SC} \approx 1.0$ dominating other spin and charge correlations, which are the quasi-1D descendent states of 2D topological superconductors. For even larger NNN couplings, a nematic $d$-wave SC phase emerges with anisotropic pairing correlations breaking rotational symmetry, which belongs to the same SC phase found in the doped $J_1$-$J_2$ QSL~\cite{jiang2021superconductivity}.
Our results establish a new route to the TSC by doping either a magnetic Mott insulator or a QSL with TRS, in which hole dynamics and geometrical frustrations play essential roles to suppress  magnetic correlations  and induce  the TSC.

\textit{Theoretical model and method.}---We study the following extended $t$-$J$ model on the triangular lattice  
\begin{eqnarray}
\label{eq_Hamiltonian}
H &=& \sum\limits_{\left \{ ij \right \},\sigma
}-t_{ij}(\hat{c}^{\dagger}_{i,\sigma }\hat{c}_{j,\sigma }+H.c.) + \sum\limits_{ \left \{ ij \right \} } J_{ij}(\hat{\boldsymbol{S}}_{i}\cdot \hat{\boldsymbol{S}}_{j}-\frac{1}{4}\hat{n}_{i}\hat{n}_{j}) \nonumber,
\end{eqnarray} 
where $\hat{c}_{i,\sigma}^{\dagger}$ ($\hat{c}_{i,\sigma}$) creates (annihilates) an electron on site $i$ with spin $\sigma=\pm 1/2$, $\hat{\boldsymbol{S}}_{i}$ is the spin-$1/2$ operator, $\hat{n}_{i}=\sum_{\sigma}\hat{c}_{i,\sigma}^{\dagger}\hat{c}_{i,\sigma}$ is the electron number operator.
We tune the ratios of neighboring couplings $t_{2}/t_{1}$ and $J_{2}/J_{1}$ to explore their  interplay in driving different phases in the system.
We set $J_{1}=1$ as the energy unit and $t_{1}/J_1 = 3$ to mimic a strong Hubbard interaction $U/t=12$.

We perform large scale DMRG simulations with charge $U(1)$ and spin $SU(2)$ symmetries~\cite{white1992density,McCulloch2007,gong2021robust} on a cylinder system, which has an open boundary in the $e_{a}$ or $x$ direction and periodic boundary conditions in the $e_{b}$ or $y$ direction [Fig.~\ref{Fig_1phase_diagram}(a)]. 
The number of sites along the $x$ ($y$) direction is denoted as $L_x$ ($L_y$) and the total number of sites is $N=L_{x}\times L_{y}$.
The electron number $N_{e}$ is related to hole doping level $\delta$ as $N_{e}/N=1-\delta$. 
We focus on the results on the $L_y = 6$ systems, which are supplemented with the studies on wider $L_{y}=8,9$ cylinders~\cite{2note}.
We keep up to M=20000 $SU(2)$ multiplets [equivalent to about 60000 $U(1)$ states] to obtain accurate results with the truncation error $\epsilon \lesssim 2\times10^{-5}$; see more details in Sec. I. of the Supplemental Material (SM)~\cite{SuppMaterial}.

\begin{figure}
\centering
\includegraphics[width=1\linewidth]{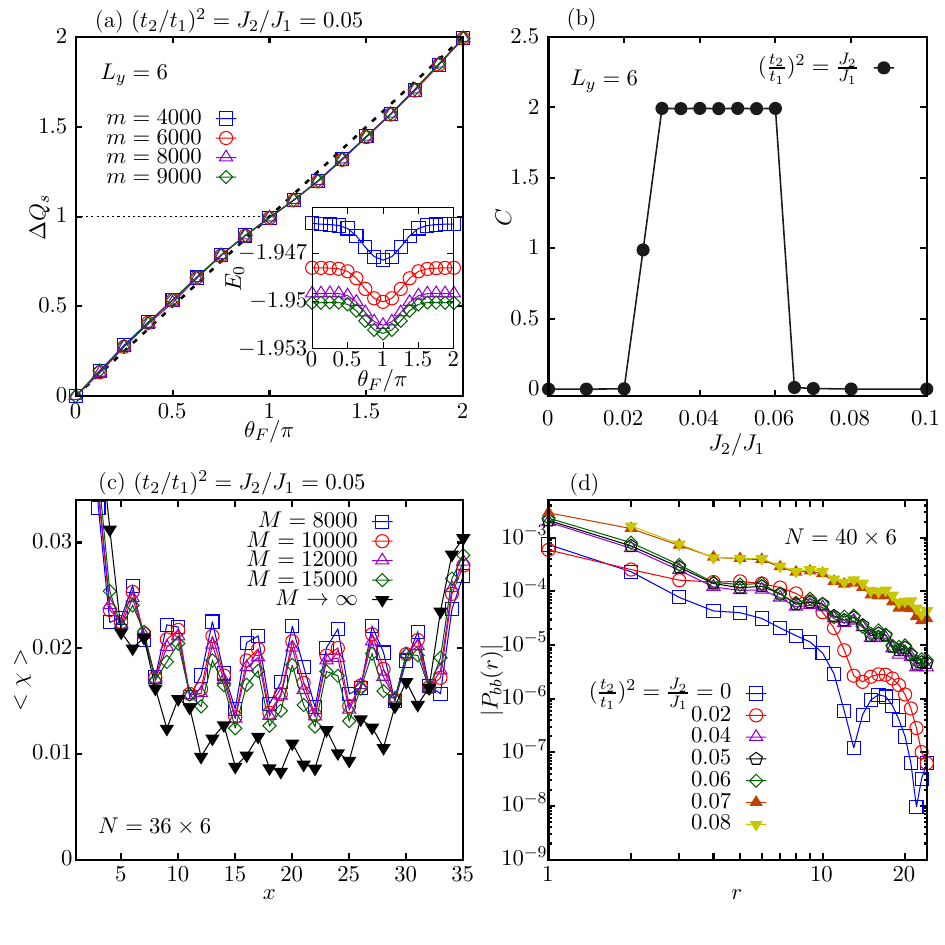}
\caption{Identifying the TSC phase and phase transitions along $(t_2/t_1)^2 = J_2 / J_1$. (a) Spin pumping simulation by adiabatically inserting flux $\theta_F$ for $J_2 / J_1 = 0.05$. $m$ is the  $U(1)$ bond dimension. By inserting a flux quantum, we obtain the Chern number $C=\Delta Q_s \approx 2$ with an error smaller than $\pm 0.03$. The inset shows the flux dependence of ground-state energy per site $E_0$. (b) Coupling dependence of the obtained Chern number with $m=8000$. (c) Spin chiral order $\langle \chi \rangle = \langle \hat{\boldsymbol{S}}_i \cdot (\hat{\boldsymbol{S}}_j \times \hat{\boldsymbol{S}}_k) \rangle$ of the triangles in each column versus the column position $x$ for $J_2/J_1 = 0.05$. $M$ is the $SU(2)$ bond dimension. (d) Double-logarithmic plot of the pairing correlation $|P_{bb}(r)|$ obtained with M=12000.}
\label{Fig_2Chern_Chiral}
\end{figure}

\textit{Phase diagram and Chern number characterization.}---We map out the phase diagram for $\delta = 1/12$ based on the results of Chern number~\cite{huang2021topological} and pairing correlation.
As shown in the phase diagram [Fig.~\ref{Fig_1phase_diagram}(c)], in the smaller $J_2$ and $t_2$ regime we identify a pseudogaplike phase~\cite{lee2014, dai2020} with dominant CDW order and short-range $d$-wave SC fluctuation.
The TSC emerges in the intermediate coupling regime while previously identified $d$-wave SC phase~\cite{jiang2021superconductivity} appears at the larger NNN couplings.

To identify the topological nature of the phases, we perform the inserting flux simulation~\cite{gong2014,huang2021topological} 
using the infinite DMRG~\cite{grushin2015characterization} with  increasing the flux adiabatically with $\theta_F \rightarrow \theta_F+\Delta \theta_F$ and $\Delta \theta_F=2\pi/16$.
We measure the accumulated spin $Q_{s}=n_{\uparrow}-n_{\downarrow}$ at left edge for each $\theta_{F}$ ($n_{\sigma}$ is the total charge with spin $\sigma$ near the edge~\cite{huang2021topological}).
For a range of intermediate NNN couplings, nonzero pumped spin $\Delta Q_{s}$ is obtained, which increases almost linearly with $\theta_F$ [Fig.~\ref{Fig_2Chern_Chiral}(a)], indicating the uniform Berry curvature~\cite{sheng2006}. 
By threading a flux quantum ($\theta_F=0\rightarrow 2\pi$), the Chern number $C=\Delta Q_{s}\approx 2.0$ characterizes a robust TRS-breaking topological state.  
The energy per site $E_0$ varies smoothly with $\theta_F$ [the inset of Fig.~\ref{Fig_2Chern_Chiral}(a)],
indicating a gapped spectrum flow and robust topological quantization~\cite{note}.
Here $C=2$ identifies the number of chiral Majorana edge modes~\cite{read2000paired,senthil1999spin}.
In Fig.~\ref{Fig_2Chern_Chiral}(b), we show the obtained Chern number along $(t_2/t_1)^{2}=J_{2} / J_{1}$,
where the quantized $C=2$ clearly distinguishes the TSC from the topologically trivial phases with $C=0$  nearby (see more results in SM Sec. II.~\cite{SuppMaterial}).
We further show the chiral order $\left \langle \chi \right \rangle = \langle \hat{\boldsymbol{S}}_{i}\cdot (\hat{\boldsymbol{S}}_{j}\times \hat{\boldsymbol{S}}_{k}) \rangle$ (the sites $i,j,k$ belong to the smallest triangle) along the $x$ direction [Fig.~\ref{Fig_2Chern_Chiral}(c)].
The chiral orders after bond-dimension scaling to $M\rightarrow \infty$ limit remain finite, supporting the spontaneous TRS breaking in the TSC.

Next, we show the evolution of the dominant spin-singlet pairing correlations $P_{\alpha \beta }(\mathbf{r}) = \langle \hat{\Delta} ^{\dagger }_{\alpha }(\mathbf{r}_{0})\hat{\Delta}_{\beta }(\mathbf{r}_{0}+\mathbf{r}) \rangle$ where the pairing order is defined as $\hat{\Delta}_{\alpha }(\mathbf{r})=(\hat{c}_{\mathbf{r}\uparrow}\hat{c}_{\mathbf{r}+e_{\alpha }\downarrow}-\hat{c}_{\mathbf{r}\downarrow}\hat{c}_{\mathbf{r}+e_{\alpha }\uparrow})/\sqrt{2}$ ($\alpha = a,b,c$). The pairing correlation $|P_{bb}(r)|$ decays very fast for $t_2=J_2=0$ and is enhanced at short distance for $(t_2/t_1)^2=J_2/J_1=0.02$ inside the CDW + SDWF phase [Fig.~\ref{Fig_2Chern_Chiral}(d)].
With larger NNN couplings in the TSC and $d$-wave SC phases, pairing correlations are strongly enhanced at all distances. 

\begin{figure}
\centering
\includegraphics[width=0.325\linewidth]{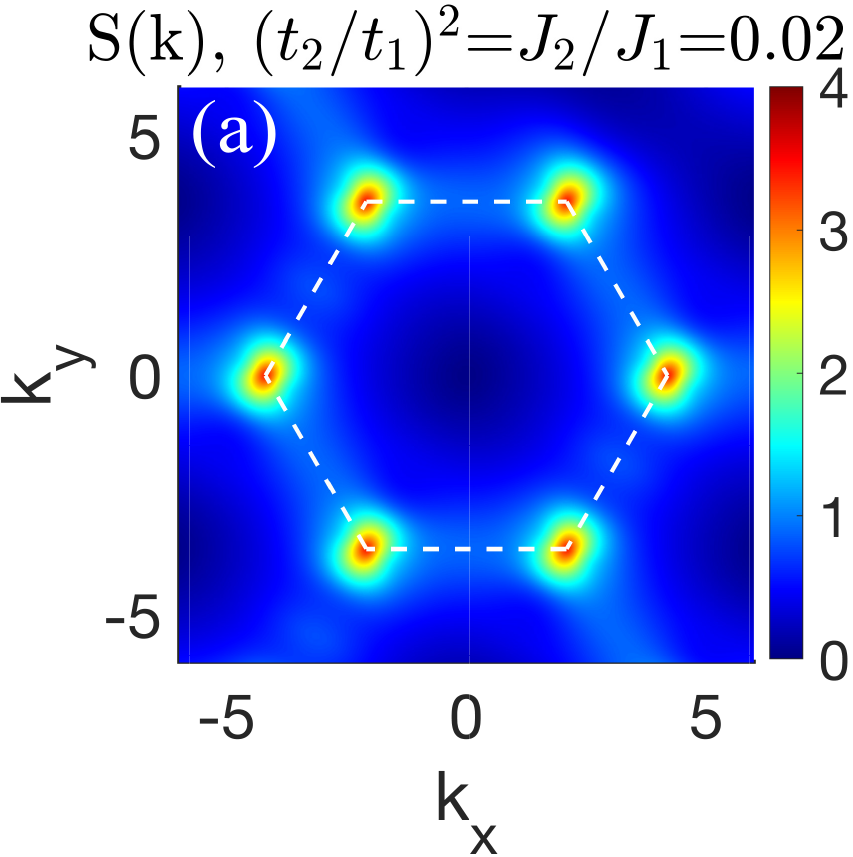}
\includegraphics[width=0.325\linewidth]{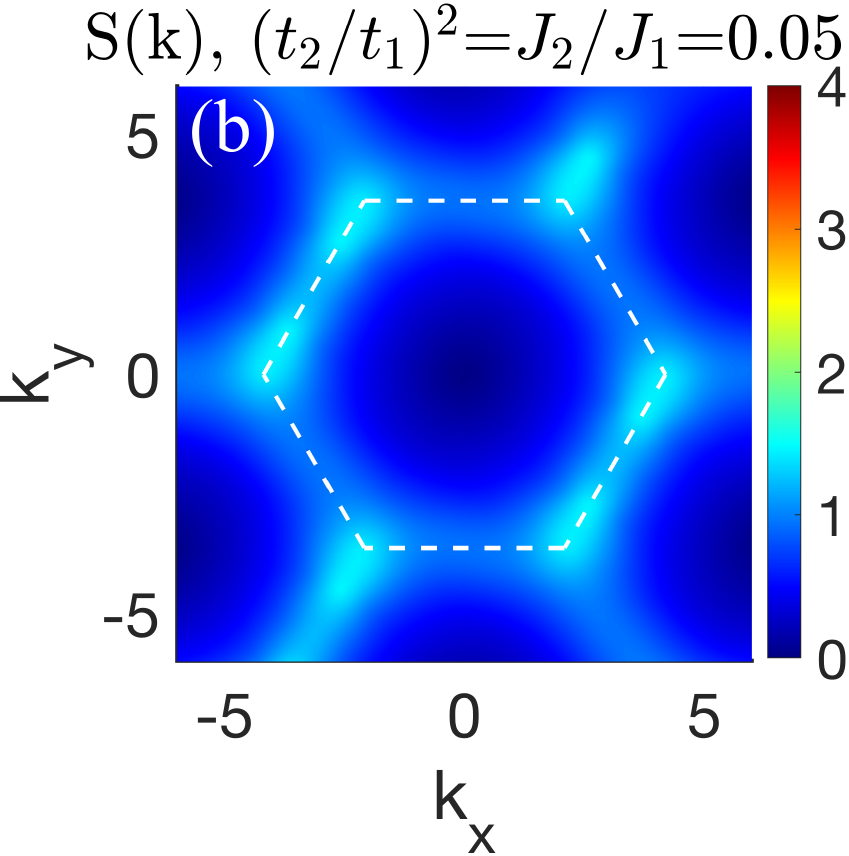}
\includegraphics[width=0.325\linewidth]{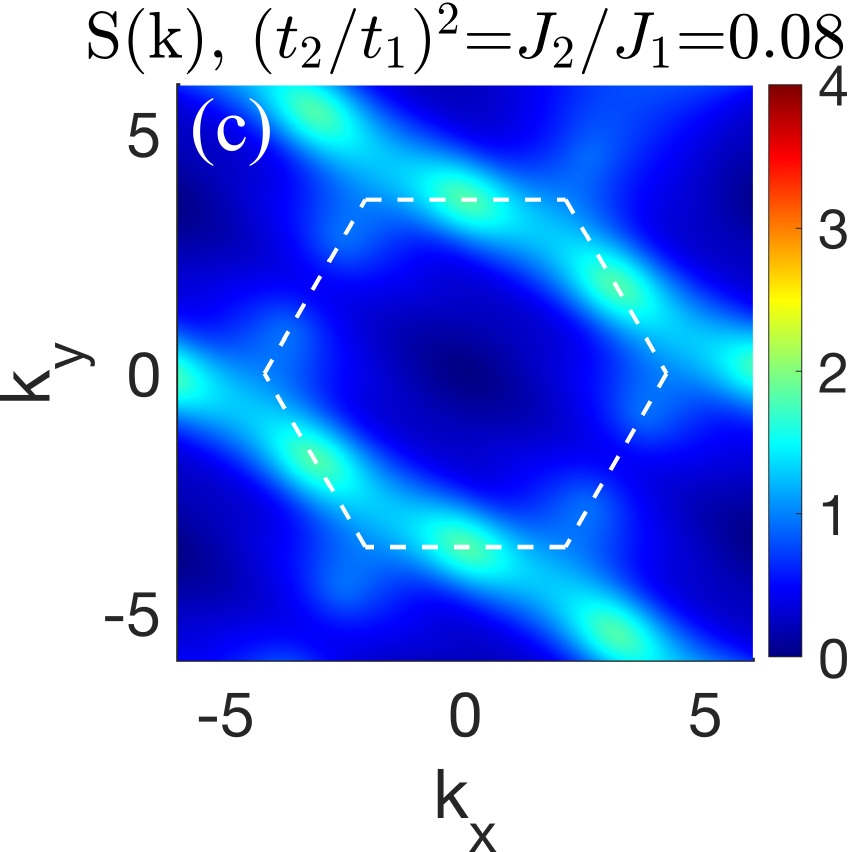}
\includegraphics[width=0.3259\linewidth]{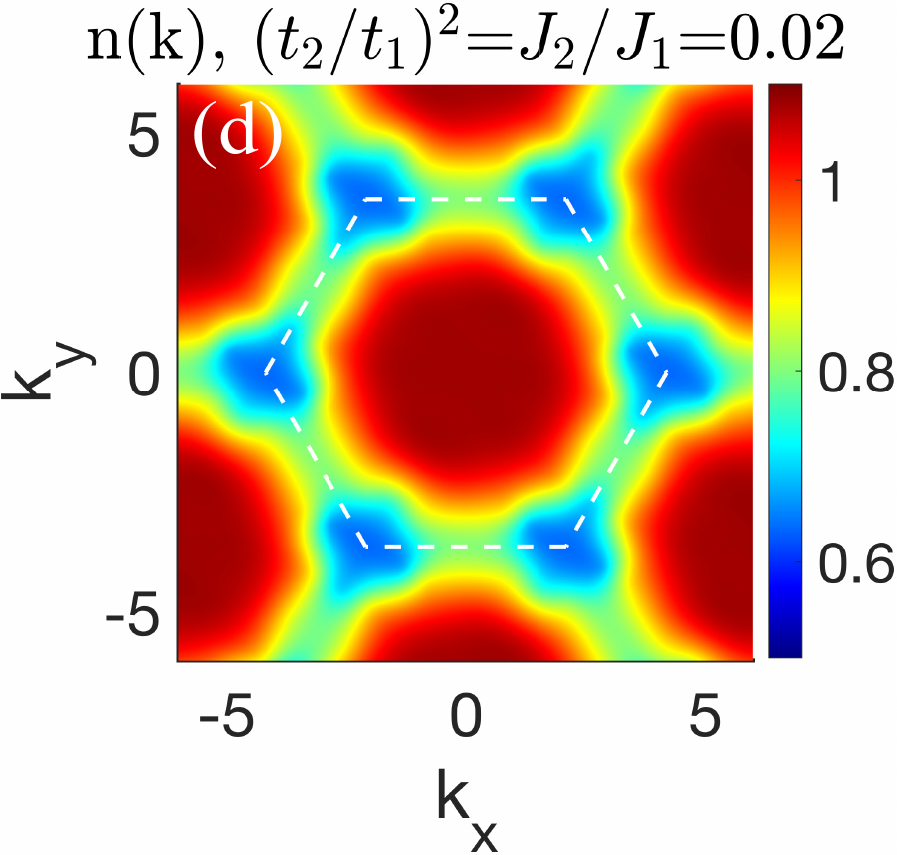}
\includegraphics[width=0.3259\linewidth]{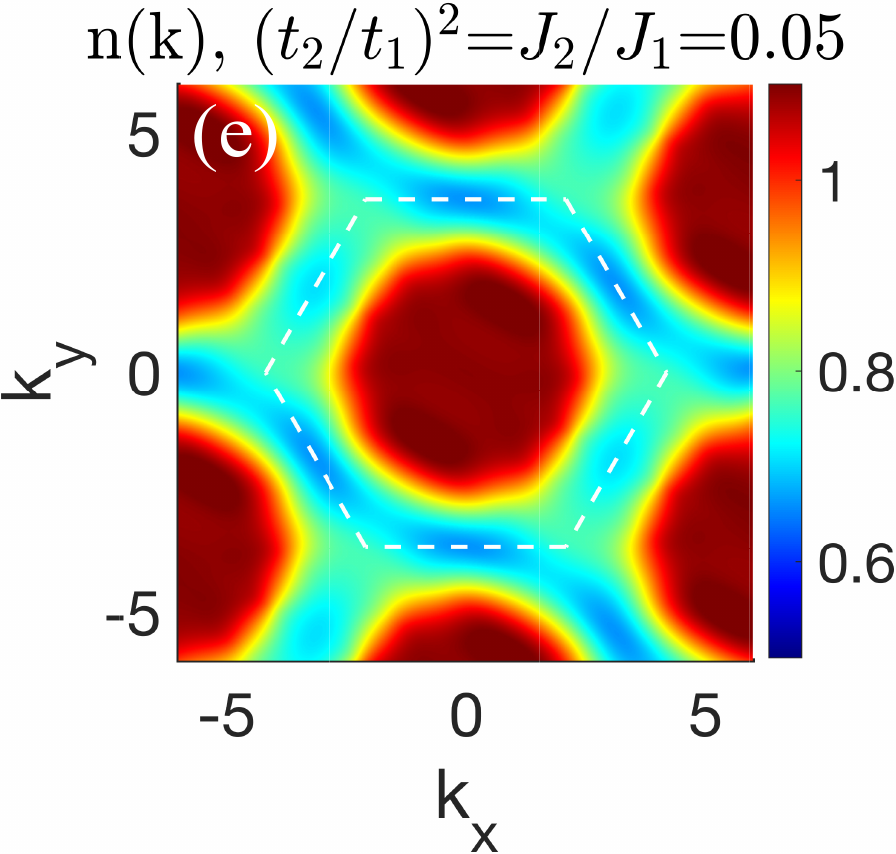}
\includegraphics[width=0.3259\linewidth]{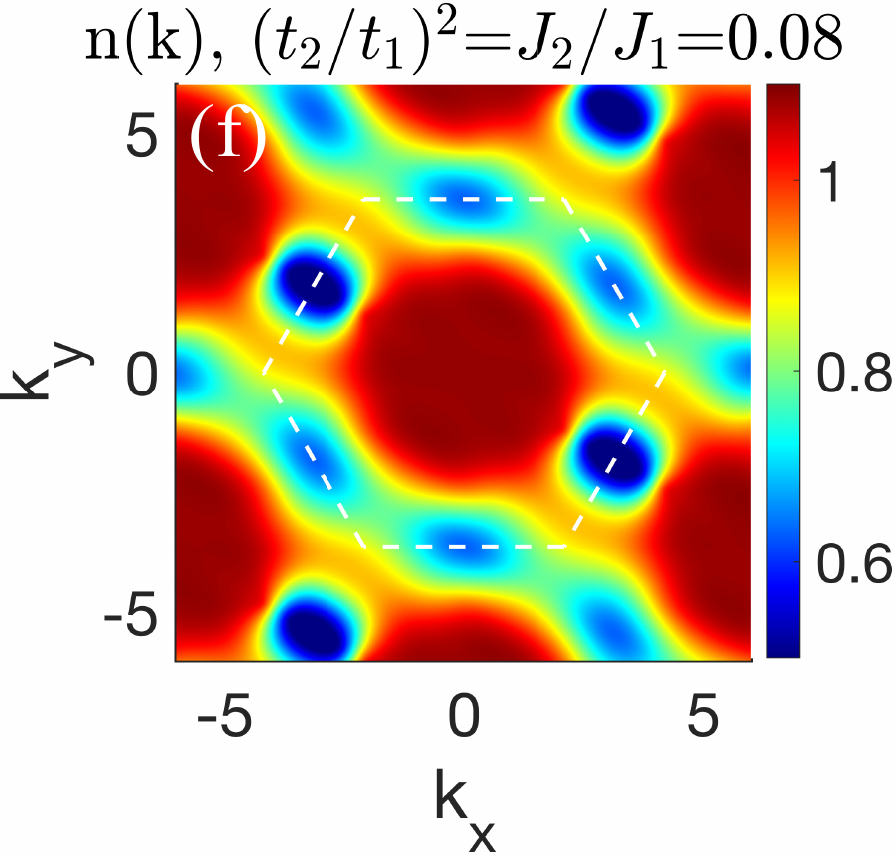}
\caption{Spin structure factor $S({\bf k})$ and electron density in momentum space $n({\bf k})$ in the three phases. The results are obtained using the middle $N_{m}=24\times 6$ sites of a long cylinder, which are calculated with $M=12000$ and well converged. The dashed hexagon denotes the Brillouin zone. (a) and (d) belong to the CDW + SDWF phase, (b) and (e) belong to the TSC phase, (c) and (f) belong to the $d$-wave SC phase.}
\label{Fig_5_structure_factor}
\end{figure}

\begin{figure*}
\centering
\includegraphics[width=1\linewidth]{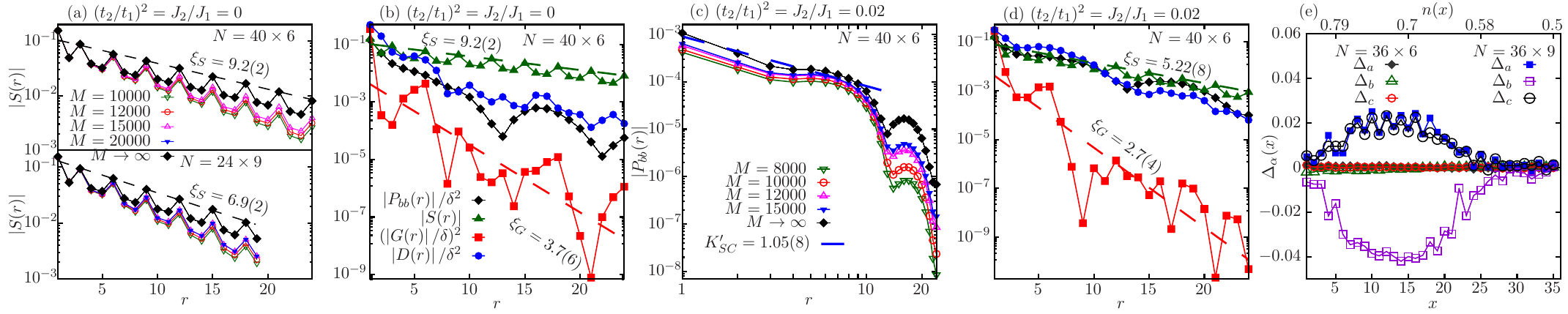}
\caption{Correlation functions and SC orders in the CDW + SDWF phase with extrapolated $M \rightarrow \infty$ data for (a) - (d). (a) Logarithmic-linear plots of the spin correlations on the $40\times 6$ and $24\times 9$ cylinders, with the correlation length $\xi_{S}=9.2(2)$ ($6.9(2)$) for $L_{y}=6$ ($9$). The number in the bracket gives the standard deviation from linear fitting. (b) Comparing correlations which are rescaled with doping ratio for a direct comparison. The fittings give $\xi_{S}=9.2(2)$ and $\xi_{G} = 3.7(6)$. (c) Double-logarithmic plot of pairing correlation $|P_{bb}(r)|$. The extrapolated results with $r \leq 10$ can be fitted algebraically with $K'_{SC} = 1.05(8)$. (d) Comparing correlations where the fittings give $\xi_{S}=5.22(8)$ and $\xi_{G} = 2.7(4)$. We choose the reference site at $x_0=L_x/4$ for demonstrating correlations. (e) Different SC orders $\Delta_{\alpha}$ versus each column $x$ for a system in a grand canonical ensemble with $M=8000$ and the averaged electron density $n(x)$. The coupling parameters are the same as (d). A varying chemical potential  $\mu_i=\mu(x)=\mu_0+x/L_x(a+b(x/L_x))$ is used to adjust the range of $n(x)$.}
\label{Fig_3SC_0_0}
\end{figure*}

\textit{Spin structure factor and charge occupation.}---Now we discuss the spin correlation and charge occupation.
In the CDW + SDWF phase, the spin structure factor $S(\mathbf{k}) = \frac{1}{N_{m}}\sum_{i,j} \langle \hat{\boldsymbol{S}}_{i} \cdot \hat{\boldsymbol{S}}_{j} \rangle e^{i\mathbf{k}\cdot (\mathbf{r}_{i}-\mathbf{r}_{j})}$ 
has prominent peaks at the $\mathbf{K}$ points representing strong $120^{\circ}$ spin fluctuation [Fig.~\ref{Fig_5_structure_factor}(a)].
In the TSC, the $\mathbf{K}$ point peaks are significantly suppressed and dispersed along one of the edges of Brillouin zone [see Fig.~\ref{Fig_5_structure_factor}(b) and SM Sec. III.~\cite{SuppMaterial}], consistent with the emergence of the CSL in spin background. In the $d$-wave SC phase, weak peaks emerge at two $\mathbf{M}$ points [Fig.~\ref{Fig_5_structure_factor}(c)], indicating nematic spin fluctuation.
Furthermore, we investigate the electron occupation number in the momentum space $n(\mathbf{k}) = \frac{1}{N_{m}}\sum_{i,j,\sigma} \langle \hat{c}^{\dagger}_{i,\sigma} \hat{c}_{j,\sigma} \rangle e^{i {\bf k}\cdot ({\bf r}_i - {\bf r}_j)}$ and find that from the CDW + SDWF phase to the TSC, the hole pockets at the $\mathbf{K}$ points disperse along the edge of the Brillouin zone, while in the $d$-wave SC phase the hole pockets concentrate at two $\mathbf{M}$ points [Figs.~\ref{Fig_5_structure_factor}(d)-\ref{Fig_5_structure_factor}(f) and SM Sec.~VII.~\cite{SuppMaterial}].
In real space, the charge density profile in the CDW + SDWF phase shows a strong stripe pattern with the wavelength $\lambda \approx 10$  
while in the SC phases, the CDW becomes much weaker with $\lambda \approx 4$ [Figs.~\ref{Fig_1phase_diagram}(d)-\ref{Fig_1phase_diagram}(f)].

\textit{Fluctuating superconductivity in the CDW + SDWF phase.}---To reveal the nature of the CDW + SDWF phase, we focus on the correlation functions. At $t_2=J_2=0$, the extrapolated spin correlations $S(\mathbf{r}) = \langle \hat{\boldsymbol{S}}_{\mathbf{r}_{0}}\cdot \hat{\boldsymbol{S}}_{\mathbf{r}_{0}+\mathbf{r}}\rangle$ decay exponentially with a large correlation length $\xi_S \approx 9.2$ ($6.9$) on the $L_y=6$ ($9$) system [Fig.~\ref{Fig_3SC_0_0}(a)], confirming the absence of magnetic order and short-range SDWF.
We further compare $S(r)$ with single-particle correlation $G(\mathbf{r})=\sum_{\sigma }\langle \hat{c}^{\dagger }_{\mathbf{r}_{0},\sigma} \hat{c}_{\mathbf{r}_{0}+\mathbf{r},\sigma}\rangle$, 
density correlation $D(\mathbf{r}) = \langle \hat{n}_{\mathbf{r}_{0}} \hat{n}_{\mathbf{r}_{0}+\mathbf{r}}\rangle - \langle \hat{n}_{\mathbf{r}_{0}}\rangle \langle \hat{n}_{\mathbf{r}_{0}+\mathbf{r}} \rangle$, and pairing correlation $\left | P_{bb}(r) \right |$ using the  extrapolated $M \rightarrow \infty$ 
data (rescaled with doping ratio for direct comparison) as shown in  Fig.~\ref{Fig_3SC_0_0}(b).
While the spin correlation is relatively strong, single-particle $|G(r)|$ decays exponentially with a short correlation length $\xi_{G} \approx 3.7$.
Although the pairing correlation also decays fast, it is much stronger compared to the two single-particle correlator $|G^2(r)|$, indicating the more suppressed single-particle channel.

At $(t_2/t_1)^2=J_2/J_1=0.02$, $|P_{bb}(r)|$ is enhanced and decays algebraically with an exponent $K_{SC}'\approx 1.05$ within short distance, 
which indicates a strong local pairing order [Fig.~\ref{Fig_3SC_0_0}(c) and Fig.~\ref{Fig_2Chern_Chiral}(d)] representing the FSC.
Remarkably, the difference between $|P_{bb}(r)|$ and $|G^2(r)|$ dramatically increases with $|P_{bb}(r)|$ larger than $|G^2(r)|$ by around 4 orders of magnitude at large distances [Fig.~\ref{Fig_3SC_0_0}(d)], unveiling the ``pseudogap'' behavior. 
To further explore the FSC, we compute the SC order in the grand canonical ensemble with varying chemical potential $H\rightarrow H-\sum_i\mu_in_i$ following the method in Ref.~\cite{jiang2021ground} (see  SM Sec.~VIII.~\cite{SuppMaterial}). 
As shown in Fig.~\ref{Fig_3SC_0_0}(e), a finite $d$-wave SC order develops with increased $L_y=9$ and the doping level over 20\%.

\begin{figure}
\centering
\includegraphics[width=1\linewidth]{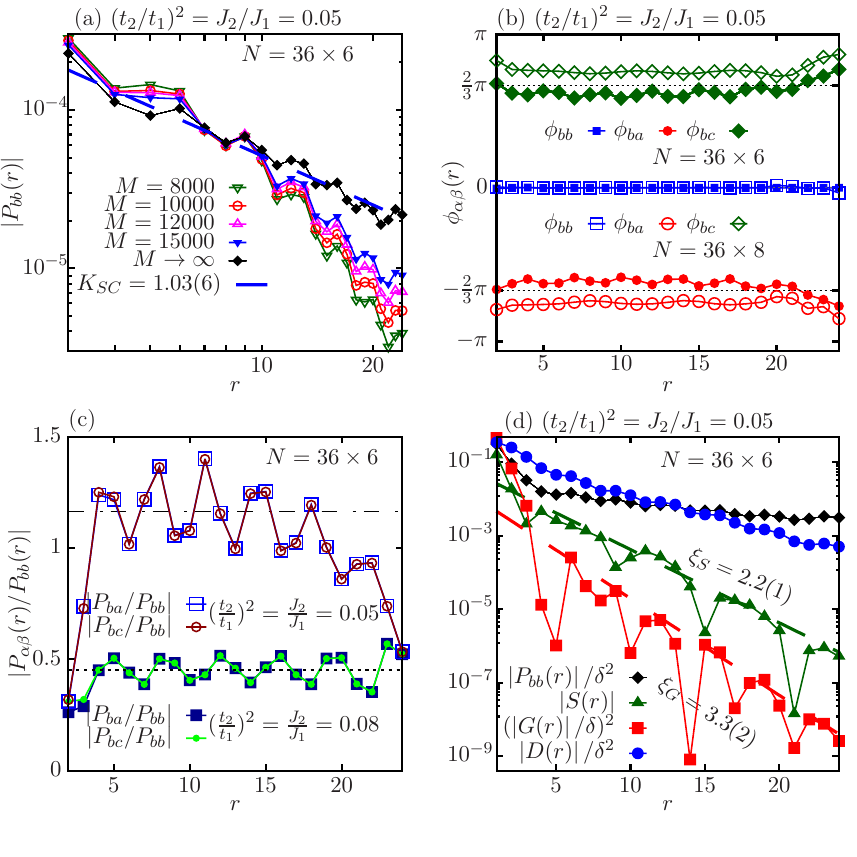}
\caption{Correlation functions for $(t_{2}/t_{1})^2=J_{2} / J_{1} =0.05$ in the TSC phase using the extrapolated data. (a) Double-logarithmic plot of the pairing correlations $|P_{bb}(r)|$ obtained by keeping different $SU(2)$ bond dimensions. The extrapolated correlations decay algebraically with $K_{SC} = 1.03(6)$. (b) $d+id$-wave pairing symmetry identified by the phase differences of pairing correlations on the $L_y = 6$ (8) cylinder using bond dimensions M=15000 (20000). (c) The ratios of the magnitudes of the pairing correlations at different bonds. The dashed dotted line indicates the averaged ratio of $1.2(1)$ for the TSC phase. The dotted line indicates the averaged ratio of $0.45(5)$ for the $d$-wave SC phase. We choose $r\leq L_{x}/2$ to calculate the averages to minimize the boundary effect. (d) Comparing the correlations which are rescaled with the doping ratio. The fittings give $\xi_{S}=2.2(1)$ and $\xi_{G} = 3.3(2)$. We choose $x_0=L_x/4$ and fit the data to the distance $r=L_x/2$ to avoid a boundary effect.}
\label{Fig_4SC_0.05_0.05}
\end{figure}

\textit{$d+id$-wave TSC phase.}---Next we turn to the characterization of the TSC phase.  
By bond-dimension extrapolation, we identify the algebraic decay of the pairing correlation.
For $(t_1/t_2)^2=J_1/J_2=0.05$ and $L_y = 6$, we find $\left | P_{bb}(r) \right |\sim r^{-K_{SC}}$ with $K_{SC} \approx 1.03$ [Fig.~\ref{Fig_4SC_0.05_0.05}(a)], indicating a divergent SC susceptibility in the zero-temperature limit~\cite{jiang2021high}.
Similar results are also obtained on the wider $L_{y}=8$ system (see SM Sec.~V.A.~\cite{SuppMaterial}), supporting the robust TSC.

To identify the pairing symmetry, we rewrite $\Delta_{\alpha}(\mathbf{r})=\left | \Delta_{\alpha}(\mathbf{r}) \right |e^{i\theta _{\alpha}(\mathbf{r})}$ and $P_{\alpha \beta }(\mathbf{r})=\left | P_{\alpha \beta }(\mathbf{r}) \right |e^{i\phi _{\alpha \beta }(\mathbf{r})}$ with the relative phases $\phi _{\alpha \beta }(\mathbf{r})=\theta _{\beta}(\mathbf{r}_{0}+\mathbf{r})-\theta _{\alpha}(\mathbf{r}_0)$. Thus, $\theta_{\alpha\beta}(\mathbf{r}) \equiv \theta_{\alpha}(\mathbf{r}) - \theta_{\beta}(\mathbf{r}) = \phi_{\alpha\alpha}(\mathbf{r}) - \phi_{\alpha\beta}(\mathbf{r})$ (see Fig.~\ref{Fig_1phase_diagram}(b)).
As shown in Fig.~\ref{Fig_4SC_0.05_0.05}(b), $\phi _{\alpha \beta }(r)$ are nearly uniform in real space and are obtained as $[{\phi }_{bb},{\phi }_{bc},{\phi }_{ba}]= [0.000(4),0.61(2)\pi, -0.61(2)\pi]   \approx [0,\frac{2}{3}\pi,-\frac{2}{3}\pi]$  for $L_y=6$, which give $\theta _{ba}=\theta_{ac}=\theta _{cb}\approx 2\pi/3$ characterizing an isotropic $d+id$-wave  pairing symmetry, while  $\theta _{ba}=\theta _{cb}=\pi$ is observed in the $d$-wave SC phase.
We also confirm this robust pairing symmetry on the wider $N=36\times 8$ 
system [see Fig.~\ref{Fig_4SC_0.05_0.05}(b) and SM Sec.~V.A.~\cite{SuppMaterial}], providing compelling evidence for the emergent TSC through spontaneous TRS breaking. 
Furthermore, 
as shown in Fig.~\ref{Fig_4SC_0.05_0.05}(c), we find that $|P_{ba}(r)/P_{bb}(r)|$ and $|P_{bc}(r)/P_{bb}(r)|$ averaged over $r$ are around $1.2$ for the near isotropic  TSC phase, while they drop to around $0.45$ in the nematic $d$-wave SC phase.

In comparison, both spin and single-particle correlations decay exponentially with small correlation lengths 
[Fig.~\ref{Fig_4SC_0.05_0.05}(d)] while 
the density correlations seem also to decay algebraically but with a large exponent $K_{CDW} \approx 2.4$, showing that the pairing correlation dominates all other correlations.

\textit{Summary and discussion.}---Through DMRG simulation on the extended triangular $t$-$J$ model, we identify a $d+id$-wave TSC through spontaneous TRS breaking, by doping either a magnetic order state or a time-reversal symmetric QSL.
The driving mechanism is the balanced spin frustrations and hole dynamics induced 
by NNN couplings, which suppress magnetic correlations and lead to the TSC for doping level $\delta=1/12-1/8$ (see additional results  in SM Sec.~V.B.~\cite{SuppMaterial}). Physically, frustration to spin background can be built up by  NNN coupling $J_2$, or $t_2$, or both terms acting jointly.  
Our findings  open a new route for discovering TSC in correlated materials, with the TMD Moir\'e superlattices~\cite{an2020interaction,schrade2021nematic,scherer2021mathcal,Tang2020} being the most promising platform~\cite{wu2018hubbard}.

We also reveal the pseudogaplike physics in the CDW + SDWF phase, which demonstrates a tendency to evolve into $d$-wave SC by increasing the phase coherence of pairing correlations. 
Our work suggests a new direction for future studies on doped Mott insulators~\cite{white2009pairing,corboz2014competing,zheng2017stripe,huang2017,jiang2020ground,qin2020,jiang2021high,gong2021robust,jiang2021ground,wu2020superconductivity,yang2021site}, which may provide insights to   the  challenging issues related to the normal states of the high-$T_c$ cuprate superconductors~\cite{lee2014, dai2020}.

Data and simulation code are available from the corresponding author upon reasonable request.

We thank Z. Y. Weng, Q. H. Wang and  F. Wang for stimulating discussions. The work done by Y.H. and D.N.S. was supported by  the U.S. Department of Energy, Office of Basic Energy Sciences under Grant No. DE-FG02-06ER46305 for large scale simulations of TSC. S.S.G. was supported by the National Natural Science Foundation of China Grants No. 12274014 and No. 11834014.


\textit{Note added.}---Recently, we noticed a related work, Ref.~\cite{zhu2022superconductivity}, which studies possible superconductivity with different hopping signs.


\bibliography{Tr_t12_J12}

\begin{thebibliography}{90}%
\makeatletter
\providecommand \@ifxundefined [1]{%
 \@ifx{#1\undefined}
}%
\providecommand \@ifnum [1]{%
 \ifnum #1\expandafter \@firstoftwo
 \else \expandafter \@secondoftwo
 \fi
}%
\providecommand \@ifx [1]{%
 \ifx #1\expandafter \@firstoftwo
 \else \expandafter \@secondoftwo
 \fi
}%
\providecommand \natexlab [1]{#1}%
\providecommand \enquote  [1]{``#1''}%
\providecommand \bibnamefont  [1]{#1}%
\providecommand \bibfnamefont [1]{#1}%
\providecommand \citenamefont [1]{#1}%
\providecommand \href@noop [0]{\@secondoftwo}%
\providecommand \href [0]{\begingroup \@sanitize@url \@href}%
\providecommand \@href[1]{\@@startlink{#1}\@@href}%
\providecommand \@@href[1]{\endgroup#1\@@endlink}%
\providecommand \@sanitize@url [0]{\catcode `\\12\catcode `\$12\catcode
  `\&12\catcode `\#12\catcode `\^12\catcode `\_12\catcode `\%12\relax}%
\providecommand \@@startlink[1]{}%
\providecommand \@@endlink[0]{}%
\providecommand \url  [0]{\begingroup\@sanitize@url \@url }%
\providecommand \@url [1]{\endgroup\@href {#1}{\urlprefix }}%
\providecommand \urlprefix  [0]{URL }%
\providecommand \Eprint [0]{\href }%
\providecommand \doibase [0]{http://dx.doi.org/}%
\providecommand \selectlanguage [0]{\@gobble}%
\providecommand \bibinfo  [0]{\@secondoftwo}%
\providecommand \bibfield  [0]{\@secondoftwo}%
\providecommand \translation [1]{[#1]}%
\providecommand \BibitemOpen [0]{}%
\providecommand \bibitemStop [0]{}%
\providecommand \bibitemNoStop [0]{.\EOS\space}%
\providecommand \EOS [0]{\spacefactor3000\relax}%
\providecommand \BibitemShut  [1]{\csname bibitem#1\endcsname}%
\let\auto@bib@innerbib\@empty
\bibitem [{\citenamefont {Tsui}\ \emph {et~al.}(1982)\citenamefont {Tsui},
  \citenamefont {Stormer},\ and\ \citenamefont {Gossard}}]{tsui1982}%
  \BibitemOpen
  \bibfield  {author} {\bibinfo {author} {\bibfnamefont {D.~C.}\ \bibnamefont
  {Tsui}}, \bibinfo {author} {\bibfnamefont {H.~L.}\ \bibnamefont {Stormer}}, \
  and\ \bibinfo {author} {\bibfnamefont {A.~C.}\ \bibnamefont {Gossard}},\
  }\href {\doibase 10.1103/PhysRevLett.48.1559} {\bibfield  {journal} {\bibinfo
   {journal} {Phys. Rev. Lett.}\ }\textbf {\bibinfo {volume} {48}},\ \bibinfo
  {pages} {1559} (\bibinfo {year} {1982})}\BibitemShut {NoStop}%
\bibitem [{\citenamefont {Laughlin}(1983)}]{laughlin1983}%
  \BibitemOpen
  \bibfield  {author} {\bibinfo {author} {\bibfnamefont {R.~B.}\ \bibnamefont
  {Laughlin}},\ }\href {\doibase 10.1103/PhysRevLett.50.1395} {\bibfield
  {journal} {\bibinfo  {journal} {Phys. Rev. Lett.}\ }\textbf {\bibinfo
  {volume} {50}},\ \bibinfo {pages} {1395} (\bibinfo {year}
  {1983})}\BibitemShut {NoStop}%
\bibitem [{\citenamefont {Halperin}(1984)}]{halperin1984}%
  \BibitemOpen
  \bibfield  {author} {\bibinfo {author} {\bibfnamefont {B.~I.}\ \bibnamefont
  {Halperin}},\ }\href {\doibase 10.1103/PhysRevLett.52.1583} {\bibfield
  {journal} {\bibinfo  {journal} {Phys. Rev. Lett.}\ }\textbf {\bibinfo
  {volume} {52}},\ \bibinfo {pages} {1583} (\bibinfo {year}
  {1984})}\BibitemShut {NoStop}%
\bibitem [{\citenamefont {Wen}(1990)}]{wen1990topological}%
  \BibitemOpen
  \bibfield  {author} {\bibinfo {author} {\bibfnamefont {X.-G.}\ \bibnamefont
  {Wen}},\ }\href
  {https://www.worldscientific.com/doi/abs/10.1142/S0217979290000139}
  {\bibfield  {journal} {\bibinfo  {journal} {International Journal of Modern
  Physics B}\ }\textbf {\bibinfo {volume} {04}},\ \bibinfo {pages} {239}
  (\bibinfo {year} {1990})}\BibitemShut {NoStop}%
\bibitem [{\citenamefont {Wen}(1991)}]{wen1991mean}%
  \BibitemOpen
  \bibfield  {author} {\bibinfo {author} {\bibfnamefont {X.~G.}\ \bibnamefont
  {Wen}},\ }\href {\doibase 10.1103/PhysRevB.44.2664} {\bibfield  {journal}
  {\bibinfo  {journal} {Phys. Rev. B}\ }\textbf {\bibinfo {volume} {44}},\
  \bibinfo {pages} {2664} (\bibinfo {year} {1991})}\BibitemShut {NoStop}%
\bibitem [{\citenamefont {Balents}(2010)}]{balents2010spin}%
  \BibitemOpen
  \bibfield  {author} {\bibinfo {author} {\bibfnamefont {L.}~\bibnamefont
  {Balents}},\ }\href {https://www.nature.com/articles/nature08917} {\bibfield
  {journal} {\bibinfo  {journal} {Nature}\ }\textbf {\bibinfo {volume} {464}},\
  \bibinfo {pages} {199} (\bibinfo {year} {2010})}\BibitemShut {NoStop}%
\bibitem [{\citenamefont {Zhou}\ \emph {et~al.}(2017)\citenamefont {Zhou},
  \citenamefont {Kanoda},\ and\ \citenamefont {Ng}}]{zhou2017quantum}%
  \BibitemOpen
  \bibfield  {author} {\bibinfo {author} {\bibfnamefont {Y.}~\bibnamefont
  {Zhou}}, \bibinfo {author} {\bibfnamefont {K.}~\bibnamefont {Kanoda}}, \ and\
  \bibinfo {author} {\bibfnamefont {T.-K.}\ \bibnamefont {Ng}},\ }\href
  {\doibase 10.1103/RevModPhys.89.025003} {\bibfield  {journal} {\bibinfo
  {journal} {Rev. Mod. Phys.}\ }\textbf {\bibinfo {volume} {89}},\ \bibinfo
  {pages} {025003} (\bibinfo {year} {2017})}\BibitemShut {NoStop}%
\bibitem [{\citenamefont {Broholm}\ \emph {et~al.}(2020)\citenamefont
  {Broholm}, \citenamefont {Cava}, \citenamefont {Kivelson}, \citenamefont
  {Nocera}, \citenamefont {Norman},\ and\ \citenamefont
  {Senthil}}]{broholm2020quantum}%
  \BibitemOpen
  \bibfield  {author} {\bibinfo {author} {\bibfnamefont {C.}~\bibnamefont
  {Broholm}}, \bibinfo {author} {\bibfnamefont {R.}~\bibnamefont {Cava}},
  \bibinfo {author} {\bibfnamefont {S.}~\bibnamefont {Kivelson}}, \bibinfo
  {author} {\bibfnamefont {D.}~\bibnamefont {Nocera}}, \bibinfo {author}
  {\bibfnamefont {M.}~\bibnamefont {Norman}}, \ and\ \bibinfo {author}
  {\bibfnamefont {T.}~\bibnamefont {Senthil}},\ }\href
  {https://www.science.org/doi/10.1126/science.aay0668} {\bibfield  {journal}
  {\bibinfo  {journal} {Science}\ }\textbf {\bibinfo {volume} {367}},\ \bibinfo
  {pages} {6475} (\bibinfo {year} {2020})}\BibitemShut {NoStop}%
\bibitem [{\citenamefont {Anderson}(1987)}]{anderson1987resonating}%
  \BibitemOpen
  \bibfield  {author} {\bibinfo {author} {\bibfnamefont {P.~W.}\ \bibnamefont
  {Anderson}},\ }\href
  {https://www.science.org/doi/10.1126/science.235.4793.1196} {\bibfield
  {journal} {\bibinfo  {journal} {Science}\ }\textbf {\bibinfo {volume}
  {235}},\ \bibinfo {pages} {1196} (\bibinfo {year} {1987})}\BibitemShut
  {NoStop}%
\bibitem [{\citenamefont {Lee}\ \emph {et~al.}(2006)\citenamefont {Lee},
  \citenamefont {Nagaosa},\ and\ \citenamefont {Wen}}]{lee2006doping}%
  \BibitemOpen
  \bibfield  {author} {\bibinfo {author} {\bibfnamefont {P.~A.}\ \bibnamefont
  {Lee}}, \bibinfo {author} {\bibfnamefont {N.}~\bibnamefont {Nagaosa}}, \ and\
  \bibinfo {author} {\bibfnamefont {X.-G.}\ \bibnamefont {Wen}},\ }\href
  {\doibase 10.1103/RevModPhys.78.17} {\bibfield  {journal} {\bibinfo
  {journal} {Rev. Mod. Phys.}\ }\textbf {\bibinfo {volume} {78}},\ \bibinfo
  {pages} {17} (\bibinfo {year} {2006})}\BibitemShut {NoStop}%
\bibitem [{\citenamefont {Keimer}\ \emph {et~al.}(2015)\citenamefont {Keimer},
  \citenamefont {Kivelson}, \citenamefont {Norman}, \citenamefont {Uchida},\
  and\ \citenamefont {Zaanen}}]{keimer2015quantum}%
  \BibitemOpen
  \bibfield  {author} {\bibinfo {author} {\bibfnamefont {B.}~\bibnamefont
  {Keimer}}, \bibinfo {author} {\bibfnamefont {S.~A.}\ \bibnamefont
  {Kivelson}}, \bibinfo {author} {\bibfnamefont {M.~R.}\ \bibnamefont
  {Norman}}, \bibinfo {author} {\bibfnamefont {S.}~\bibnamefont {Uchida}}, \
  and\ \bibinfo {author} {\bibfnamefont {J.}~\bibnamefont {Zaanen}},\ }\href
  {https://www.nature.com/articles/nature14165} {\bibfield  {journal} {\bibinfo
   {journal} {Nature}\ }\textbf {\bibinfo {volume} {518}},\ \bibinfo {pages}
  {179} (\bibinfo {year} {2015})}\BibitemShut {NoStop}%
\bibitem [{\citenamefont {Proust}\ and\ \citenamefont
  {Taillefer}(2019)}]{proust2019remarkable}%
  \BibitemOpen
  \bibfield  {author} {\bibinfo {author} {\bibfnamefont {C.}~\bibnamefont
  {Proust}}\ and\ \bibinfo {author} {\bibfnamefont {L.}~\bibnamefont
  {Taillefer}},\ }\href
  {https://www.annualreviews.org/doi/abs/10.1146/annurev-conmatphys-031218-013210}
  {\bibfield  {journal} {\bibinfo  {journal} {Annual Review of Condensed Matter
  Physics}\ }\textbf {\bibinfo {volume} {10}},\ \bibinfo {pages} {409}
  (\bibinfo {year} {2019})}\BibitemShut {NoStop}%
\bibitem [{\citenamefont {Wen}\ and\ \citenamefont
  {Lee}(1996)}]{wen1996theory}%
  \BibitemOpen
  \bibfield  {author} {\bibinfo {author} {\bibfnamefont {X.-G.}\ \bibnamefont
  {Wen}}\ and\ \bibinfo {author} {\bibfnamefont {P.~A.}\ \bibnamefont {Lee}},\
  }\href {\doibase 10.1103/PhysRevLett.76.503} {\bibfield  {journal} {\bibinfo
  {journal} {Phys. Rev. Lett.}\ }\textbf {\bibinfo {volume} {76}},\ \bibinfo
  {pages} {503} (\bibinfo {year} {1996})}\BibitemShut {NoStop}%
\bibitem [{\citenamefont {Fradkin}\ \emph {et~al.}(2015)\citenamefont
  {Fradkin}, \citenamefont {Kivelson},\ and\ \citenamefont
  {Tranquada}}]{fradkin2015colloquium}%
  \BibitemOpen
  \bibfield  {author} {\bibinfo {author} {\bibfnamefont {E.}~\bibnamefont
  {Fradkin}}, \bibinfo {author} {\bibfnamefont {S.~A.}\ \bibnamefont
  {Kivelson}}, \ and\ \bibinfo {author} {\bibfnamefont {J.~M.}\ \bibnamefont
  {Tranquada}},\ }\href {\doibase 10.1103/RevModPhys.87.457} {\bibfield
  {journal} {\bibinfo  {journal} {Rev. Mod. Phys.}\ }\textbf {\bibinfo {volume}
  {87}},\ \bibinfo {pages} {457} (\bibinfo {year} {2015})}\BibitemShut
  {NoStop}%
\bibitem [{\citenamefont {Senthil}\ and\ \citenamefont
  {Lee}(2005)}]{senthil2005cup}%
  \BibitemOpen
  \bibfield  {author} {\bibinfo {author} {\bibfnamefont {T.}~\bibnamefont
  {Senthil}}\ and\ \bibinfo {author} {\bibfnamefont {P.~A.}\ \bibnamefont
  {Lee}},\ }\href {\doibase 10.1103/PhysRevB.71.174515} {\bibfield  {journal}
  {\bibinfo  {journal} {Phys. Rev. B}\ }\textbf {\bibinfo {volume} {71}},\
  \bibinfo {pages} {174515} (\bibinfo {year} {2005})}\BibitemShut {NoStop}%
\bibitem [{\citenamefont {Balents}\ and\ \citenamefont
  {Sachdev}(2007)}]{balents2007dual}%
  \BibitemOpen
  \bibfield  {author} {\bibinfo {author} {\bibfnamefont {L.}~\bibnamefont
  {Balents}}\ and\ \bibinfo {author} {\bibfnamefont {S.}~\bibnamefont
  {Sachdev}},\ }\href {\doibase 10.1016/j.aop.2007.02.001} {\bibfield
  {journal} {\bibinfo  {journal} {Annals of Physics}\ }\textbf {\bibinfo
  {volume} {322}},\ \bibinfo {pages} {2635–2664} (\bibinfo {year}
  {2007})}\BibitemShut {NoStop}%
\bibitem [{\citenamefont {Sachdev}(2010)}]{sachdev2010}%
  \BibitemOpen
  \bibfield  {author} {\bibinfo {author} {\bibfnamefont {S.}~\bibnamefont
  {Sachdev}},\ }\href {\doibase 10.1103/PhysRevLett.105.151602} {\bibfield
  {journal} {\bibinfo  {journal} {Phys. Rev. Lett.}\ }\textbf {\bibinfo
  {volume} {105}},\ \bibinfo {pages} {151602} (\bibinfo {year}
  {2010})}\BibitemShut {NoStop}%
\bibitem [{\citenamefont {Kalmeyer}\ and\ \citenamefont
  {Laughlin}(1987)}]{kalmeyer1987equivalence}%
  \BibitemOpen
  \bibfield  {author} {\bibinfo {author} {\bibfnamefont {V.}~\bibnamefont
  {Kalmeyer}}\ and\ \bibinfo {author} {\bibfnamefont {R.~B.}\ \bibnamefont
  {Laughlin}},\ }\href {\doibase 10.1103/PhysRevLett.59.2095} {\bibfield
  {journal} {\bibinfo  {journal} {Phys. Rev. Lett.}\ }\textbf {\bibinfo
  {volume} {59}},\ \bibinfo {pages} {2095} (\bibinfo {year}
  {1987})}\BibitemShut {NoStop}%
\bibitem [{\citenamefont {Laughlin}(1988)}]{laughlin1988}%
  \BibitemOpen
  \bibfield  {author} {\bibinfo {author} {\bibfnamefont {R.~B.}\ \bibnamefont
  {Laughlin}},\ }\href {\doibase 10.1103/PhysRevLett.60.2677} {\bibfield
  {journal} {\bibinfo  {journal} {Phys. Rev. Lett.}\ }\textbf {\bibinfo
  {volume} {60}},\ \bibinfo {pages} {2677} (\bibinfo {year}
  {1988})}\BibitemShut {NoStop}%
\bibitem [{\citenamefont {Wen}\ \emph {et~al.}(1989)\citenamefont {Wen},
  \citenamefont {Wilczek},\ and\ \citenamefont {Zee}}]{wen1989chiral}%
  \BibitemOpen
  \bibfield  {author} {\bibinfo {author} {\bibfnamefont {X.~G.}\ \bibnamefont
  {Wen}}, \bibinfo {author} {\bibfnamefont {F.}~\bibnamefont {Wilczek}}, \ and\
  \bibinfo {author} {\bibfnamefont {A.}~\bibnamefont {Zee}},\ }\href {\doibase
  10.1103/PhysRevB.39.11413} {\bibfield  {journal} {\bibinfo  {journal} {Phys.
  Rev. B}\ }\textbf {\bibinfo {volume} {39}},\ \bibinfo {pages} {11413}
  (\bibinfo {year} {1989})}\BibitemShut {NoStop}%
\bibitem [{\citenamefont {Lee}\ and\ \citenamefont {Fisher}(1989)}]{lee1989}%
  \BibitemOpen
  \bibfield  {author} {\bibinfo {author} {\bibfnamefont {D.-H.}\ \bibnamefont
  {Lee}}\ and\ \bibinfo {author} {\bibfnamefont {M.~P.~A.}\ \bibnamefont
  {Fisher}},\ }\href {\doibase 10.1103/PhysRevLett.63.903} {\bibfield
  {journal} {\bibinfo  {journal} {Phys. Rev. Lett.}\ }\textbf {\bibinfo
  {volume} {63}},\ \bibinfo {pages} {903} (\bibinfo {year} {1989})}\BibitemShut
  {NoStop}%
\bibitem [{\citenamefont {He}\ \emph {et~al.}(2014)\citenamefont {He},
  \citenamefont {Sheng},\ and\ \citenamefont {Chen}}]{he2014}%
  \BibitemOpen
  \bibfield  {author} {\bibinfo {author} {\bibfnamefont {Y.-C.}\ \bibnamefont
  {He}}, \bibinfo {author} {\bibfnamefont {D.~N.}\ \bibnamefont {Sheng}}, \
  and\ \bibinfo {author} {\bibfnamefont {Y.}~\bibnamefont {Chen}},\ }\href
  {\doibase 10.1103/PhysRevLett.112.137202} {\bibfield  {journal} {\bibinfo
  {journal} {Phys. Rev. Lett.}\ }\textbf {\bibinfo {volume} {112}},\ \bibinfo
  {pages} {137202} (\bibinfo {year} {2014})}\BibitemShut {NoStop}%
\bibitem [{\citenamefont {Gong}\ \emph {et~al.}(2014)\citenamefont {Gong},
  \citenamefont {Zhu},\ and\ \citenamefont {Sheng}}]{gong2014}%
  \BibitemOpen
  \bibfield  {author} {\bibinfo {author} {\bibfnamefont {S.-S.}\ \bibnamefont
  {Gong}}, \bibinfo {author} {\bibfnamefont {W.}~\bibnamefont {Zhu}}, \ and\
  \bibinfo {author} {\bibfnamefont {D.~N.}\ \bibnamefont {Sheng}},\ }\href
  {\doibase 10.1038/srep06317} {\bibfield  {journal} {\bibinfo  {journal}
  {Scientific Reports}\ }\textbf {\bibinfo {volume} {4}},\ \bibinfo {pages}
  {6317} (\bibinfo {year} {2014})}\BibitemShut {NoStop}%
\bibitem [{\citenamefont {Bauer}\ \emph {et~al.}(2014)\citenamefont {Bauer},
  \citenamefont {Cincio}, \citenamefont {Keller}, \citenamefont {Dolfi},
  \citenamefont {Vidal}, \citenamefont {Trebst},\ and\ \citenamefont
  {Ludwig}}]{bauer2014}%
  \BibitemOpen
  \bibfield  {author} {\bibinfo {author} {\bibfnamefont {B.}~\bibnamefont
  {Bauer}}, \bibinfo {author} {\bibfnamefont {L.}~\bibnamefont {Cincio}},
  \bibinfo {author} {\bibfnamefont {B.~P.}\ \bibnamefont {Keller}}, \bibinfo
  {author} {\bibfnamefont {M.}~\bibnamefont {Dolfi}}, \bibinfo {author}
  {\bibfnamefont {G.}~\bibnamefont {Vidal}}, \bibinfo {author} {\bibfnamefont
  {S.}~\bibnamefont {Trebst}}, \ and\ \bibinfo {author} {\bibfnamefont
  {A.~W.~W.}\ \bibnamefont {Ludwig}},\ }\href {\doibase 10.1038/ncomms6137}
  {\bibfield  {journal} {\bibinfo  {journal} {Nature Communications}\ }\textbf
  {\bibinfo {volume} {5}},\ \bibinfo {pages} {5137} (\bibinfo {year}
  {2014})}\BibitemShut {NoStop}%
\bibitem [{\citenamefont {Gong}\ \emph {et~al.}(2015)\citenamefont {Gong},
  \citenamefont {Zhu}, \citenamefont {Balents},\ and\ \citenamefont
  {Sheng}}]{gong2015}%
  \BibitemOpen
  \bibfield  {author} {\bibinfo {author} {\bibfnamefont {S.-S.}\ \bibnamefont
  {Gong}}, \bibinfo {author} {\bibfnamefont {W.}~\bibnamefont {Zhu}}, \bibinfo
  {author} {\bibfnamefont {L.}~\bibnamefont {Balents}}, \ and\ \bibinfo
  {author} {\bibfnamefont {D.~N.}\ \bibnamefont {Sheng}},\ }\href {\doibase
  10.1103/PhysRevB.91.075112} {\bibfield  {journal} {\bibinfo  {journal} {Phys.
  Rev. B}\ }\textbf {\bibinfo {volume} {91}},\ \bibinfo {pages} {075112}
  (\bibinfo {year} {2015})}\BibitemShut {NoStop}%
\bibitem [{\citenamefont {Szasz}\ \emph {et~al.}(2020)\citenamefont {Szasz},
  \citenamefont {Motruk}, \citenamefont {Zaletel},\ and\ \citenamefont
  {Moore}}]{szasz2020chiral}%
  \BibitemOpen
  \bibfield  {author} {\bibinfo {author} {\bibfnamefont {A.}~\bibnamefont
  {Szasz}}, \bibinfo {author} {\bibfnamefont {J.}~\bibnamefont {Motruk}},
  \bibinfo {author} {\bibfnamefont {M.~P.}\ \bibnamefont {Zaletel}}, \ and\
  \bibinfo {author} {\bibfnamefont {J.~E.}\ \bibnamefont {Moore}},\ }\href
  {\doibase 10.1103/PhysRevX.10.021042} {\bibfield  {journal} {\bibinfo
  {journal} {Phys. Rev. X}\ }\textbf {\bibinfo {volume} {10}},\ \bibinfo
  {pages} {021042} (\bibinfo {year} {2020})}\BibitemShut {NoStop}%
\bibitem [{\citenamefont {Chen}\ \emph {et~al.}(2022)\citenamefont {Chen},
  \citenamefont {Chen}, \citenamefont {Gong}, \citenamefont {Sheng},
  \citenamefont {Li},\ and\ \citenamefont {Weichselbaum}}]{chen2021quantum}%
  \BibitemOpen
  \bibfield  {author} {\bibinfo {author} {\bibfnamefont {B.-B.}\ \bibnamefont
  {Chen}}, \bibinfo {author} {\bibfnamefont {Z.}~\bibnamefont {Chen}}, \bibinfo
  {author} {\bibfnamefont {S.-S.}\ \bibnamefont {Gong}}, \bibinfo {author}
  {\bibfnamefont {D.}~\bibnamefont {Sheng}}, \bibinfo {author} {\bibfnamefont
  {W.}~\bibnamefont {Li}}, \ and\ \bibinfo {author} {\bibfnamefont
  {A.}~\bibnamefont {Weichselbaum}},\ }\href@noop {} {\bibfield  {journal}
  {\bibinfo  {journal} {Physical Review B}\ }\textbf {\bibinfo {volume}
  {106}},\ \bibinfo {pages} {094420} (\bibinfo {year} {2022})}\BibitemShut
  {NoStop}%
\bibitem [{\citenamefont {Wietek}\ \emph {et~al.}(2021)\citenamefont {Wietek},
  \citenamefont {Rossi}, \citenamefont {\ifmmode~\check{S}\else
  \v{S}\fi{}imkovic}, \citenamefont {Klett}, \citenamefont {Hansmann},
  \citenamefont {Ferrero}, \citenamefont {Stoudenmire}, \citenamefont
  {Sch\"afer},\ and\ \citenamefont {Georges}}]{wietek2021}%
  \BibitemOpen
  \bibfield  {author} {\bibinfo {author} {\bibfnamefont {A.}~\bibnamefont
  {Wietek}}, \bibinfo {author} {\bibfnamefont {R.}~\bibnamefont {Rossi}},
  \bibinfo {author} {\bibfnamefont {F.}~\bibnamefont {\ifmmode~\check{S}\else
  \v{S}\fi{}imkovic}}, \bibinfo {author} {\bibfnamefont {M.}~\bibnamefont
  {Klett}}, \bibinfo {author} {\bibfnamefont {P.}~\bibnamefont {Hansmann}},
  \bibinfo {author} {\bibfnamefont {M.}~\bibnamefont {Ferrero}}, \bibinfo
  {author} {\bibfnamefont {E.~M.}\ \bibnamefont {Stoudenmire}}, \bibinfo
  {author} {\bibfnamefont {T.}~\bibnamefont {Sch\"afer}}, \ and\ \bibinfo
  {author} {\bibfnamefont {A.}~\bibnamefont {Georges}},\ }\href {\doibase
  10.1103/PhysRevX.11.041013} {\bibfield  {journal} {\bibinfo  {journal} {Phys.
  Rev. X}\ }\textbf {\bibinfo {volume} {11}},\ \bibinfo {pages} {041013}
  (\bibinfo {year} {2021})}\BibitemShut {NoStop}%
\bibitem [{\citenamefont {Jiang}\ \emph {et~al.}(2017)\citenamefont {Jiang},
  \citenamefont {Devereaux},\ and\ \citenamefont {Kivelson}}]{jiang2017holon}%
  \BibitemOpen
  \bibfield  {author} {\bibinfo {author} {\bibfnamefont {H.-C.}\ \bibnamefont
  {Jiang}}, \bibinfo {author} {\bibfnamefont {T.}~\bibnamefont {Devereaux}}, \
  and\ \bibinfo {author} {\bibfnamefont {S.~A.}\ \bibnamefont {Kivelson}},\
  }\href {\doibase 10.1103/PhysRevLett.119.067002} {\bibfield  {journal}
  {\bibinfo  {journal} {Phys. Rev. Lett.}\ }\textbf {\bibinfo {volume} {119}},\
  \bibinfo {pages} {067002} (\bibinfo {year} {2017})}\BibitemShut {NoStop}%
\bibitem [{\citenamefont {Peng}\ \emph
  {et~al.}(2021{\natexlab{a}})\citenamefont {Peng}, \citenamefont {Jiang},
  \citenamefont {Sheng},\ and\ \citenamefont {Jiang}}]{peng2021doping}%
  \BibitemOpen
  \bibfield  {author} {\bibinfo {author} {\bibfnamefont {C.}~\bibnamefont
  {Peng}}, \bibinfo {author} {\bibfnamefont {Y.-F.}\ \bibnamefont {Jiang}},
  \bibinfo {author} {\bibfnamefont {D.-N.}\ \bibnamefont {Sheng}}, \ and\
  \bibinfo {author} {\bibfnamefont {H.-C.}\ \bibnamefont {Jiang}},\ }\href
  {https://onlinelibrary.wiley.com/doi/abs/10.1002/qute.202000126} {\bibfield
  {journal} {\bibinfo  {journal} {Advanced Quantum Technologies}\ }\textbf
  {\bibinfo {volume} {4}},\ \bibinfo {pages} {2000126} (\bibinfo {year}
  {2021}{\natexlab{a}})}\BibitemShut {NoStop}%
\bibitem [{\citenamefont {Zhu}\ \emph {et~al.}(2022)\citenamefont {Zhu},
  \citenamefont {Sheng},\ and\ \citenamefont {Vishwanath}}]{zhu2022doped}%
  \BibitemOpen
  \bibfield  {author} {\bibinfo {author} {\bibfnamefont {Z.}~\bibnamefont
  {Zhu}}, \bibinfo {author} {\bibfnamefont {D.~N.}\ \bibnamefont {Sheng}}, \
  and\ \bibinfo {author} {\bibfnamefont {A.}~\bibnamefont {Vishwanath}},\
  }\href {\doibase 10.1103/PhysRevB.105.205110} {\bibfield  {journal} {\bibinfo
   {journal} {Phys. Rev. B}\ }\textbf {\bibinfo {volume} {105}},\ \bibinfo
  {pages} {205110} (\bibinfo {year} {2022})}\BibitemShut {NoStop}%
\bibitem [{\citenamefont {Song}\ \emph {et~al.}(2021)\citenamefont {Song},
  \citenamefont {Vishwanath},\ and\ \citenamefont {Zhang}}]{song2021doping}%
  \BibitemOpen
  \bibfield  {author} {\bibinfo {author} {\bibfnamefont {X.-Y.}\ \bibnamefont
  {Song}}, \bibinfo {author} {\bibfnamefont {A.}~\bibnamefont {Vishwanath}}, \
  and\ \bibinfo {author} {\bibfnamefont {Y.-H.}\ \bibnamefont {Zhang}},\ }\href
  {\doibase 10.1103/PhysRevB.103.165138} {\bibfield  {journal} {\bibinfo
  {journal} {Phys. Rev. B}\ }\textbf {\bibinfo {volume} {103}},\ \bibinfo
  {pages} {165138} (\bibinfo {year} {2021})}\BibitemShut {NoStop}%
\bibitem [{\citenamefont {Baskaran}(2003)}]{baskaran2003electronic}%
  \BibitemOpen
  \bibfield  {author} {\bibinfo {author} {\bibfnamefont {G.}~\bibnamefont
  {Baskaran}},\ }\href {\doibase 10.1103/PhysRevLett.91.097003} {\bibfield
  {journal} {\bibinfo  {journal} {Phys. Rev. Lett.}\ }\textbf {\bibinfo
  {volume} {91}},\ \bibinfo {pages} {097003} (\bibinfo {year}
  {2003})}\BibitemShut {NoStop}%
\bibitem [{\citenamefont {Kumar}\ and\ \citenamefont
  {Shastry}(2003)}]{kumar2003superconductivity}%
  \BibitemOpen
  \bibfield  {author} {\bibinfo {author} {\bibfnamefont {B.}~\bibnamefont
  {Kumar}}\ and\ \bibinfo {author} {\bibfnamefont {B.~S.}\ \bibnamefont
  {Shastry}},\ }\href
  {https://journals.aps.org/prb/abstract/10.1103/PhysRevB.68.104508} {\bibfield
   {journal} {\bibinfo  {journal} {Physical Review B}\ }\textbf {\bibinfo
  {volume} {68}},\ \bibinfo {pages} {104508} (\bibinfo {year}
  {2003})}\BibitemShut {NoStop}%
\bibitem [{\citenamefont {Wang}\ \emph {et~al.}(2004)\citenamefont {Wang},
  \citenamefont {Lee},\ and\ \citenamefont {Lee}}]{wang2004doped}%
  \BibitemOpen
  \bibfield  {author} {\bibinfo {author} {\bibfnamefont {Q.-H.}\ \bibnamefont
  {Wang}}, \bibinfo {author} {\bibfnamefont {D.-H.}\ \bibnamefont {Lee}}, \
  and\ \bibinfo {author} {\bibfnamefont {P.~A.}\ \bibnamefont {Lee}},\ }\href
  {https://journals.aps.org/prb/abstract/10.1103/PhysRevB.69.092504} {\bibfield
   {journal} {\bibinfo  {journal} {Physical Review B}\ }\textbf {\bibinfo
  {volume} {69}},\ \bibinfo {pages} {092504} (\bibinfo {year}
  {2004})}\BibitemShut {NoStop}%
\bibitem [{\citenamefont {{Watanabe}}\ \emph {et~al.}(2004)\citenamefont
  {{Watanabe}}, \citenamefont {{Yokoyama}}, \citenamefont {{Tanaka}},
  \citenamefont {{Inoue}},\ and\ \citenamefont {{Ogata}}}]{watanabe2004}%
  \BibitemOpen
  \bibfield  {author} {\bibinfo {author} {\bibfnamefont {T.}~\bibnamefont
  {{Watanabe}}}, \bibinfo {author} {\bibfnamefont {H.}~\bibnamefont
  {{Yokoyama}}}, \bibinfo {author} {\bibfnamefont {Y.}~\bibnamefont
  {{Tanaka}}}, \bibinfo {author} {\bibfnamefont {J.-i.}\ \bibnamefont
  {{Inoue}}}, \ and\ \bibinfo {author} {\bibfnamefont {M.}~\bibnamefont
  {{Ogata}}},\ }\href {\doibase 10.1143/JPSJ.73.3404} {\bibfield  {journal}
  {\bibinfo  {journal} {Journal of the Physical Society of Japan}\ }\textbf
  {\bibinfo {volume} {73}},\ \bibinfo {pages} {3404} (\bibinfo {year}
  {2004})}\BibitemShut {NoStop}%
\bibitem [{\citenamefont {Braunecker}\ \emph {et~al.}(2005)\citenamefont
  {Braunecker}, \citenamefont {Lee},\ and\ \citenamefont
  {Wang}}]{braunecker2005edge}%
  \BibitemOpen
  \bibfield  {author} {\bibinfo {author} {\bibfnamefont {B.}~\bibnamefont
  {Braunecker}}, \bibinfo {author} {\bibfnamefont {P.~A.}\ \bibnamefont {Lee}},
  \ and\ \bibinfo {author} {\bibfnamefont {Z.}~\bibnamefont {Wang}},\ }\href
  {\doibase 10.1103/PhysRevLett.95.017004} {\bibfield  {journal} {\bibinfo
  {journal} {Phys. Rev. Lett.}\ }\textbf {\bibinfo {volume} {95}},\ \bibinfo
  {pages} {017004} (\bibinfo {year} {2005})}\BibitemShut {NoStop}%
\bibitem [{\citenamefont {Weber}\ \emph {et~al.}(2006)\citenamefont {Weber},
  \citenamefont {L\"auchli}, \citenamefont {Mila},\ and\ \citenamefont
  {Giamarchi}}]{weber2006magnetism}%
  \BibitemOpen
  \bibfield  {author} {\bibinfo {author} {\bibfnamefont {C.}~\bibnamefont
  {Weber}}, \bibinfo {author} {\bibfnamefont {A.}~\bibnamefont {L\"auchli}},
  \bibinfo {author} {\bibfnamefont {F.}~\bibnamefont {Mila}}, \ and\ \bibinfo
  {author} {\bibfnamefont {T.}~\bibnamefont {Giamarchi}},\ }\href {\doibase
  10.1103/PhysRevB.73.014519} {\bibfield  {journal} {\bibinfo  {journal} {Phys.
  Rev. B}\ }\textbf {\bibinfo {volume} {73}},\ \bibinfo {pages} {014519}
  (\bibinfo {year} {2006})}\BibitemShut {NoStop}%
\bibitem [{\citenamefont {Gan}\ \emph {et~al.}(2006)\citenamefont {Gan},
  \citenamefont {Chen},\ and\ \citenamefont {Zhang}}]{gan2006superconducting}%
  \BibitemOpen
  \bibfield  {author} {\bibinfo {author} {\bibfnamefont {J.~Y.}\ \bibnamefont
  {Gan}}, \bibinfo {author} {\bibfnamefont {Y.}~\bibnamefont {Chen}}, \ and\
  \bibinfo {author} {\bibfnamefont {F.~C.}\ \bibnamefont {Zhang}},\ }\href
  {\doibase 10.1103/PhysRevB.74.094515} {\bibfield  {journal} {\bibinfo
  {journal} {Phys. Rev. B}\ }\textbf {\bibinfo {volume} {74}},\ \bibinfo
  {pages} {094515} (\bibinfo {year} {2006})}\BibitemShut {NoStop}%
\bibitem [{\citenamefont {Zhou}\ and\ \citenamefont
  {Wang}(2008)}]{zhou2008nodal}%
  \BibitemOpen
  \bibfield  {author} {\bibinfo {author} {\bibfnamefont {S.}~\bibnamefont
  {Zhou}}\ and\ \bibinfo {author} {\bibfnamefont {Z.}~\bibnamefont {Wang}},\
  }\href {https://journals.aps.org/prl/abstract/10.1103/PhysRevLett.100.217002}
  {\bibfield  {journal} {\bibinfo  {journal} {Physical review letters}\
  }\textbf {\bibinfo {volume} {100}},\ \bibinfo {pages} {217002} (\bibinfo
  {year} {2008})}\BibitemShut {NoStop}%
\bibitem [{\citenamefont {Chen}\ \emph {et~al.}(2013)\citenamefont {Chen},
  \citenamefont {Meng}, \citenamefont {Yu}, \citenamefont {Yang}, \citenamefont
  {Jarrell},\ and\ \citenamefont {Moreno}}]{chen2013unconventional}%
  \BibitemOpen
  \bibfield  {author} {\bibinfo {author} {\bibfnamefont {K.~S.}\ \bibnamefont
  {Chen}}, \bibinfo {author} {\bibfnamefont {Z.~Y.}\ \bibnamefont {Meng}},
  \bibinfo {author} {\bibfnamefont {U.}~\bibnamefont {Yu}}, \bibinfo {author}
  {\bibfnamefont {S.}~\bibnamefont {Yang}}, \bibinfo {author} {\bibfnamefont
  {M.}~\bibnamefont {Jarrell}}, \ and\ \bibinfo {author} {\bibfnamefont
  {J.}~\bibnamefont {Moreno}},\ }\href
  {https://journals.aps.org/prb/abstract/10.1103/PhysRevB.88.041103} {\bibfield
   {journal} {\bibinfo  {journal} {Physical Review B}\ }\textbf {\bibinfo
  {volume} {88}},\ \bibinfo {pages} {041103(R)} (\bibinfo {year}
  {2013})}\BibitemShut {NoStop}%
\bibitem [{\citenamefont {Motrunich}\ and\ \citenamefont
  {Lee}(2004)}]{motrunich2004}%
  \BibitemOpen
  \bibfield  {author} {\bibinfo {author} {\bibfnamefont {O.~I.}\ \bibnamefont
  {Motrunich}}\ and\ \bibinfo {author} {\bibfnamefont {P.~A.}\ \bibnamefont
  {Lee}},\ }\href {\doibase 10.1103/PhysRevB.69.214516} {\bibfield  {journal}
  {\bibinfo  {journal} {Phys. Rev. B}\ }\textbf {\bibinfo {volume} {69}},\
  \bibinfo {pages} {214516} (\bibinfo {year} {2004})}\BibitemShut {NoStop}%
\bibitem [{\citenamefont {Kiesel}\ \emph {et~al.}(2013)\citenamefont {Kiesel},
  \citenamefont {Platt}, \citenamefont {Hanke},\ and\ \citenamefont
  {Thomale}}]{kiesel2013}%
  \BibitemOpen
  \bibfield  {author} {\bibinfo {author} {\bibfnamefont {M.~L.}\ \bibnamefont
  {Kiesel}}, \bibinfo {author} {\bibfnamefont {C.}~\bibnamefont {Platt}},
  \bibinfo {author} {\bibfnamefont {W.}~\bibnamefont {Hanke}}, \ and\ \bibinfo
  {author} {\bibfnamefont {R.}~\bibnamefont {Thomale}},\ }\href {\doibase
  10.1103/PhysRevLett.111.097001} {\bibfield  {journal} {\bibinfo  {journal}
  {Phys. Rev. Lett.}\ }\textbf {\bibinfo {volume} {111}},\ \bibinfo {pages}
  {097001} (\bibinfo {year} {2013})}\BibitemShut {NoStop}%
\bibitem [{\citenamefont {Arovas}\ \emph {et~al.}(2022)\citenamefont {Arovas},
  \citenamefont {Berg}, \citenamefont {Kivelson},\ and\ \citenamefont
  {Raghu}}]{arovas2022hubbard}%
  \BibitemOpen
  \bibfield  {author} {\bibinfo {author} {\bibfnamefont {D.~P.}\ \bibnamefont
  {Arovas}}, \bibinfo {author} {\bibfnamefont {E.}~\bibnamefont {Berg}},
  \bibinfo {author} {\bibfnamefont {S.~A.}\ \bibnamefont {Kivelson}}, \ and\
  \bibinfo {author} {\bibfnamefont {S.}~\bibnamefont {Raghu}},\ }\href
  {https://www.annualreviews.org/doi/abs/10.1146/annurev-conmatphys-031620-102024}
  {\bibfield  {journal} {\bibinfo  {journal} {Annual Review of Condensed Matter
  Physics}\ }\textbf {\bibinfo {volume} {13}},\ \bibinfo {pages} {239}
  (\bibinfo {year} {2022})}\BibitemShut {NoStop}%
\bibitem [{\citenamefont {Gannot}\ \emph {et~al.}(2020)\citenamefont {Gannot},
  \citenamefont {Jiang},\ and\ \citenamefont {Kivelson}}]{gannot2020SU}%
  \BibitemOpen
  \bibfield  {author} {\bibinfo {author} {\bibfnamefont {Y.}~\bibnamefont
  {Gannot}}, \bibinfo {author} {\bibfnamefont {Y.-F.}\ \bibnamefont {Jiang}}, \
  and\ \bibinfo {author} {\bibfnamefont {S.~A.}\ \bibnamefont {Kivelson}},\
  }\href {\doibase 10.1103/PhysRevB.102.115136} {\bibfield  {journal} {\bibinfo
   {journal} {Phys. Rev. B}\ }\textbf {\bibinfo {volume} {102}},\ \bibinfo
  {pages} {115136} (\bibinfo {year} {2020})}\BibitemShut {NoStop}%
\bibitem [{\citenamefont {Peng}\ \emph
  {et~al.}(2021{\natexlab{b}})\citenamefont {Peng}, \citenamefont {Jiang},
  \citenamefont {Wang},\ and\ \citenamefont {Jiang}}]{peng2021gapless}%
  \BibitemOpen
  \bibfield  {author} {\bibinfo {author} {\bibfnamefont {C.}~\bibnamefont
  {Peng}}, \bibinfo {author} {\bibfnamefont {Y.-F.}\ \bibnamefont {Jiang}},
  \bibinfo {author} {\bibfnamefont {Y.}~\bibnamefont {Wang}}, \ and\ \bibinfo
  {author} {\bibfnamefont {H.-C.}\ \bibnamefont {Jiang}},\ }\href
  {https://iopscience.iop.org/article/10.1088/1367-2630/ac3a83} {\bibfield
  {journal} {\bibinfo  {journal} {New Journal of Physics}\ }\textbf {\bibinfo
  {volume} {23}},\ \bibinfo {pages} {123004} (\bibinfo {year}
  {2021}{\natexlab{b}})}\BibitemShut {NoStop}%
\bibitem [{\citenamefont {Aghaei}\ \emph {et~al.}(2020)\citenamefont {Aghaei},
  \citenamefont {Bauer}, \citenamefont {Shtengel},\ and\ \citenamefont
  {Mishmash}}]{aghaei2020efficient}%
  \BibitemOpen
  \bibfield  {author} {\bibinfo {author} {\bibfnamefont {A.~M.}\ \bibnamefont
  {Aghaei}}, \bibinfo {author} {\bibfnamefont {B.}~\bibnamefont {Bauer}},
  \bibinfo {author} {\bibfnamefont {K.}~\bibnamefont {Shtengel}}, \ and\
  \bibinfo {author} {\bibfnamefont {R.~V.}\ \bibnamefont {Mishmash}},\
  }\href@noop {} {\bibfield  {journal} {\bibinfo  {journal} {arXiv preprint
  arXiv:2009.12435}\ } (\bibinfo {year} {2020})}\BibitemShut {NoStop}%
\bibitem [{\citenamefont {Jiang}\ \emph
  {et~al.}(2021{\natexlab{a}})\citenamefont {Jiang}, \citenamefont {Yao},\ and\
  \citenamefont {Yang}}]{jiang2021possible}%
  \BibitemOpen
  \bibfield  {author} {\bibinfo {author} {\bibfnamefont {Y.-F.}\ \bibnamefont
  {Jiang}}, \bibinfo {author} {\bibfnamefont {H.}~\bibnamefont {Yao}}, \ and\
  \bibinfo {author} {\bibfnamefont {F.}~\bibnamefont {Yang}},\ }\href {\doibase
  10.1103/PhysRevLett.127.187003} {\bibfield  {journal} {\bibinfo  {journal}
  {Phys. Rev. Lett.}\ }\textbf {\bibinfo {volume} {127}},\ \bibinfo {pages}
  {187003} (\bibinfo {year} {2021}{\natexlab{a}})}\BibitemShut {NoStop}%
\bibitem [{\citenamefont {Jiang}\ and\ \citenamefont
  {Jiang}(2020)}]{jiang2020topological}%
  \BibitemOpen
  \bibfield  {author} {\bibinfo {author} {\bibfnamefont {Y.-F.}\ \bibnamefont
  {Jiang}}\ and\ \bibinfo {author} {\bibfnamefont {H.-C.}\ \bibnamefont
  {Jiang}},\ }\href {\doibase 10.1103/PhysRevLett.125.157002} {\bibfield
  {journal} {\bibinfo  {journal} {Phys. Rev. Lett.}\ }\textbf {\bibinfo
  {volume} {125}},\ \bibinfo {pages} {157002} (\bibinfo {year}
  {2020})}\BibitemShut {NoStop}%
\bibitem [{\citenamefont {Huang}\ and\ \citenamefont
  {Sheng}(2022)}]{huang2021topological}%
  \BibitemOpen
  \bibfield  {author} {\bibinfo {author} {\bibfnamefont {Y.}~\bibnamefont
  {Huang}}\ and\ \bibinfo {author} {\bibfnamefont {D.~N.}\ \bibnamefont
  {Sheng}},\ }\href {\doibase 10.1103/PhysRevX.12.031009} {\bibfield  {journal}
  {\bibinfo  {journal} {Phys. Rev. X}\ }\textbf {\bibinfo {volume} {12}},\
  \bibinfo {pages} {031009} (\bibinfo {year} {2022})}\BibitemShut {NoStop}%
\bibitem [{\citenamefont {Gu}\ \emph {et~al.}(2013)\citenamefont {Gu},
  \citenamefont {Jiang}, \citenamefont {Sheng}, \citenamefont {Yao},
  \citenamefont {Balents},\ and\ \citenamefont {Wen}}]{Gu2013Time}%
  \BibitemOpen
  \bibfield  {author} {\bibinfo {author} {\bibfnamefont {Z.-C.}\ \bibnamefont
  {Gu}}, \bibinfo {author} {\bibfnamefont {H.-C.}\ \bibnamefont {Jiang}},
  \bibinfo {author} {\bibfnamefont {D.~N.}\ \bibnamefont {Sheng}}, \bibinfo
  {author} {\bibfnamefont {H.}~\bibnamefont {Yao}}, \bibinfo {author}
  {\bibfnamefont {L.}~\bibnamefont {Balents}}, \ and\ \bibinfo {author}
  {\bibfnamefont {X.-G.}\ \bibnamefont {Wen}},\ }\href {\doibase
  10.1103/PhysRevB.88.155112} {\bibfield  {journal} {\bibinfo  {journal} {Phys.
  Rev. B}\ }\textbf {\bibinfo {volume} {88}},\ \bibinfo {pages} {155112}
  (\bibinfo {year} {2013})}\BibitemShut {NoStop}%
\bibitem [{\citenamefont {Xu}\ and\ \citenamefont
  {Balents}(2018)}]{xu2018topological}%
  \BibitemOpen
  \bibfield  {author} {\bibinfo {author} {\bibfnamefont {C.}~\bibnamefont
  {Xu}}\ and\ \bibinfo {author} {\bibfnamefont {L.}~\bibnamefont {Balents}},\
  }\href {\doibase 10.1103/PhysRevLett.121.087001} {\bibfield  {journal}
  {\bibinfo  {journal} {Phys. Rev. Lett.}\ }\textbf {\bibinfo {volume} {121}},\
  \bibinfo {pages} {087001} (\bibinfo {year} {2018})}\BibitemShut {NoStop}%
\bibitem [{\citenamefont {Zhou}\ and\ \citenamefont
  {Zhang}(2022)}]{zhou2022chiral}%
  \BibitemOpen
  \bibfield  {author} {\bibinfo {author} {\bibfnamefont {B.}~\bibnamefont
  {Zhou}}\ and\ \bibinfo {author} {\bibfnamefont {Y.-H.}\ \bibnamefont
  {Zhang}},\ }\href@noop {} {\bibfield  {journal} {\bibinfo  {journal} {arXiv
  preprint arXiv:2209.10023}\ } (\bibinfo {year} {2022})}\BibitemShut {NoStop}%
\bibitem [{\citenamefont {B{\'e}langer}\ \emph {et~al.}(2022)\citenamefont
  {B{\'e}langer}, \citenamefont {Fournier},\ and\ \citenamefont
  {S{\'e}n{\'e}chal}}]{belanger2022superconductivity}%
  \BibitemOpen
  \bibfield  {author} {\bibinfo {author} {\bibfnamefont {M.}~\bibnamefont
  {B{\'e}langer}}, \bibinfo {author} {\bibfnamefont {J.}~\bibnamefont
  {Fournier}}, \ and\ \bibinfo {author} {\bibfnamefont {D.}~\bibnamefont
  {S{\'e}n{\'e}chal}},\ }\href@noop {} {\bibfield  {journal} {\bibinfo
  {journal} {Physical Review B}\ }\textbf {\bibinfo {volume} {106}},\ \bibinfo
  {pages} {235135} (\bibinfo {year} {2022})}\BibitemShut {NoStop}%
\bibitem [{\citenamefont {Jiang}(2021)}]{jiang2021superconductivity}%
  \BibitemOpen
  \bibfield  {author} {\bibinfo {author} {\bibfnamefont {H.-C.}\ \bibnamefont
  {Jiang}},\ }\href {https://www.nature.com/articles/s41535-021-00375-w}
  {\bibfield  {journal} {\bibinfo  {journal} {npj Quantum Materials}\ }\textbf
  {\bibinfo {volume} {6}},\ \bibinfo {pages} {1} (\bibinfo {year}
  {2021})}\BibitemShut {NoStop}%
\bibitem [{\citenamefont {Kurosaki}\ \emph {et~al.}(2005)\citenamefont
  {Kurosaki}, \citenamefont {Shimizu}, \citenamefont {Miyagawa}, \citenamefont
  {Kanoda},\ and\ \citenamefont {Saito}}]{kurosaki2005mott}%
  \BibitemOpen
  \bibfield  {author} {\bibinfo {author} {\bibfnamefont {Y.}~\bibnamefont
  {Kurosaki}}, \bibinfo {author} {\bibfnamefont {Y.}~\bibnamefont {Shimizu}},
  \bibinfo {author} {\bibfnamefont {K.}~\bibnamefont {Miyagawa}}, \bibinfo
  {author} {\bibfnamefont {K.}~\bibnamefont {Kanoda}}, \ and\ \bibinfo {author}
  {\bibfnamefont {G.}~\bibnamefont {Saito}},\ }\href {\doibase
  10.1103/PhysRevLett.95.177001} {\bibfield  {journal} {\bibinfo  {journal}
  {Phys. Rev. Lett.}\ }\textbf {\bibinfo {volume} {95}},\ \bibinfo {pages}
  {177001} (\bibinfo {year} {2005})}\BibitemShut {NoStop}%
\bibitem [{\citenamefont {Itou}\ \emph {et~al.}(2007)\citenamefont {Itou},
  \citenamefont {Oyamada}, \citenamefont {Maegawa}, \citenamefont {Tamura},\
  and\ \citenamefont {Kato}}]{itou2007spin}%
  \BibitemOpen
  \bibfield  {author} {\bibinfo {author} {\bibfnamefont {T.}~\bibnamefont
  {Itou}}, \bibinfo {author} {\bibfnamefont {A.}~\bibnamefont {Oyamada}},
  \bibinfo {author} {\bibfnamefont {S.}~\bibnamefont {Maegawa}}, \bibinfo
  {author} {\bibfnamefont {M.}~\bibnamefont {Tamura}}, \ and\ \bibinfo {author}
  {\bibfnamefont {R.}~\bibnamefont {Kato}},\ }\href
  {https://iopscience.iop.org/article/10.1088/0953-8984/19/14/145247}
  {\bibfield  {journal} {\bibinfo  {journal} {Journal of Physics: Condensed
  Matter}\ }\textbf {\bibinfo {volume} {19}},\ \bibinfo {pages} {145247}
  (\bibinfo {year} {2007})}\BibitemShut {NoStop}%
\bibitem [{\citenamefont {Yamashita}\ \emph {et~al.}(2008)\citenamefont
  {Yamashita}, \citenamefont {Nakazawa}, \citenamefont {Oguni}, \citenamefont
  {Oshima}, \citenamefont {Nojiri}, \citenamefont {Shimizu}, \citenamefont
  {Miyagawa},\ and\ \citenamefont {Kanoda}}]{yamashita2008thermodynamic}%
  \BibitemOpen
  \bibfield  {author} {\bibinfo {author} {\bibfnamefont {S.}~\bibnamefont
  {Yamashita}}, \bibinfo {author} {\bibfnamefont {Y.}~\bibnamefont {Nakazawa}},
  \bibinfo {author} {\bibfnamefont {M.}~\bibnamefont {Oguni}}, \bibinfo
  {author} {\bibfnamefont {Y.}~\bibnamefont {Oshima}}, \bibinfo {author}
  {\bibfnamefont {H.}~\bibnamefont {Nojiri}}, \bibinfo {author} {\bibfnamefont
  {Y.}~\bibnamefont {Shimizu}}, \bibinfo {author} {\bibfnamefont
  {K.}~\bibnamefont {Miyagawa}}, \ and\ \bibinfo {author} {\bibfnamefont
  {K.}~\bibnamefont {Kanoda}},\ }\href
  {https://www.nature.com/articles/nphys942} {\bibfield  {journal} {\bibinfo
  {journal} {Nature Physics}\ }\textbf {\bibinfo {volume} {4}},\ \bibinfo
  {pages} {459} (\bibinfo {year} {2008})}\BibitemShut {NoStop}%
\bibitem [{\citenamefont {Takada}\ \emph {et~al.}(2003)\citenamefont {Takada},
  \citenamefont {Sakurai}, \citenamefont {Takayama-Muromachi}, \citenamefont
  {Izumi}, \citenamefont {Dilanian},\ and\ \citenamefont
  {Sasaki}}]{takada2003superconductivity}%
  \BibitemOpen
  \bibfield  {author} {\bibinfo {author} {\bibfnamefont {K.}~\bibnamefont
  {Takada}}, \bibinfo {author} {\bibfnamefont {H.}~\bibnamefont {Sakurai}},
  \bibinfo {author} {\bibfnamefont {E.}~\bibnamefont {Takayama-Muromachi}},
  \bibinfo {author} {\bibfnamefont {F.}~\bibnamefont {Izumi}}, \bibinfo
  {author} {\bibfnamefont {R.~A.}\ \bibnamefont {Dilanian}}, \ and\ \bibinfo
  {author} {\bibfnamefont {T.}~\bibnamefont {Sasaki}},\ }\href {\doibase
  10.1038/nature01450} {\bibfield  {journal} {\bibinfo  {journal} {Nature}\
  }\textbf {\bibinfo {volume} {422}},\ \bibinfo {pages} {53} (\bibinfo {year}
  {2003})}\BibitemShut {NoStop}%
\bibitem [{\citenamefont {Schaak}\ \emph {et~al.}(2003)\citenamefont {Schaak},
  \citenamefont {Klimczuk}, \citenamefont {Foo},\ and\ \citenamefont
  {Cava}}]{schaak2003}%
  \BibitemOpen
  \bibfield  {author} {\bibinfo {author} {\bibfnamefont {R.~E.}\ \bibnamefont
  {Schaak}}, \bibinfo {author} {\bibfnamefont {T.}~\bibnamefont {Klimczuk}},
  \bibinfo {author} {\bibfnamefont {M.~L.}\ \bibnamefont {Foo}}, \ and\
  \bibinfo {author} {\bibfnamefont {R.~J.}\ \bibnamefont {Cava}},\ }\href
  {\doibase 10.1038/nature01877} {\bibfield  {journal} {\bibinfo  {journal}
  {Nature}\ }\textbf {\bibinfo {volume} {424}},\ \bibinfo {pages} {527}
  (\bibinfo {year} {2003})}\BibitemShut {NoStop}%
\bibitem [{\citenamefont {Fujimoto}\ \emph {et~al.}(2004)\citenamefont
  {Fujimoto}, \citenamefont {Zheng}, \citenamefont {Kitaoka}, \citenamefont
  {Meng}, \citenamefont {Cmaidalka},\ and\ \citenamefont {Chu}}]{fujimoto2004}%
  \BibitemOpen
  \bibfield  {author} {\bibinfo {author} {\bibfnamefont {T.}~\bibnamefont
  {Fujimoto}}, \bibinfo {author} {\bibfnamefont {G.-q.}\ \bibnamefont {Zheng}},
  \bibinfo {author} {\bibfnamefont {Y.}~\bibnamefont {Kitaoka}}, \bibinfo
  {author} {\bibfnamefont {R.~L.}\ \bibnamefont {Meng}}, \bibinfo {author}
  {\bibfnamefont {J.}~\bibnamefont {Cmaidalka}}, \ and\ \bibinfo {author}
  {\bibfnamefont {C.~W.}\ \bibnamefont {Chu}},\ }\href {\doibase
  10.1103/PhysRevLett.92.047004} {\bibfield  {journal} {\bibinfo  {journal}
  {Phys. Rev. Lett.}\ }\textbf {\bibinfo {volume} {92}},\ \bibinfo {pages}
  {047004} (\bibinfo {year} {2004})}\BibitemShut {NoStop}%
\bibitem [{\citenamefont {Ming}\ \emph {et~al.}(2023)\citenamefont {Ming},
  \citenamefont {Wu}, \citenamefont {Chen}, \citenamefont {Wang}, \citenamefont
  {Mai}, \citenamefont {Maier}, \citenamefont {Strockoz}, \citenamefont
  {Venderbos}, \citenamefont {Gonz{\'a}lez}, \citenamefont {Ortega} \emph
  {et~al.}}]{ming2022evidence}%
  \BibitemOpen
  \bibfield  {author} {\bibinfo {author} {\bibfnamefont {F.}~\bibnamefont
  {Ming}}, \bibinfo {author} {\bibfnamefont {X.}~\bibnamefont {Wu}}, \bibinfo
  {author} {\bibfnamefont {C.}~\bibnamefont {Chen}}, \bibinfo {author}
  {\bibfnamefont {K.}~\bibnamefont {Wang}}, \bibinfo {author} {\bibfnamefont
  {P.}~\bibnamefont {Mai}}, \bibinfo {author} {\bibfnamefont {T.}~\bibnamefont
  {Maier}}, \bibinfo {author} {\bibfnamefont {J.}~\bibnamefont {Strockoz}},
  \bibinfo {author} {\bibfnamefont {J.}~\bibnamefont {Venderbos}}, \bibinfo
  {author} {\bibfnamefont {C.}~\bibnamefont {Gonz{\'a}lez}}, \bibinfo {author}
  {\bibfnamefont {J.}~\bibnamefont {Ortega}},  \emph {et~al.},\ }\href@noop {}
  {\bibfield  {journal} {\bibinfo  {journal} {Nature Physics}\ ,\ \bibinfo
  {pages} {1}} (\bibinfo {year} {2023})}\BibitemShut {NoStop}%
\bibitem [{\citenamefont {Wu}\ \emph {et~al.}(2018)\citenamefont {Wu},
  \citenamefont {Lovorn}, \citenamefont {Tutuc},\ and\ \citenamefont
  {MacDonald}}]{wu2018hubbard}%
  \BibitemOpen
  \bibfield  {author} {\bibinfo {author} {\bibfnamefont {F.}~\bibnamefont
  {Wu}}, \bibinfo {author} {\bibfnamefont {T.}~\bibnamefont {Lovorn}}, \bibinfo
  {author} {\bibfnamefont {E.}~\bibnamefont {Tutuc}}, \ and\ \bibinfo {author}
  {\bibfnamefont {A.~H.}\ \bibnamefont {MacDonald}},\ }\href
  {https://journals.aps.org/prl/abstract/10.1103/PhysRevLett.121.026402}
  {\bibfield  {journal} {\bibinfo  {journal} {Physical review letters}\
  }\textbf {\bibinfo {volume} {121}},\ \bibinfo {pages} {026402} (\bibinfo
  {year} {2018})}\BibitemShut {NoStop}%
\bibitem [{\citenamefont {Tang}\ \emph {et~al.}(2020)\citenamefont {Tang},
  \citenamefont {Li}, \citenamefont {Li}, \citenamefont {Xu}, \citenamefont
  {Liu}, \citenamefont {Barmak}, \citenamefont {Watanabe}, \citenamefont
  {Taniguchi}, \citenamefont {MacDonald}, \citenamefont {Shan},\ and\
  \citenamefont {Mak}}]{Tang2020}%
  \BibitemOpen
  \bibfield  {author} {\bibinfo {author} {\bibfnamefont {Y.}~\bibnamefont
  {Tang}}, \bibinfo {author} {\bibfnamefont {L.}~\bibnamefont {Li}}, \bibinfo
  {author} {\bibfnamefont {T.}~\bibnamefont {Li}}, \bibinfo {author}
  {\bibfnamefont {Y.}~\bibnamefont {Xu}}, \bibinfo {author} {\bibfnamefont
  {S.}~\bibnamefont {Liu}}, \bibinfo {author} {\bibfnamefont {K.}~\bibnamefont
  {Barmak}}, \bibinfo {author} {\bibfnamefont {K.}~\bibnamefont {Watanabe}},
  \bibinfo {author} {\bibfnamefont {T.}~\bibnamefont {Taniguchi}}, \bibinfo
  {author} {\bibfnamefont {A.~H.}\ \bibnamefont {MacDonald}}, \bibinfo {author}
  {\bibfnamefont {J.}~\bibnamefont {Shan}}, \ and\ \bibinfo {author}
  {\bibfnamefont {K.~F.}\ \bibnamefont {Mak}},\ }\href {\doibase
  10.1038/s41586-020-2085-3} {\bibfield  {journal} {\bibinfo  {journal}
  {Nature}\ }\textbf {\bibinfo {volume} {579}},\ \bibinfo {pages} {353}
  (\bibinfo {year} {2020})}\BibitemShut {NoStop}%
\bibitem [{\citenamefont {An}\ \emph {et~al.}(2020)\citenamefont {An},
  \citenamefont {Cai}, \citenamefont {Pei}, \citenamefont {Huang},
  \citenamefont {Wu}, \citenamefont {Zhou}, \citenamefont {Lin}, \citenamefont
  {Ying}, \citenamefont {Ye}, \citenamefont {Feng} \emph
  {et~al.}}]{an2020interaction}%
  \BibitemOpen
  \bibfield  {author} {\bibinfo {author} {\bibfnamefont {L.}~\bibnamefont
  {An}}, \bibinfo {author} {\bibfnamefont {X.}~\bibnamefont {Cai}}, \bibinfo
  {author} {\bibfnamefont {D.}~\bibnamefont {Pei}}, \bibinfo {author}
  {\bibfnamefont {M.}~\bibnamefont {Huang}}, \bibinfo {author} {\bibfnamefont
  {Z.}~\bibnamefont {Wu}}, \bibinfo {author} {\bibfnamefont {Z.}~\bibnamefont
  {Zhou}}, \bibinfo {author} {\bibfnamefont {J.}~\bibnamefont {Lin}}, \bibinfo
  {author} {\bibfnamefont {Z.}~\bibnamefont {Ying}}, \bibinfo {author}
  {\bibfnamefont {Z.}~\bibnamefont {Ye}}, \bibinfo {author} {\bibfnamefont
  {X.}~\bibnamefont {Feng}},  \emph {et~al.},\ }\href
  {https://pubs.rsc.org/en/content/articlelanding/2020/nh/d0nh00248h}
  {\bibfield  {journal} {\bibinfo  {journal} {Nanoscale horizons}\ }\textbf
  {\bibinfo {volume} {5}},\ \bibinfo {pages} {1309} (\bibinfo {year}
  {2020})}\BibitemShut {NoStop}%
\bibitem [{\citenamefont {Schrade}\ and\ \citenamefont
  {Fu}(2021)}]{schrade2021nematic}%
  \BibitemOpen
  \bibfield  {author} {\bibinfo {author} {\bibfnamefont {C.}~\bibnamefont
  {Schrade}}\ and\ \bibinfo {author} {\bibfnamefont {L.}~\bibnamefont {Fu}},\
  }\href@noop {} {\bibfield  {journal} {\bibinfo  {journal} {arXiv preprint
  arXiv:2110.10172}\ } (\bibinfo {year} {2021})}\BibitemShut {NoStop}%
\bibitem [{\citenamefont {Scherer}\ \emph {et~al.}(2022)\citenamefont
  {Scherer}, \citenamefont {Kennes},\ and\ \citenamefont
  {Classen}}]{scherer2021mathcal}%
  \BibitemOpen
  \bibfield  {author} {\bibinfo {author} {\bibfnamefont {M.~M.}\ \bibnamefont
  {Scherer}}, \bibinfo {author} {\bibfnamefont {D.~M.}\ \bibnamefont {Kennes}},
  \ and\ \bibinfo {author} {\bibfnamefont {L.}~\bibnamefont {Classen}},\
  }\href@noop {} {\bibfield  {journal} {\bibinfo  {journal} {npj Quantum
  Materials}\ }\textbf {\bibinfo {volume} {7}},\ \bibinfo {pages} {100}
  (\bibinfo {year} {2022})}\BibitemShut {NoStop}%
\bibitem [{\citenamefont {Read}\ and\ \citenamefont
  {Green}(2000)}]{read2000paired}%
  \BibitemOpen
  \bibfield  {author} {\bibinfo {author} {\bibfnamefont {N.}~\bibnamefont
  {Read}}\ and\ \bibinfo {author} {\bibfnamefont {D.}~\bibnamefont {Green}},\
  }\href {https://journals.aps.org/prb/abstract/10.1103/PhysRevB.61.10267}
  {\bibfield  {journal} {\bibinfo  {journal} {Physical Review B}\ }\textbf
  {\bibinfo {volume} {61}},\ \bibinfo {pages} {10267} (\bibinfo {year}
  {2000})}\BibitemShut {NoStop}%
\bibitem [{\citenamefont {Senthil}\ \emph {et~al.}(1999)\citenamefont
  {Senthil}, \citenamefont {Marston},\ and\ \citenamefont
  {Fisher}}]{senthil1999spin}%
  \BibitemOpen
  \bibfield  {author} {\bibinfo {author} {\bibfnamefont {T.}~\bibnamefont
  {Senthil}}, \bibinfo {author} {\bibfnamefont {J.~B.}\ \bibnamefont
  {Marston}}, \ and\ \bibinfo {author} {\bibfnamefont {M.~P.~A.}\ \bibnamefont
  {Fisher}},\ }\href
  {https://journals.aps.org/prb/abstract/10.1103/PhysRevB.60.4245} {\bibfield
  {journal} {\bibinfo  {journal} {Physical Review B}\ }\textbf {\bibinfo
  {volume} {60}},\ \bibinfo {pages} {4245} (\bibinfo {year}
  {1999})}\BibitemShut {NoStop}%
\bibitem [{\citenamefont {White}(1992)}]{white1992density}%
  \BibitemOpen
  \bibfield  {author} {\bibinfo {author} {\bibfnamefont {S.~R.}\ \bibnamefont
  {White}},\ }\href
  {https://journals.aps.org/prl/abstract/10.1103/PhysRevLett.69.2863}
  {\bibfield  {journal} {\bibinfo  {journal} {Physical review letters}\
  }\textbf {\bibinfo {volume} {69}},\ \bibinfo {pages} {2863} (\bibinfo {year}
  {1992})}\BibitemShut {NoStop}%
\bibitem [{\citenamefont {McCulloch}(2007)}]{McCulloch2007}%
  \BibitemOpen
  \bibfield  {author} {\bibinfo {author} {\bibfnamefont {I.~P.}\ \bibnamefont
  {McCulloch}},\ }\href {\doibase 10.1088/1742-5468/2007/10/p10014} {\bibfield
  {journal} {\bibinfo  {journal} {Journal of Statistical Mechanics: Theory and
  Experiment}\ }\textbf {\bibinfo {volume} {2007}},\ \bibinfo {pages} {P10014}
  (\bibinfo {year} {2007})}\BibitemShut {NoStop}%
\bibitem [{\citenamefont {Gong}\ \emph {et~al.}(2021)\citenamefont {Gong},
  \citenamefont {Zhu},\ and\ \citenamefont {Sheng}}]{gong2021robust}%
  \BibitemOpen
  \bibfield  {author} {\bibinfo {author} {\bibfnamefont {S.}~\bibnamefont
  {Gong}}, \bibinfo {author} {\bibfnamefont {W.}~\bibnamefont {Zhu}}, \ and\
  \bibinfo {author} {\bibfnamefont {D.~N.}\ \bibnamefont {Sheng}},\ }\href
  {\doibase 10.1103/PhysRevLett.127.097003} {\bibfield  {journal} {\bibinfo
  {journal} {Phys. Rev. Lett.}\ }\textbf {\bibinfo {volume} {127}},\ \bibinfo
  {pages} {097003} (\bibinfo {year} {2021})}\BibitemShut {NoStop}%
\bibitem [{2no()}]{2note}%
  \BibitemOpen
  \href@noop {} {}\bibinfo {howpublished} {The DMRG simulations on the $L_y=4$
  system have not found the topological superconductivity with spontaneous
  time-reversal symmetry breaking in the intermediate-coupling regime. Instead,
  a phase separation between the hole rich region and electron rich region is
  found due to the relatively small system circumference; see the supplementary
  information of Ref.~\cite{huang2021topological}}\BibitemShut {NoStop}%
\bibitem [{Sup()}]{SuppMaterial}%
  \BibitemOpen
  \href@noop {} {}\bibinfo {howpublished} {See Supplemental Material at [URL]
  for detailed numerical results and discussions.}\BibitemShut {Stop}%
\bibitem [{\citenamefont {Lee}(2014)}]{lee2014}%
  \BibitemOpen
  \bibfield  {author} {\bibinfo {author} {\bibfnamefont {P.~A.}\ \bibnamefont
  {Lee}},\ }\href {\doibase 10.1103/PhysRevX.4.031017} {\bibfield  {journal}
  {\bibinfo  {journal} {Phys. Rev. X}\ }\textbf {\bibinfo {volume} {4}},\
  \bibinfo {pages} {031017} (\bibinfo {year} {2014})}\BibitemShut {NoStop}%
\bibitem [{\citenamefont {Dai}\ \emph {et~al.}(2020)\citenamefont {Dai},
  \citenamefont {Senthil},\ and\ \citenamefont {Lee}}]{dai2020}%
  \BibitemOpen
  \bibfield  {author} {\bibinfo {author} {\bibfnamefont {Z.}~\bibnamefont
  {Dai}}, \bibinfo {author} {\bibfnamefont {T.}~\bibnamefont {Senthil}}, \ and\
  \bibinfo {author} {\bibfnamefont {P.~A.}\ \bibnamefont {Lee}},\ }\href
  {\doibase 10.1103/PhysRevB.101.064502} {\bibfield  {journal} {\bibinfo
  {journal} {Phys. Rev. B}\ }\textbf {\bibinfo {volume} {101}},\ \bibinfo
  {pages} {064502} (\bibinfo {year} {2020})}\BibitemShut {NoStop}%
\bibitem [{\citenamefont {Grushin}\ \emph {et~al.}(2015)\citenamefont
  {Grushin}, \citenamefont {Motruk}, \citenamefont {Zaletel},\ and\
  \citenamefont {Pollmann}}]{grushin2015characterization}%
  \BibitemOpen
  \bibfield  {author} {\bibinfo {author} {\bibfnamefont {A.~G.}\ \bibnamefont
  {Grushin}}, \bibinfo {author} {\bibfnamefont {J.}~\bibnamefont {Motruk}},
  \bibinfo {author} {\bibfnamefont {M.~P.}\ \bibnamefont {Zaletel}}, \ and\
  \bibinfo {author} {\bibfnamefont {F.}~\bibnamefont {Pollmann}},\ }\href
  {https://journals.aps.org/prb/abstract/10.1103/PhysRevB.91.035136} {\bibfield
   {journal} {\bibinfo  {journal} {Physical Review B}\ }\textbf {\bibinfo
  {volume} {91}},\ \bibinfo {pages} {035136} (\bibinfo {year}
  {2015})}\BibitemShut {NoStop}%
\bibitem [{\citenamefont {Sheng}\ \emph {et~al.}(2006)\citenamefont {Sheng},
  \citenamefont {Weng}, \citenamefont {Sheng},\ and\ \citenamefont
  {Haldane}}]{sheng2006}%
  \BibitemOpen
  \bibfield  {author} {\bibinfo {author} {\bibfnamefont {D.~N.}\ \bibnamefont
  {Sheng}}, \bibinfo {author} {\bibfnamefont {Z.~Y.}\ \bibnamefont {Weng}},
  \bibinfo {author} {\bibfnamefont {L.}~\bibnamefont {Sheng}}, \ and\ \bibinfo
  {author} {\bibfnamefont {F.~D.~M.}\ \bibnamefont {Haldane}},\ }\href
  {\doibase 10.1103/PhysRevLett.97.036808} {\bibfield  {journal} {\bibinfo
  {journal} {Phys. Rev. Lett.}\ }\textbf {\bibinfo {volume} {97}},\ \bibinfo
  {pages} {036808} (\bibinfo {year} {2006})}\BibitemShut {NoStop}%
\bibitem [{not()}]{note}%
  \BibitemOpen
  \href@noop {} {}\bibinfo {howpublished} {Because the time-revesal symmetry is
  breaking spontaneously, we can identify nonzero Chern number $C=\pm 2$ with
  equal probability in our DMRG simulation with random initial complex
  wavefunction.}\BibitemShut {Stop}%
\bibitem [{\citenamefont {Jiang}\ \emph
  {et~al.}(2021{\natexlab{b}})\citenamefont {Jiang}, \citenamefont
  {Scalapino},\ and\ \citenamefont {White}}]{jiang2021ground}%
  \BibitemOpen
  \bibfield  {author} {\bibinfo {author} {\bibfnamefont {S.}~\bibnamefont
  {Jiang}}, \bibinfo {author} {\bibfnamefont {D.~J.}\ \bibnamefont
  {Scalapino}}, \ and\ \bibinfo {author} {\bibfnamefont {S.~R.}\ \bibnamefont
  {White}},\ }\href {https://www.pnas.org/doi/abs/10.1073/pnas.2109978118}
  {\bibfield  {journal} {\bibinfo  {journal} {Proceedings of the National
  Academy of Sciences}\ }\textbf {\bibinfo {volume} {118}},\ \bibinfo {pages}
  {e2109978118} (\bibinfo {year} {2021}{\natexlab{b}})}\BibitemShut {NoStop}%
\bibitem [{\citenamefont {Jiang}\ and\ \citenamefont
  {Kivelson}(2021)}]{jiang2021high}%
  \BibitemOpen
  \bibfield  {author} {\bibinfo {author} {\bibfnamefont {H.-C.}\ \bibnamefont
  {Jiang}}\ and\ \bibinfo {author} {\bibfnamefont {S.~A.}\ \bibnamefont
  {Kivelson}},\ }\href {\doibase 10.1103/PhysRevLett.127.097002} {\bibfield
  {journal} {\bibinfo  {journal} {Phys. Rev. Lett.}\ }\textbf {\bibinfo
  {volume} {127}},\ \bibinfo {pages} {097002} (\bibinfo {year}
  {2021})}\BibitemShut {NoStop}%
\bibitem [{\citenamefont {White}\ and\ \citenamefont
  {Scalapino}(2009)}]{white2009pairing}%
  \BibitemOpen
  \bibfield  {author} {\bibinfo {author} {\bibfnamefont {S.~R.}\ \bibnamefont
  {White}}\ and\ \bibinfo {author} {\bibfnamefont {D.~J.}\ \bibnamefont
  {Scalapino}},\ }\href
  {https://journals.aps.org/prb/abstract/10.1103/PhysRevB.79.220504} {\bibfield
   {journal} {\bibinfo  {journal} {Physical Review B}\ }\textbf {\bibinfo
  {volume} {79}},\ \bibinfo {pages} {220504(R)} (\bibinfo {year}
  {2009})}\BibitemShut {NoStop}%
\bibitem [{\citenamefont {Corboz}\ \emph {et~al.}(2014)\citenamefont {Corboz},
  \citenamefont {Rice},\ and\ \citenamefont {Troyer}}]{corboz2014competing}%
  \BibitemOpen
  \bibfield  {author} {\bibinfo {author} {\bibfnamefont {P.}~\bibnamefont
  {Corboz}}, \bibinfo {author} {\bibfnamefont {T.~M.}\ \bibnamefont {Rice}}, \
  and\ \bibinfo {author} {\bibfnamefont {M.}~\bibnamefont {Troyer}},\ }\href
  {https://journals.aps.org/prl/abstract/10.1103/PhysRevLett.113.046402}
  {\bibfield  {journal} {\bibinfo  {journal} {Physical review letters}\
  }\textbf {\bibinfo {volume} {113}},\ \bibinfo {pages} {046402} (\bibinfo
  {year} {2014})}\BibitemShut {NoStop}%
\bibitem [{\citenamefont {Zheng}\ \emph {et~al.}(2017)\citenamefont {Zheng},
  \citenamefont {Chung}, \citenamefont {Corboz}, \citenamefont {Ehlers},
  \citenamefont {Qin}, \citenamefont {Noack}, \citenamefont {Shi},
  \citenamefont {White}, \citenamefont {Zhang},\ and\ \citenamefont
  {Chan}}]{zheng2017stripe}%
  \BibitemOpen
  \bibfield  {author} {\bibinfo {author} {\bibfnamefont {B.-X.}\ \bibnamefont
  {Zheng}}, \bibinfo {author} {\bibfnamefont {C.-M.}\ \bibnamefont {Chung}},
  \bibinfo {author} {\bibfnamefont {P.}~\bibnamefont {Corboz}}, \bibinfo
  {author} {\bibfnamefont {G.}~\bibnamefont {Ehlers}}, \bibinfo {author}
  {\bibfnamefont {M.-P.}\ \bibnamefont {Qin}}, \bibinfo {author} {\bibfnamefont
  {R.~M.}\ \bibnamefont {Noack}}, \bibinfo {author} {\bibfnamefont
  {H.}~\bibnamefont {Shi}}, \bibinfo {author} {\bibfnamefont {S.~R.}\
  \bibnamefont {White}}, \bibinfo {author} {\bibfnamefont {S.}~\bibnamefont
  {Zhang}}, \ and\ \bibinfo {author} {\bibfnamefont {G.~K.-L.}\ \bibnamefont
  {Chan}},\ }\href {https://www.science.org/doi/10.1126/science.aam7127}
  {\bibfield  {journal} {\bibinfo  {journal} {Science}\ }\textbf {\bibinfo
  {volume} {358}},\ \bibinfo {pages} {1155} (\bibinfo {year}
  {2017})}\BibitemShut {NoStop}%
\bibitem [{\citenamefont {Huang}\ \emph {et~al.}(2017)\citenamefont {Huang},
  \citenamefont {Mendl}, \citenamefont {Liu}, \citenamefont {Johnston},
  \citenamefont {Jiang}, \citenamefont {Moritz},\ and\ \citenamefont
  {Devereaux}}]{huang2017}%
  \BibitemOpen
  \bibfield  {author} {\bibinfo {author} {\bibfnamefont {E.~W.}\ \bibnamefont
  {Huang}}, \bibinfo {author} {\bibfnamefont {C.~B.}\ \bibnamefont {Mendl}},
  \bibinfo {author} {\bibfnamefont {S.}~\bibnamefont {Liu}}, \bibinfo {author}
  {\bibfnamefont {S.}~\bibnamefont {Johnston}}, \bibinfo {author}
  {\bibfnamefont {H.-C.}\ \bibnamefont {Jiang}}, \bibinfo {author}
  {\bibfnamefont {B.}~\bibnamefont {Moritz}}, \ and\ \bibinfo {author}
  {\bibfnamefont {T.~P.}\ \bibnamefont {Devereaux}},\ }\href
  {https://www.science.org/doi/abs/10.1126/science.aak9546} {\bibfield
  {journal} {\bibinfo  {journal} {Science}\ }\textbf {\bibinfo {volume}
  {358}},\ \bibinfo {pages} {1161} (\bibinfo {year} {2017})}\BibitemShut
  {NoStop}%
\bibitem [{\citenamefont {Jiang}\ \emph {et~al.}(2020)\citenamefont {Jiang},
  \citenamefont {Zaanen}, \citenamefont {Devereaux},\ and\ \citenamefont
  {Jiang}}]{jiang2020ground}%
  \BibitemOpen
  \bibfield  {author} {\bibinfo {author} {\bibfnamefont {Y.-F.}\ \bibnamefont
  {Jiang}}, \bibinfo {author} {\bibfnamefont {J.}~\bibnamefont {Zaanen}},
  \bibinfo {author} {\bibfnamefont {T.~P.}\ \bibnamefont {Devereaux}}, \ and\
  \bibinfo {author} {\bibfnamefont {H.-C.}\ \bibnamefont {Jiang}},\ }\href
  {https://journals.aps.org/prresearch/abstract/10.1103/PhysRevResearch.2.033073}
  {\bibfield  {journal} {\bibinfo  {journal} {Physical Review Research}\
  }\textbf {\bibinfo {volume} {2}},\ \bibinfo {pages} {033073} (\bibinfo {year}
  {2020})}\BibitemShut {NoStop}%
\bibitem [{\citenamefont {Qin}\ \emph {et~al.}(2020)\citenamefont {Qin},
  \citenamefont {Chung}, \citenamefont {Shi}, \citenamefont {Vitali},
  \citenamefont {Hubig}, \citenamefont {Schollw\"ock}, \citenamefont {White},\
  and\ \citenamefont {Zhang}}]{qin2020}%
  \BibitemOpen
  \bibfield  {author} {\bibinfo {author} {\bibfnamefont {M.}~\bibnamefont
  {Qin}}, \bibinfo {author} {\bibfnamefont {C.-M.}\ \bibnamefont {Chung}},
  \bibinfo {author} {\bibfnamefont {H.}~\bibnamefont {Shi}}, \bibinfo {author}
  {\bibfnamefont {E.}~\bibnamefont {Vitali}}, \bibinfo {author} {\bibfnamefont
  {C.}~\bibnamefont {Hubig}}, \bibinfo {author} {\bibfnamefont
  {U.}~\bibnamefont {Schollw\"ock}}, \bibinfo {author} {\bibfnamefont {S.~R.}\
  \bibnamefont {White}}, \ and\ \bibinfo {author} {\bibfnamefont
  {S.}~\bibnamefont {Zhang}} (\bibinfo {collaboration} {Simons Collaboration on
  the Many-Electron Problem}),\ }\href {\doibase 10.1103/PhysRevX.10.031016}
  {\bibfield  {journal} {\bibinfo  {journal} {Phys. Rev. X}\ }\textbf {\bibinfo
  {volume} {10}},\ \bibinfo {pages} {031016} (\bibinfo {year}
  {2020})}\BibitemShut {NoStop}%
\bibitem [{\citenamefont {Wu}\ \emph {et~al.}(2020)\citenamefont {Wu},
  \citenamefont {Ming}, \citenamefont {Smith}, \citenamefont {Liu},
  \citenamefont {Ye}, \citenamefont {Wang}, \citenamefont {Johnston},\ and\
  \citenamefont {Weitering}}]{wu2020superconductivity}%
  \BibitemOpen
  \bibfield  {author} {\bibinfo {author} {\bibfnamefont {X.}~\bibnamefont
  {Wu}}, \bibinfo {author} {\bibfnamefont {F.}~\bibnamefont {Ming}}, \bibinfo
  {author} {\bibfnamefont {T.~S.}\ \bibnamefont {Smith}}, \bibinfo {author}
  {\bibfnamefont {G.}~\bibnamefont {Liu}}, \bibinfo {author} {\bibfnamefont
  {F.}~\bibnamefont {Ye}}, \bibinfo {author} {\bibfnamefont {K.}~\bibnamefont
  {Wang}}, \bibinfo {author} {\bibfnamefont {S.}~\bibnamefont {Johnston}}, \
  and\ \bibinfo {author} {\bibfnamefont {H.~H.}\ \bibnamefont {Weitering}},\
  }\href@noop {} {\bibfield  {journal} {\bibinfo  {journal} {Physical Review
  Letters}\ }\textbf {\bibinfo {volume} {125}},\ \bibinfo {pages} {117001}
  (\bibinfo {year} {2020})}\BibitemShut {NoStop}%
\bibitem [{\citenamefont {Yang}\ \emph {et~al.}(2021)\citenamefont {Yang},
  \citenamefont {Liu}, \citenamefont {Mongkolkiattichai},\ and\ \citenamefont
  {Schauss}}]{yang2021site}%
  \BibitemOpen
  \bibfield  {author} {\bibinfo {author} {\bibfnamefont {J.}~\bibnamefont
  {Yang}}, \bibinfo {author} {\bibfnamefont {L.}~\bibnamefont {Liu}}, \bibinfo
  {author} {\bibfnamefont {J.}~\bibnamefont {Mongkolkiattichai}}, \ and\
  \bibinfo {author} {\bibfnamefont {P.}~\bibnamefont {Schauss}},\ }\href@noop
  {} {\bibfield  {journal} {\bibinfo  {journal} {PRX Quantum}\ }\textbf
  {\bibinfo {volume} {2}},\ \bibinfo {pages} {020344} (\bibinfo {year}
  {2021})}\BibitemShut {NoStop}%
\bibitem [{\citenamefont {Zhu}\ and\ \citenamefont
  {Chen}(2022)}]{zhu2022superconductivity}%
  \BibitemOpen
  \bibfield  {author} {\bibinfo {author} {\bibfnamefont {Z.}~\bibnamefont
  {Zhu}}\ and\ \bibinfo {author} {\bibfnamefont {Q.}~\bibnamefont {Chen}},\
  }\href@noop {} {\bibfield  {journal} {\bibinfo  {journal} {arXiv preprint
  arXiv:2210.06847}\ } (\bibinfo {year} {2022})}\BibitemShut {NoStop}%
\end{thebibliography}%


\newcommand{\beginsupplement}{%
        \setcounter{table}{0}
        \renewcommand{\thetable}{S\arabic{table}}%
        \setcounter{figure}{0}
        \renewcommand{\thefigure}{S\arabic{figure}}%
        \setcounter{section}{0}
        \renewcommand{\thesection}{\Roman{section}}%
        \setcounter{equation}{0}
        \renewcommand{\theequation}{S\arabic{equation}}%
        }
\clearpage
\onecolumngrid
\beginsupplement
\setcounter{secnumdepth}{2}

\begin{center}
\Large {\bf Supplemental Material for ``Quantum phase diagram and spontaneously emergent topological chiral superconductivity in the doped triangular lattice Mott insulators''}
\end{center}

In the Supplemental Materials, we provide more numerical details to support the conclusions we have discussed in the main text. 
In Sec.~\ref{convergence}, we show the good convergence of density matrix renormalization group (DMRG) calculations and the details of the finite bond-dimension extrapolation of physical quantities.
In Sec.~\ref{pumping}, we present more data of the inserting flux simulation.
In Sec.~\ref{spin}, we discuss the common nature of spin correlation functions in different phases.
In Sec.~\ref{compare}, we present more results of the various correlation functions to characterize the quantum phase transition from the charge density wave (CDW) phase with strong spin density wave fluctuation (SDWF) to the topological superconducting (TSC) phase.
In Sec.~\ref{correlations_TSC}, we provide more numerical results to identify the $d+id$-wave TSC on different $L_y=6$ and $8$ systems, as well as for the doping level $\delta = 1/8$.
In Sec.~\ref{further_neighbor_correlations}, we examine and compare SC pairing correlations on further neighboring bonds.
In Sec.~\ref{electron_density}, we show more detailed results regarding the evolution of the 
electron occupation number in the momentum space with tuning the next-nearest-neighbor (NNN)  couplings.
In Sec.~\ref{grand_canonical}, we show more detailed results on the pseudogap-like (PGL) phase to $d$-wave SC phase transition by increasing $L_{y}$ and doping level, which are obtained in the grand canonical ensemble.
Sec.~\ref{data_code} contains the data availability statement.

\section{DMRG convergence and bond-dimension extrapolation}
\label{convergence}

\begin{figure}
\centering
\includegraphics[width=1\linewidth]{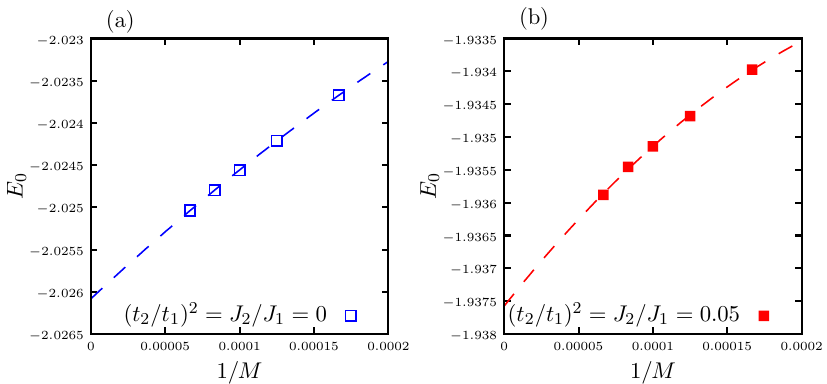}
\caption{Extrapolation of the ground-state energy per site $E_0$ versus the inverse bond dimension in DMRG calculation. (a) $(t_{2}/t_{1})^2 = J_{2}/J_1 =0$, $\delta = 1/12$ on the $L_y = 6$ cylinder. (b) $(t_{2}/t_{1})^2 = J_{2}/J_1 = 0.05$, $\delta = 1/12$ on the $L_y = 6$ cylinder. $M$ is the $SU(2)$ bond dimension, which corresponds to $M = 6000, 8000, 10000, 12000, 15000$.}
\label{FigS_energy}
\end{figure}

First of all, we show the obtained ground-state energy per site $E_0$ versus the inverse DMRG bond dimension ($1/M$), where $M$ is the number of the kept $SU(2)$ multiplets. 
For the $L_y = 6$ system, we keep the bond dimensions up to $M = 15000$.
In Fig.~\ref{FigS_energy}, we show the energies in both the CDW/SDWF and the TSC phase.
The energies converge smoothly with bond dimension and the extrapolated energies are very close to
the lowest energies we obtain, indicating the good convergence of the results.

\begin{figure}
\centering
\includegraphics[width=1\linewidth]{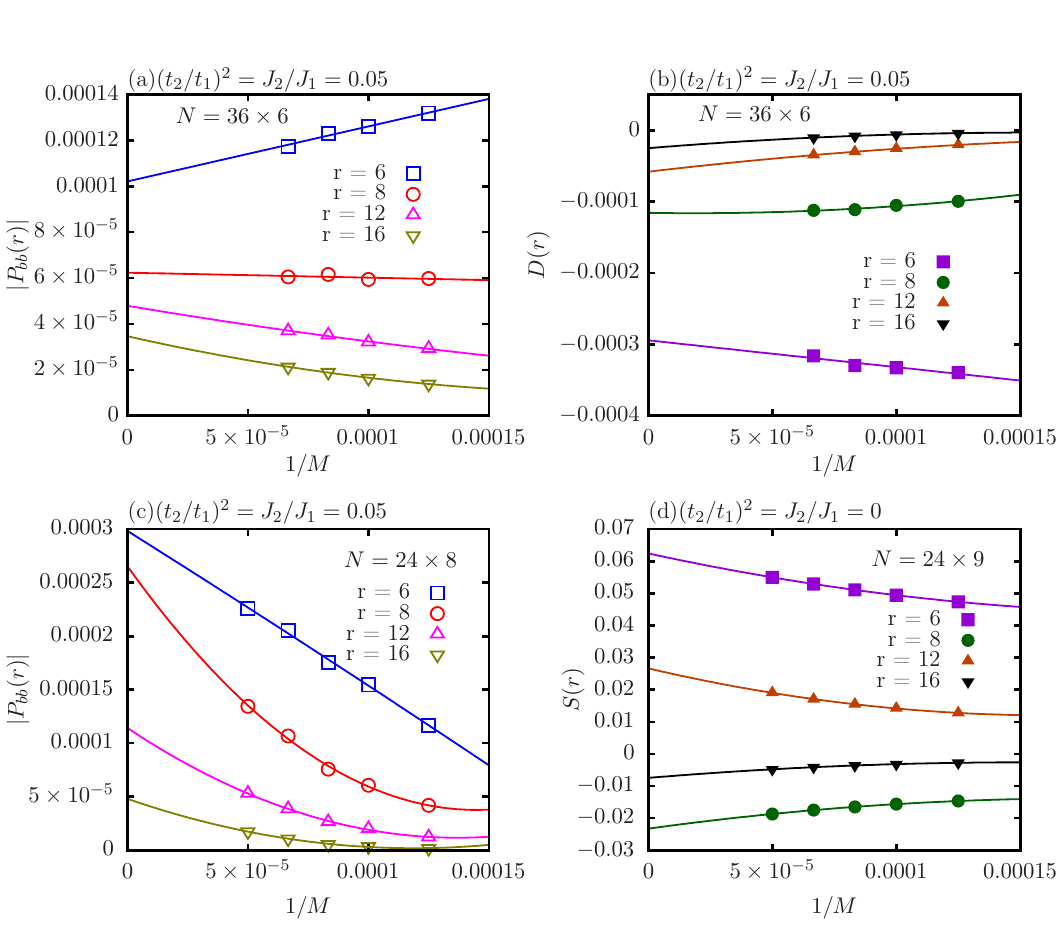}
\caption{Extrapolation of correlation functions versus the inverse bond dimension. (a) and (b) show the extrapolations of the pairing correlation
function $|P_{bb}(r)|$ and the density correlation function $D(r)$ for $(t_2/t_1)^2 = J_2/J_1 = 0.05$, $\delta = 1/12$ on the $L_y = 6$ cylinder. $M$ is the $SU(2)$ bond dimension, which corresponds to $M = 8000, 10000, 12000, 15000$ here. (c) shows the extrapolations of the pairing correlation function $|P_{bb}(r)|$ for $(t_2/t_1)^2 = J_2/J_1 = 0.05$ on the $N = 24 \times 8$ cylinder with $\delta = 1/12$. (d) shows the extrapolations of the spin correlation function $S(r)$ for $(t_2/t_1)^2 = J_2/J_1 = 0$ on the $N = 24 \times 9$ cylinder with $\delta = 1/12$. The different symbols denote the correlations at different distances $r$. For each given distance $r$, the correlations obtained by different bond dimensions are extrapolated by the second-order polynomial function $C(1/M) = C(0) + a/M + b/M^2$.}
\label{FigS_extrapolation}
\end{figure}

In the DMRG calculation of correlation functions on wide systems, it is important to perform the finite bond-dimension
scaling to extrapolate the results in the infinite-bond-dimension limit ($M \rightarrow \infty$).
Here we show the extrapolation in more details.
For each given distance $r$, the correlations
are extrapolated by the second-order polynomial function $C(1/M) = C(0)+a/M + b/M^2$, where $C(0)$ is the extrapolated result in the $M \rightarrow \infty$ limit.
Typical examples on the $L_y = 6$ cylinder are shown in Figs.~\ref{FigS_extrapolation}(a) and \ref{FigS_extrapolation}(b), for pairing and density correlation, respectively.

For the calculations of the $L_y = 9$ cylinder in the CDW/SDWF phase and the $L_y = 8$ cylinder in the TSC phase, we keep the bond dimensions up to $M = 20000$. 
Although the fully convergence of all the quantities is still challenging, we find that the dominant correlations converge faster. 
For example, spin correlations in the CDW/SDWF phase converge quickly, which provide strong evidence to identify the spin density wave fluctuation as shown in Fig.~4(a) of the main text.  
For the TSC phase, the pairing correlations dominate other correlations, which also converge with increasing bond dimension.
The finite bond-dimension scaling of the pairing correlations on the $L_y = 8$ cylinder and that of the spin correlations on the $L_y = 9$ cylinder are shown in Figs.~\ref{FigS_extrapolation}(c) and \ref{FigS_extrapolation}(d), respectively.

In additional, we would like to mention that in the simulation of the CDW/SDWF phase on the $L_y = 6$ cylinder the system length $L_x$ should be compatible with the CDW wavelength $\lambda \approx 10$; otherwise, nonuniform electron density would be obtained with higher energy.
Therefore, we choose $L_x = 40$ to demonstrate our results in the CDW/SDWF phase.

\section{Inserting flux simulation and Chern number}
\label{pumping}

\begin{figure}
\centering
\includegraphics[width=0.5\linewidth]{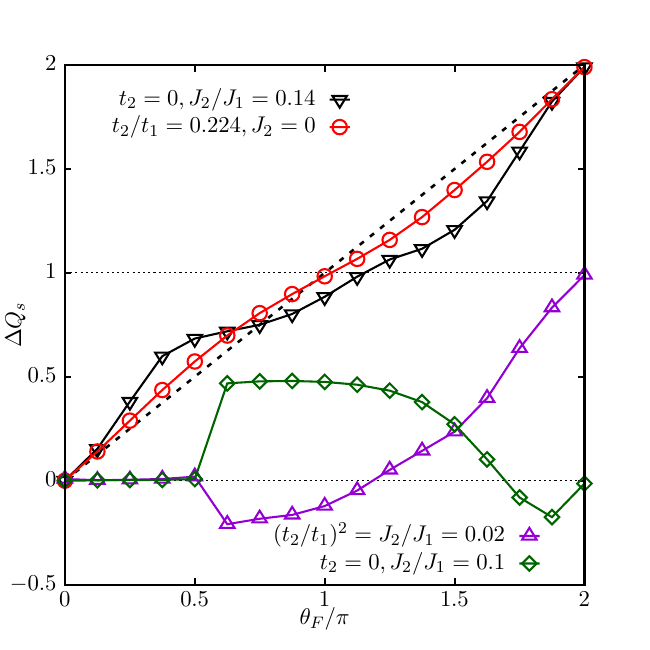}
\caption{Inserting flux simulation and spin pumping results for different couplings. Both the parameters $t_2/t_1 = 0, J_2/J_1 = 0.14$ and $t_2/t_1 = 0.224, J_2/J_1 = 0$ are in the TSC phase. With the adiabatically inserted flux, the pumped spin in a period of flux $\theta_{F}  = 0 \rightarrow 2\pi$ is $\Delta Q_{s} \approx 2$. For $t_2/t_1 = 0, J_2/J_1 = 0.1$ in the CDW/SDWF phase, the pumped spin is zero. For $(t_2/t_1)^2 = J_2/J_1 = 0.02$ in the CDW/SDWF phase but close to the phase boundary, the pumped spin is $\Delta Q_{s} \approx 1$. All the calculations are performed on the $L_y = 6$ cylinders with $\delta = 1/12$ and the $U(1)$ bond dimension $m=8000$.}
\label{FigS_Chern}
\end{figure}

In DMRG simulation, the flux $\theta_{F}$ is introduced by using the twisted boundary conditions along the circumference direction of the cylinder.
Different from the periodic boundary conditions $\hat{c}_{x,y,\sigma} = \hat{c}_{x,y+L_y, \sigma}$, the twisted boundaries require $\hat{c}_{x,y+L_y,\sigma} = e^{i \theta_F \sigma} \hat{c}_{x,y, \sigma}$, where $\sigma$ takes $+1$ for spin up and $-1$ for spin down.
Therefore, the spin flip terms couple to doubled flux $2\theta_F$.
In the main text, we have shown the results of spin pumping simulation by adiabatically threading a flux in the cylinder, from which one can obtain the quantized Chern number.
We have also shown how to distinguish the three phases along the line with $(t_2/t_1)^2 = J_2/J_1$ by using the obtained Chern number.
Here, we show the spin pumping results for more parameter points.
By tuning either $J_2/J_1$ or $t_2/t_1$ to enter the TSC phase, the spin pumping curves are always smooth and give the quantized Chern number $C = 2$, as shown in Fig.~\ref{FigS_Chern} for $t_2/t_1 = 0, J_2/J_1 = 0.14$ and $t_2/t_1 = 0.224, J_2/J_1 = 0$.
For the parameter points in the CDW/SDWF phase and away from the phase boundary, Chern number $C = 0$ is always obtained. 
Near the phase boundary to the TSC phase, we obtain $C = 1$ which may indicate a tiny transition region with averaged nonzero Chern number.  We present more details about the quantum phase transition in Sec.~\ref{compare}.

\section{Spin structure factor and spin correlation function}
\label{spin}

\begin{figure}
\centering
\includegraphics[width=0.315\linewidth]{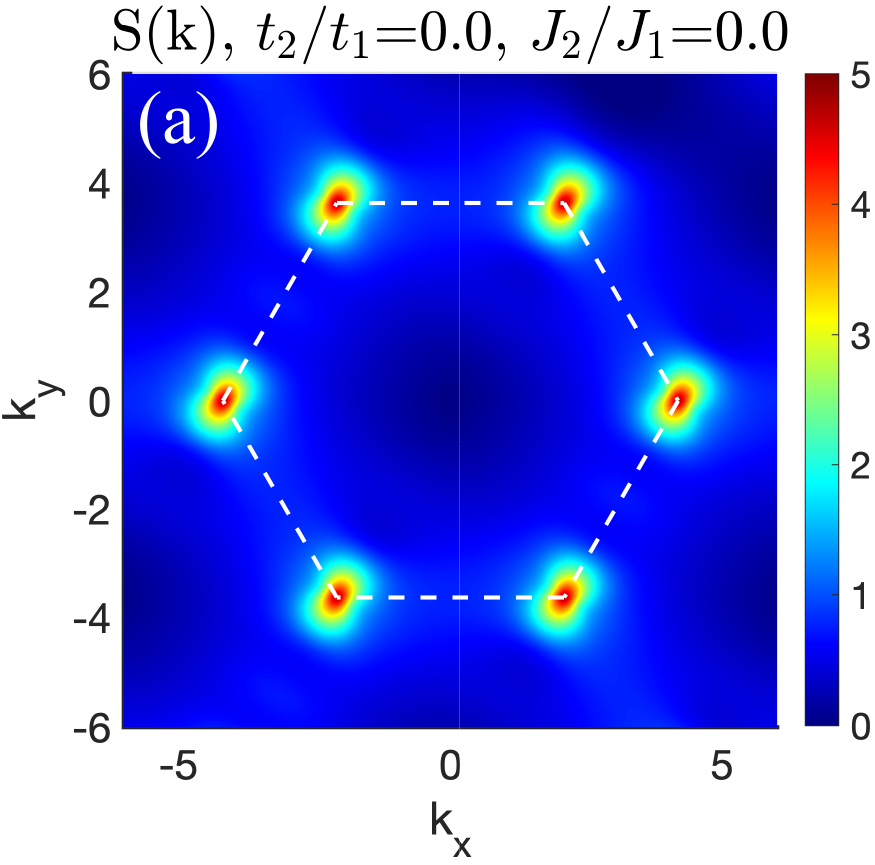}
\includegraphics[width=0.325\linewidth]{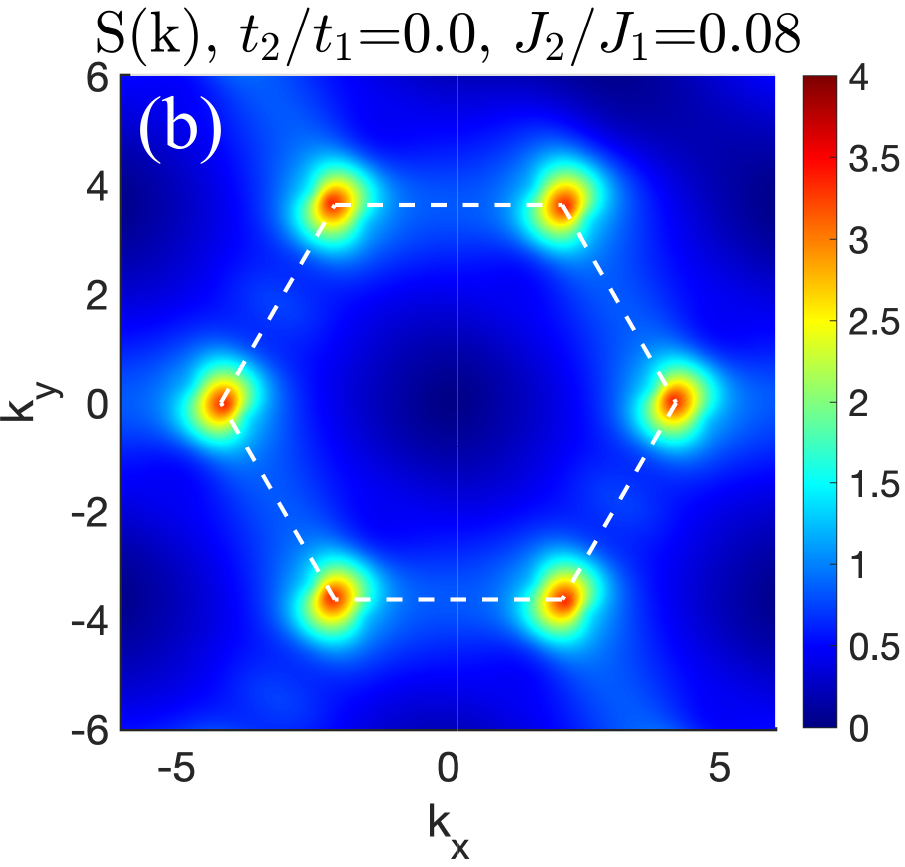}
\includegraphics[width=0.325\linewidth]{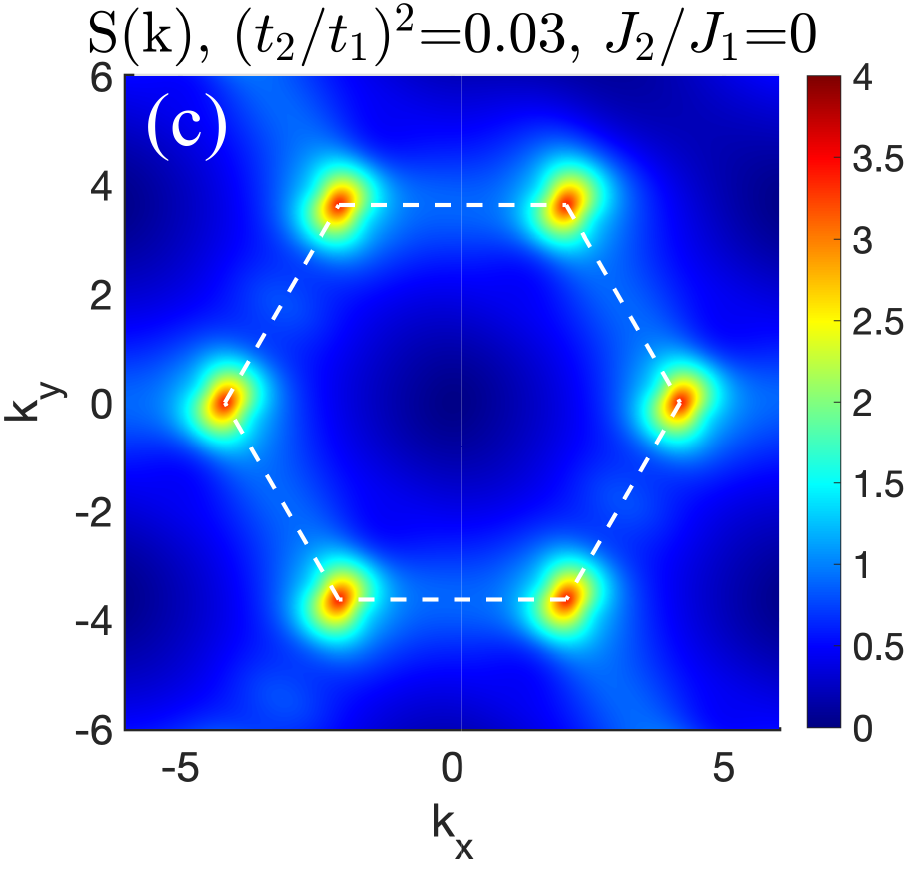}
\includegraphics[width=0.325\linewidth]{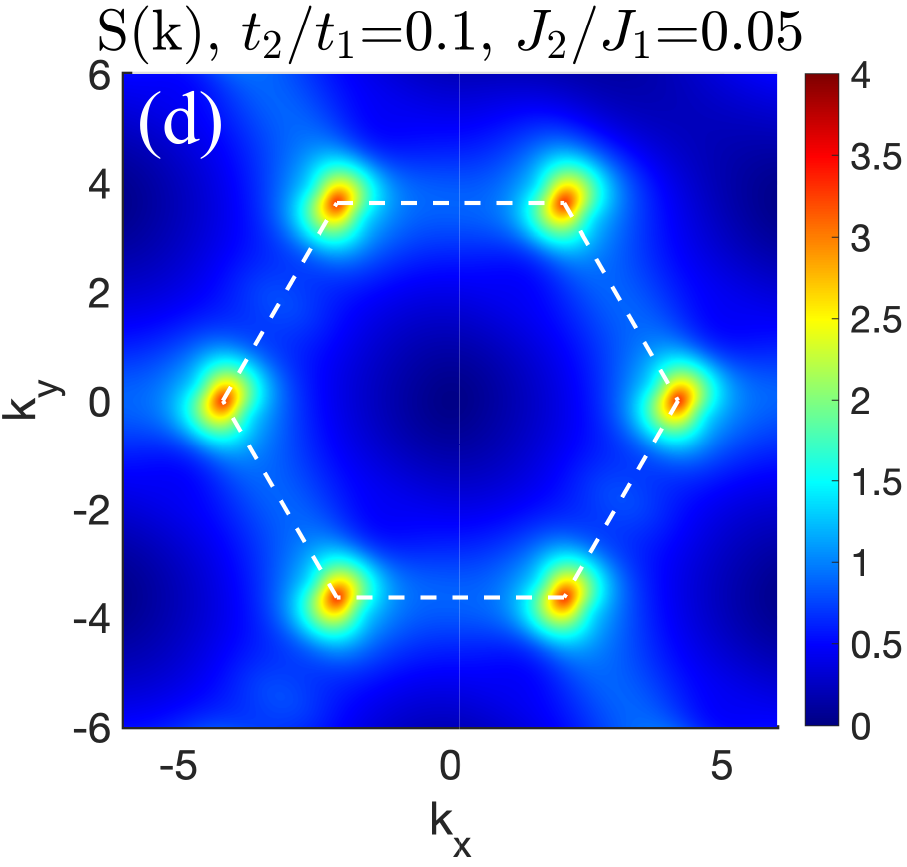}
\includegraphics[width=0.325\linewidth]{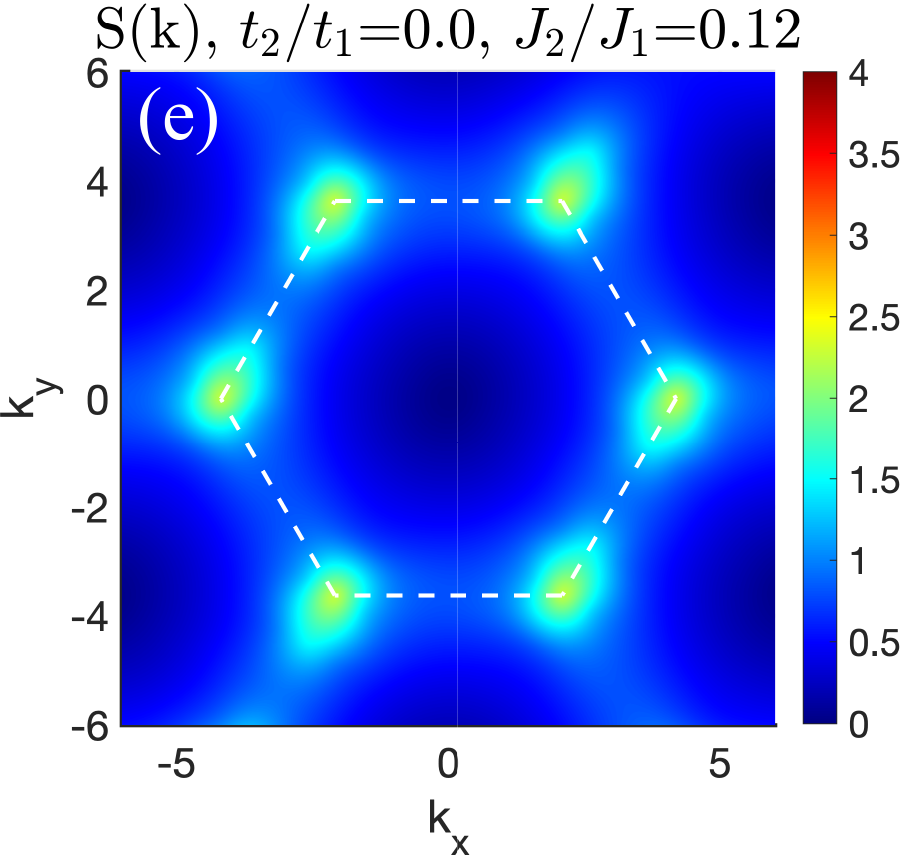}
\includegraphics[width=0.325\linewidth]{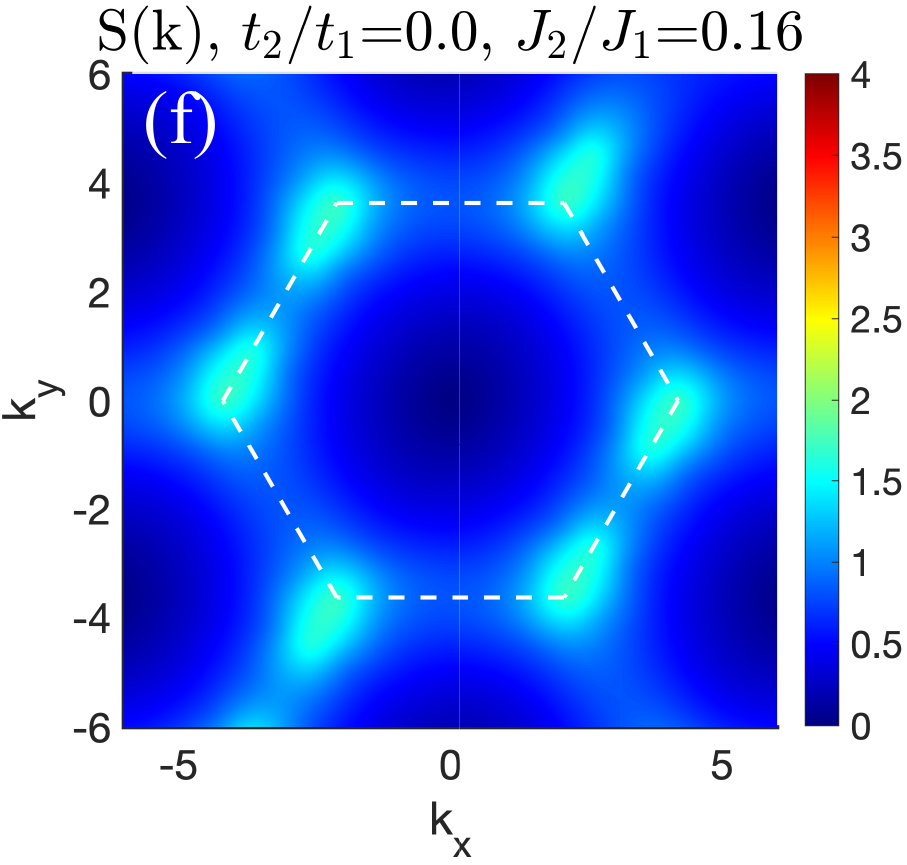}
\includegraphics[width=0.325\linewidth]{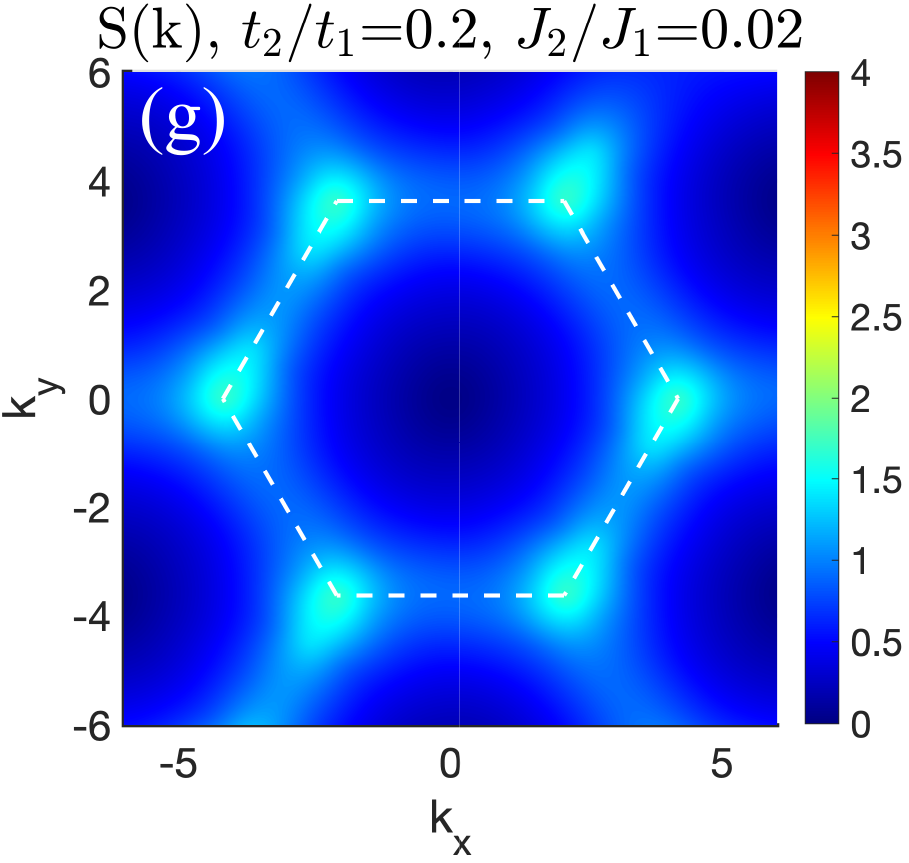}
\includegraphics[width=0.325\linewidth]{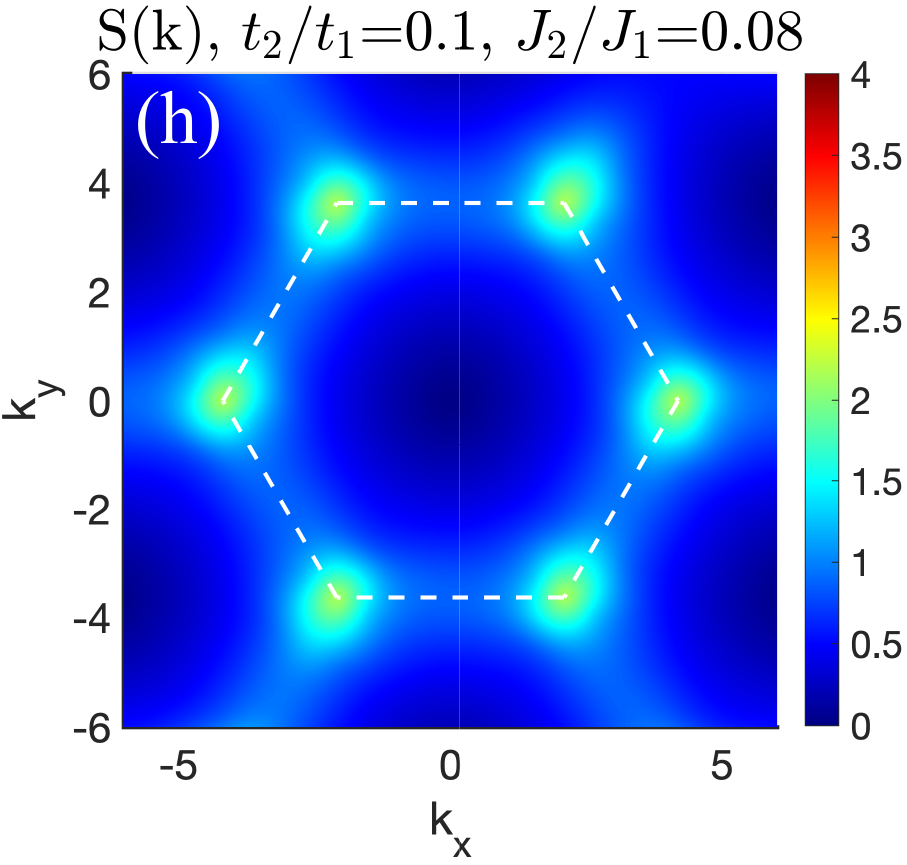}
\includegraphics[width=0.325\linewidth]{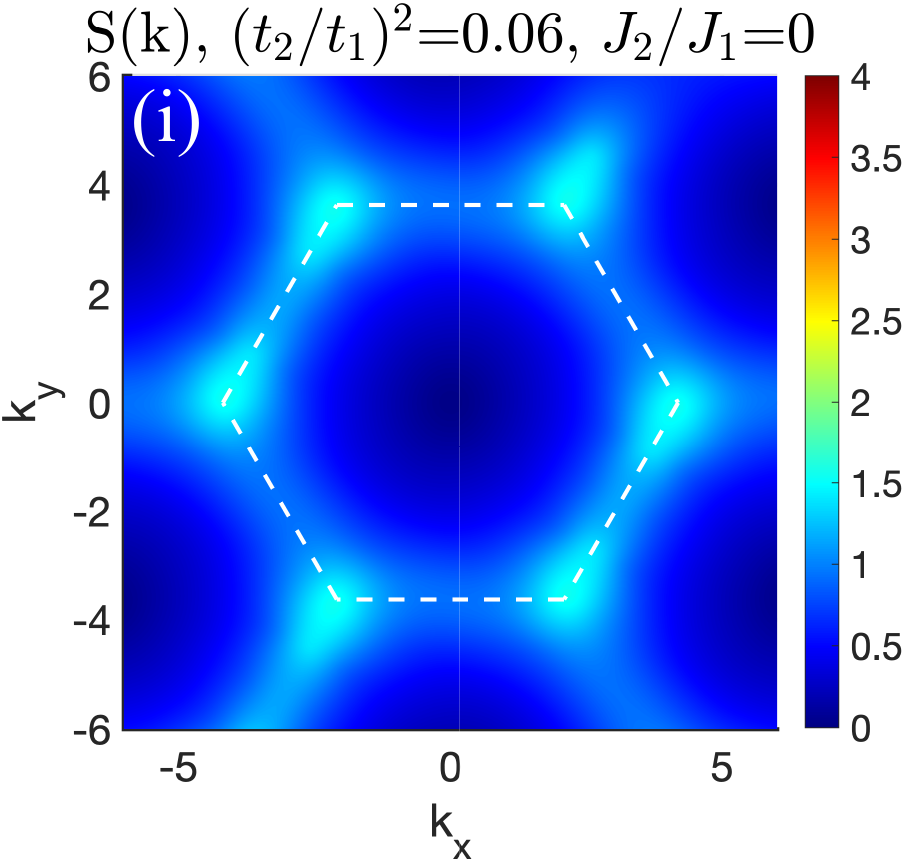}
\caption{Spin structure factor $S({\bf k})$ at different couplings. The results are obtained using the middle $24 \times 6$ sites on the $L_y = 6$ long cylinder with doping ratio $\delta = 1/12$. The dashed hexagon denotes the Brillouin zone. The parameter points in (a)-(d) locate in the CDW/SDWF phase. (e)-(i) belong to the TSC phase. Here we use the $M=10000$ data, which are well converged.}
\label{FigS_spin_str}
\end{figure}

\begin{figure}
\centering
\includegraphics[width=0.9\linewidth]{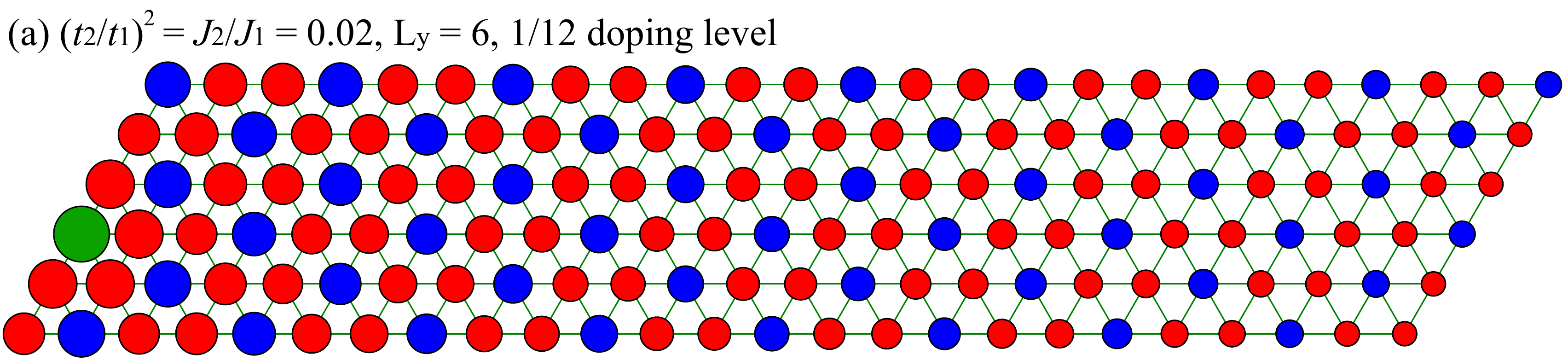}
\includegraphics[width=0.9\linewidth]{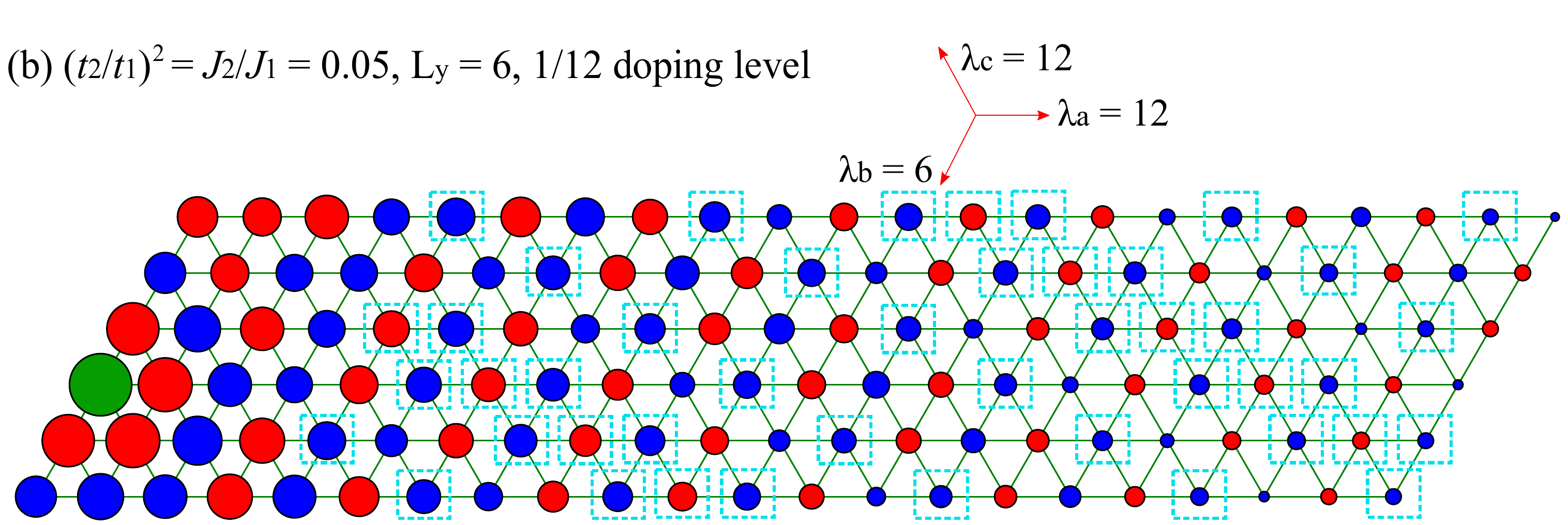}
\caption{Spin correlation functions in the CDW/SDWF phase and the TSC phase. The green circle denotes the reference site near the left boundary of the cylinder. The blue and red circles indicate the positive and negative values of the spin correlations. Here we do not show the sites on the left of the reference site. (a) $(t_2/t_1)^2 = J_2/J_1 = 0.02$ in the CDW/SDWF phase. (b) $(t_2/t_1)^2 = J_2/J_1 = 0.05$ in the TSC phase. The results are obtained on the $L_y = 6$ cylinder with $\delta = 1/12$ doping level. The blue squares in (b) denote the long-distance sites in which the spin correlations have the opposite sign compared with the same-site correlations in (a). $\lambda_{a,b,c}$ denote the periods of the spin correlation pattern along the three bond directions. We use $M=12000$ for obtaining these data.}
\label{FigS_spin_corr}
\end{figure}

In the main text, we have demonstrated the spin structure factor $S(\bf k)$ in the different phases, along the parameter line with $(t_2/t_1)^2 = J_2/J_1$.
Here, we show $S(\bf k)$ at more parameter points in Fig.~\ref{FigS_spin_str}.
In the CDW/SDWF phase [Figs.~\ref{FigS_spin_str}(a)-\ref{FigS_spin_str}(d)], $S(\bf k)$ always has the peaks at the ${\bf K}$ points, which can also be verified by the spin correlations in real space.
As shown in Fig.~\ref{FigS_spin_corr}(a), the reference site is denoted by the green circle, and the blue and red circles indicate the positive and negative spin correlations, respectively.
The spin correlation of the $120^{\circ}$ configuration is unveiled by the same sign of the spin correlations in each sublattice, in which the sites are connected by the NNN bonds.
These results indicate that although the doping suppresses long-range magnetic order, the short-range magnetic pattern in spin background is preserved.

With growing $t_2/t_1$ or (and) $J_2/J_1$, the system has a transition to the TSC phase, which is accompanied with a remarkable change of spin correlation.
While the peaks of $S(\bf k)$ at the ${\bf K}$ points are strongly suppressed, the intensities tend to extend along one of the boundaries of the Brillouin zone, as shown in Figs.~\ref{FigS_spin_str}(e)-\ref{FigS_spin_str}(i). This feature of $S(\bf k)$ seems to be common in the TSC phase.
We further analyze the spin correlation functions in real space, and we find that in most region of the TSC phase the spin correlations have a common pattern as shown in Fig.~\ref{FigS_spin_corr}(b), which suggests that tuning either $t_2/t_1$ or $J_2/J_1$ plays the similar role in the suppression of the $120^{\circ}$ SDWF.
We compare this correlation pattern with that of the $120^{\circ}$ SDWF in Fig.~\ref{FigS_spin_corr}(a), and we mark the different signs of the long-distance correlations by the dashed squares. The spin correlations in the TSC phase also show a periodic pattern but with enlarged periods along all the three bond directions.

\section{Quantum phase transitions from the CDW/SDWF to the TSC phase along different parameter lines}
\label{compare}

\begin{figure}
\centering
\includegraphics[width=0.8\linewidth]{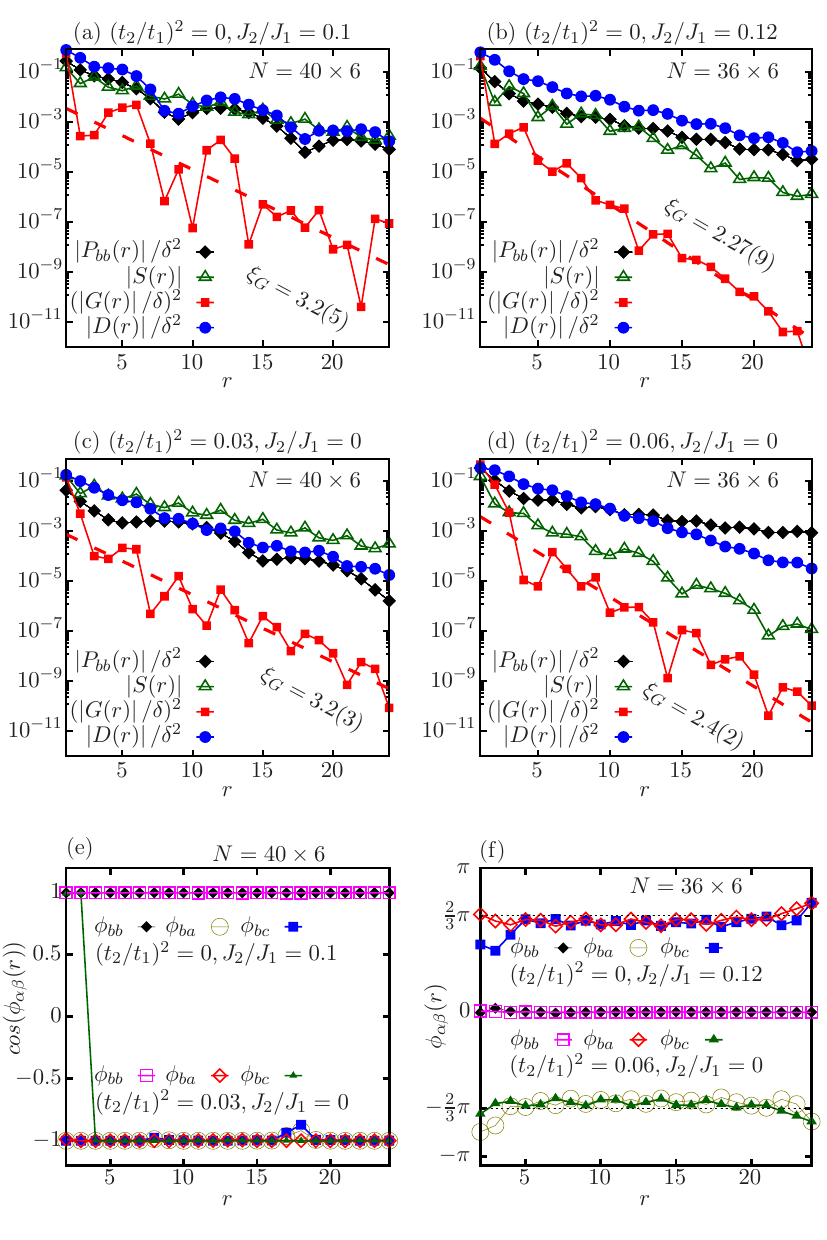}
\caption{Comparing the correlation functions with the quantum phase transition from the CDW/SDFW to the TSC by tuning either $t_2/t_1$ or $J_2/J_1$. (a) and (b) show the transition with tuning $J_2/J_1$. (c) and (d) show the transition with tuning $t_2/t_1$. (e) and (f) show the SC pairing symmetries on different bonds as defined in the main text for the CDW/SDWF and TSC, respectively. All the results are obtained on the $L_y = 6$ cylinders with doping ratio $\delta = 1/12$. We use the $M=12000$ data.}
\label{FigS_Correlations_compare}
\end{figure}

In the main text, we have shown the fluctuating superconductivity in the CDW/SDWF phase and the dominant SC pairing correlations in the TSC phase with $d+id$-wave pairing symmetry, along the parameter line of $J_2/J_1 = (t_2/t_1)^2$.
Here in Fig.~\ref{FigS_Correlations_compare}, we demonstrate more numerical results of correlation functions regarding the quantum phase transition from the CDW/SDWF to the TSC by tuning either $J_2/J_1$ or $t_2/t_1$.  
We observe the characteristic features of the two phases by tuning either $J_2/J_1$ from $0.1$ to $0.12$ [Figs.~\ref{FigS_Correlations_compare}(a) and \ref{FigS_Correlations_compare}(b)] or $(t_2/t_1)^2$ from $0.03$ to $0.06$ [Figs.~\ref{FigS_Correlations_compare}(c) and \ref{FigS_Correlations_compare}(d)], respectively.
In the CDW/SDWF phase, the spin correlations $S(\mathbf{r})
 = \langle \hat{\boldsymbol{S}}_{\mathbf{r}_{0}}\cdot \hat{\boldsymbol{S}}_{\mathbf{r}_{0}+\mathbf{r}}\rangle$, 
charge density correlation $D(\mathbf{r}) = \langle \hat{n}_{\mathbf{r}_{0}} \hat{n}_{\mathbf{r}_{0}+\mathbf{r}}\rangle - \langle \hat{n}_{\mathbf{r}_{0}}\rangle \langle \hat{n}_{\mathbf{r}_{0}+\mathbf{r}} \rangle$, and SC pairing correlation $\left | P_{bb}(r) \right |$   are all relatively strong, and they decay much slower than the two single-particle correlator $G^2(r)$ ($G({\bf r}) = \langle \sum _{\sigma }\hat{c}^{\dagger}_{\mathbf{r}_{0},\sigma }\hat{c}_{\mathbf{r}_{0}+\mathbf{r},\sigma}\rangle$), which further confirm that the strong spin fluctuation, the fluctuating SC, and the more suppressed single-particle channel are common properties in the CDW/SDWF phase.  
Remarkably, the long-distance magnitudes of $\left | P_{bb}(r) \right |$ are always larger than $G^2(r)$ by more than two orders, which demonstrates that the ``pseudogap'' behavior is also universal in the CDW/SDWF phase.

With increasing either $J_2/J_1$ or $t_2/t_1$, the system has a transition to the TSC phase. The pairing correlation becomes dominant, and the single-particle correlation remains pretty weak and decays exponentially.
This phase transition can also be verified by the pairing symmetry.
In the CDW/SDWF phase, the pairing symmetry agrees with the $d_{x^2-y^2}$-wave symmetry as illustrated by the signs of pairing correlations [Fig.~\ref{FigS_Correlations_compare}(e)].
In the TSC phase, it becomes an isotropic $d+id$-wave with the relative pairing phases close to $\pm 2\pi/3$ as shown in Fig.~\ref{FigS_Correlations_compare}(f).
These features presented in Fig.~\ref{FigS_Correlations_compare} are robust for all the bond dimensions ($M=8000-12000$) we have checked.

This quantum phase transition happens with the changes of charge order, SC pairing symmetry, and topological Chern number, which imply that the transition may be first order. We leave the more quantitative understanding of the transition to future studies.
Interestingly, if we consider additional three-spin chiral interaction $J_{\chi}$, we will find a transition from the CDW/SDWF phase to a TSC phase with Chern number $C=1$. This $C=1$ TSC phase has been identified in recent DMRG study~\cite{huang2021topological}.

\section{Correlation functions in the TSC phase: on various systems sizes and doping levels}
\label{correlations_TSC}

In this part, we demonstrate more results of correlation functions in the TSC phase, including the
results on the wider systems with $N=24\times 8$ and $36\times 8$ at   the doping level $\delta=1/12$, and the results for $N=32\times 6$ at $\delta = 1/8$. 
These results further support the robust $d+id$-wave TSC phase.

\subsection{$N=36\times 8$ and $N = 24\times 8$ at $\delta = 1/12$}

\begin{figure}
\centering
\includegraphics[width=0.8\linewidth]{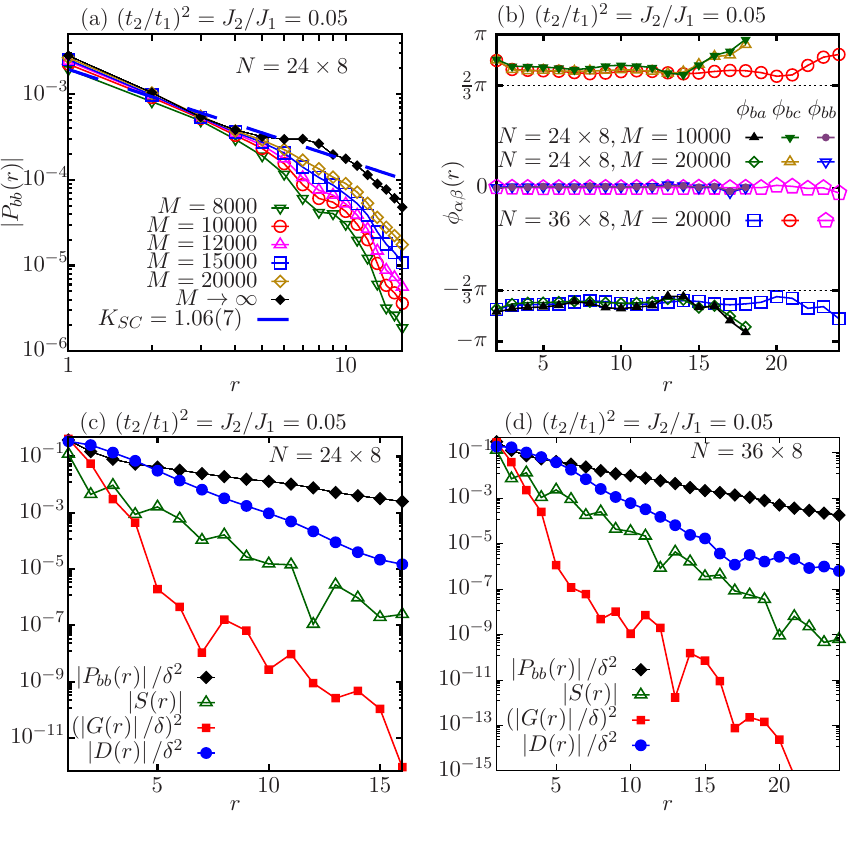}
\caption{Correlation functions for the TSC on the $N = 24\times 8$ and $N = 36\times 8$ cylinders. $(t_{2}/t_{1})^2 = J_{2}/J_1 = 0.05$ and $\delta = 1/12$. (a) Double-logarithmic plot of the SC pairing correlations $|P_{bb}(r)|$. We fit the extrapolated data from bond dimensions of $M=8000 - 20000$, which give the power exponent $K_{SC}=1.06(7)$. (b) The relative  phases of the pairing correlations for different bond dimensions and different system lengths. (c) Comparison of the rescaled correlation functions for $N=24\times 8$, which are obtained with $M=20000$. (d) Comparison of the rescaled correlation functions for $N=36\times 8$, which are obtained with $M=15000$.}
\label{FigS_0.05_0.05_24_8}
\end{figure}

To explore the size effect, we also investigate the TSC phase on the wider $L_{y}=8$ systems. 
As shown in Fig.~\ref{FigS_0.05_0.05_24_8}(a), for the bond dimensions $M=8000$ to $20000$, we find that the pairing correlations increase with $M$ relatively fast.
We also show the algebraic fitting of the extrapolated $M \rightarrow \infty$ data up to the distance $r\leq L_{x}/2$ to minimize the boundary effect.
The fitting gives the power exponent $K_{SC} \approx 1.06$, consistent with the exponent on the $L_y = 6$ system.
We also identify the SC pairing symmetry by analyzing the complex phases of the pairing correlations on different bonds, as shown in Fig.~\ref{FigS_0.05_0.05_24_8}(b).  
An important detail is that, the relative phases $-\phi_{ba}$ and $\phi_{bc}$ are moving closer to $2\pi/3$ with increased bond dimension, confirming an isotropic chiral $d+id$-wave TSC phase on these larger systems. 
By comparing the correlation functions in Figs.~\ref{FigS_0.05_0.05_24_8}(c) and \ref{FigS_0.05_0.05_24_8}(d) for the system sizes $N=24\times 8$ and $36\times 8$, we find that the SC pairing correlations strongly dominate other correlations, which agree with the results on the $L_y=6$ systems.

\subsection{$N=32\times 6$ at $\delta = 1/8$}

\begin{figure}
\centering
\includegraphics[width=0.8\linewidth]{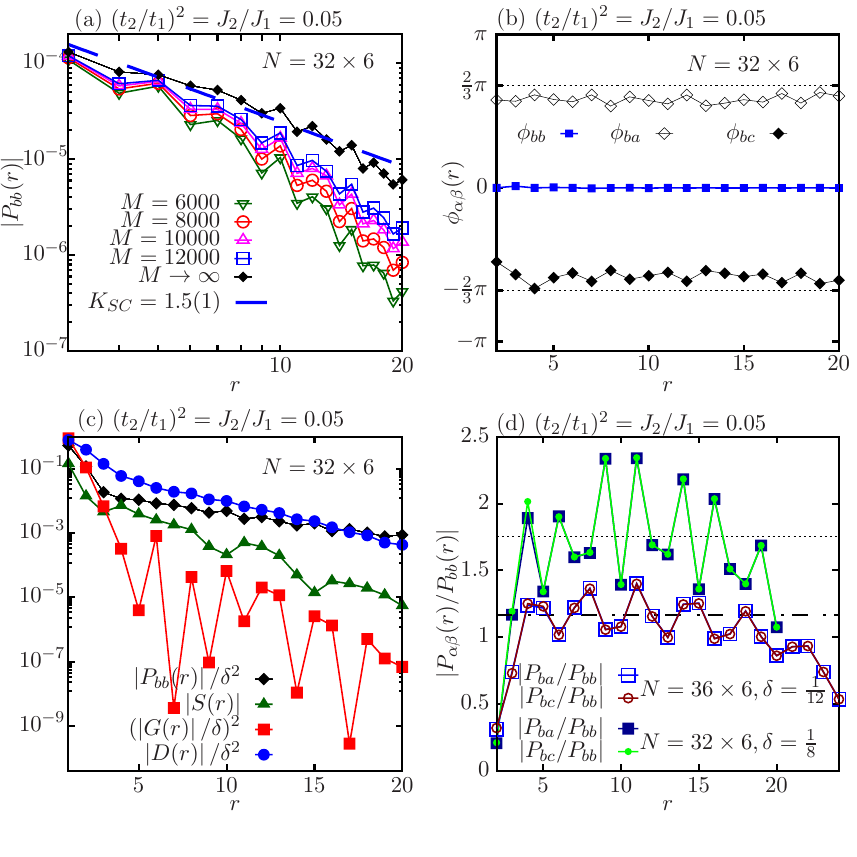}
\caption{Correlation functions for the TSC at $\delta = 1/8$ doping level. $(t_{2}/t_{1})^2 = J_{2}/J_1 = 0.05$ on the $N = 32\times 6$ cylinder. (a) Double-logarithmic plot of the SC pairing correlations $|P_{bb}(r)|$. We fit the extrapolated data from bond dimensions of $M=6000$ - $12000$, which give the power exponent $K_{SC}=1.5(1)$. (b) The relative phases of the pairing correlations for $M=12000$. (c) Comparison of the rescaled correlation functions with the extrapolated data. (d) The ratios of the magnitudes of the pairing correlations at different bonds for $M=12000$. The dotted line indicates the averaged ratio around $1.7(4)$ at $\delta=1/8$ doping. The dashed dotted line indicates the averaged ratio around $1.2(1)$ at $\delta=1/12$ doping. We choose $r\leq L_{x}/2$ to calculate the averages to minimize the boundary effect.}
\label{FigS_0.05_0.05_32_8_delta8}
\end{figure}

While we have established the phase diagram and identified the TSC phase at the doping level $\delta = 1/12$, here we provide evidence to identify the TSC at $\delta = 1/8$, showing that this TSC is robust in a range of doping level. 
As shown in Fig.~\ref{FigS_0.05_0.05_32_8_delta8}(a) for $N=32\times 6$ cylinder, the SC pairing correlations of the extrapolated $M \rightarrow \infty$ results decay algebraically with a small power exponent $K_{SC} \approx 1.5$. 
In addition, the relative phases of the different pairing correlations along different bond directions are also consistent with the complex pairing symmetry, as shown in Fig.~\ref{FigS_0.05_0.05_32_8_delta8}(b). Noticing that complex phases may take opposite signs in different runs of DMRG simulations, it may realize either $d+id$- or $d-id$-wave superconducting symmetry due to spontaneously breaking time-reversal symmetry. Furthermore, we also compare the different correlations in Fig.~\ref{FigS_0.05_0.05_32_8_delta8}(c). The behaviors of the correlations are qualitatively consistent with our observations on the $L_y = 6$ system at $\delta = 1/12$, and the SC pairing correlations still dominant over other correlations at long distance. The averaged ratios between the magnitudes of pairing correlations for different bonds in Fig.~\ref{FigS_0.05_0.05_32_8_delta8}(d) become larger than $1$, which suggests that the $d_{xy}$ component is larger than the $d_{x^2-y^2}$ component. In comparison, the ratio is closer to $1$ at $\delta = 1/12$ doping level.

\section{SC pairing correlations of the further-neighbor bonds}
\label{further_neighbor_correlations}

\begin{figure}
\centering
\includegraphics[width=0.3\linewidth]{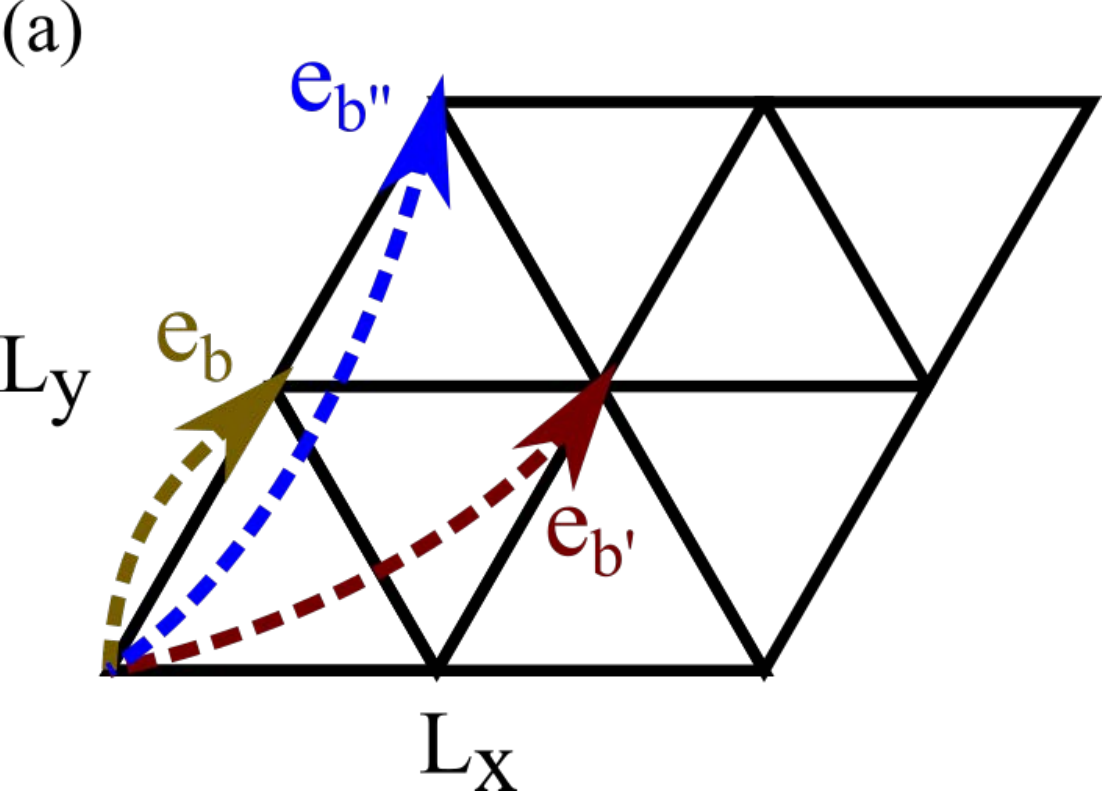}
\includegraphics[width=0.9\linewidth]{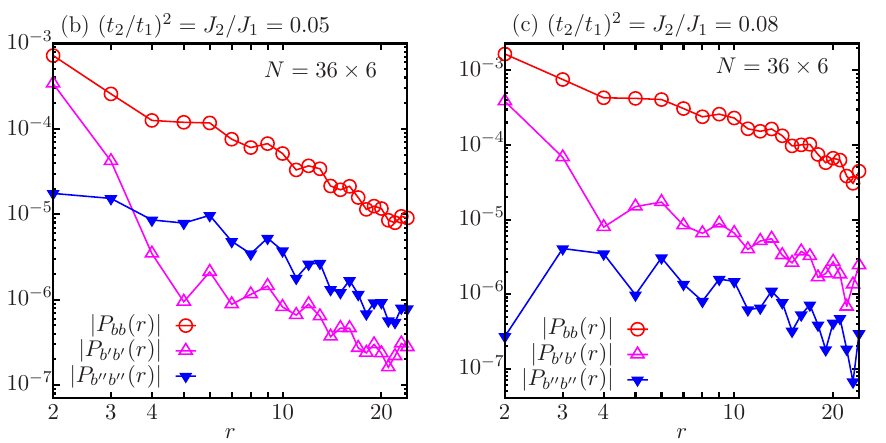}
\caption{SC pairing correlations of the further-neighbor bonds. (a) The illustration of the nearest-neighbor $b$, next-nearest-neighbor $b'$ and next-next-nearest-neighbor-bond $b''$. (b) Comparing SC pairing correlations on different bonds in the TSC phase at $ (t_2/t_1)^2 =J_2/J_1 =0.05$, which are obtained with $M=15000$. The averaged ratio of $|P_{bb}/P_{b'b'}|$ is around $62$ and the averaged ratio of $|P_{bb}/P_{b''b''}|$ is around $15$. (c) Comparing SC pairing correlations on different bonds in the $d$-wave SC phase at $ (t_2/t_1)^2 = J_2/J_1 = 0.08$, which are obtained with $M=12000$. The averaged ratio of $|P_{bb}/P_{b'b'}|$ is around $38$ and the averaged ratio of $|P_{bb}/P_{b''b''}|$ is around $214$. The averages are obtained from the data at the distance $r=5 - 24$.}
\label{FigS_further_neighbor_correlation}
\end{figure}

We examine the SC pairing correlations of the further-neighbor bonds which are illustrated in Fig.~\ref{FigS_further_neighbor_correlation}(a). As shown in Figs.~\ref{FigS_further_neighbor_correlation}(b) and \ref{FigS_further_neighbor_correlation}(c), the SC pairing correlations on the nearest-neighbor bond are over ten times larger than the ones on the next-nearest-neighbor and next-next-nearest-neighbor bonds in both the TSC and $d$-wave SC phase. Our results indicate the dominant role of the nearest-neighbor pairing induced by the stronger spin interaction.

\section{Electron occupation number in the momentum space}
\label{electron_density}

\begin{figure}
\centering
\includegraphics[width=0.325\linewidth]{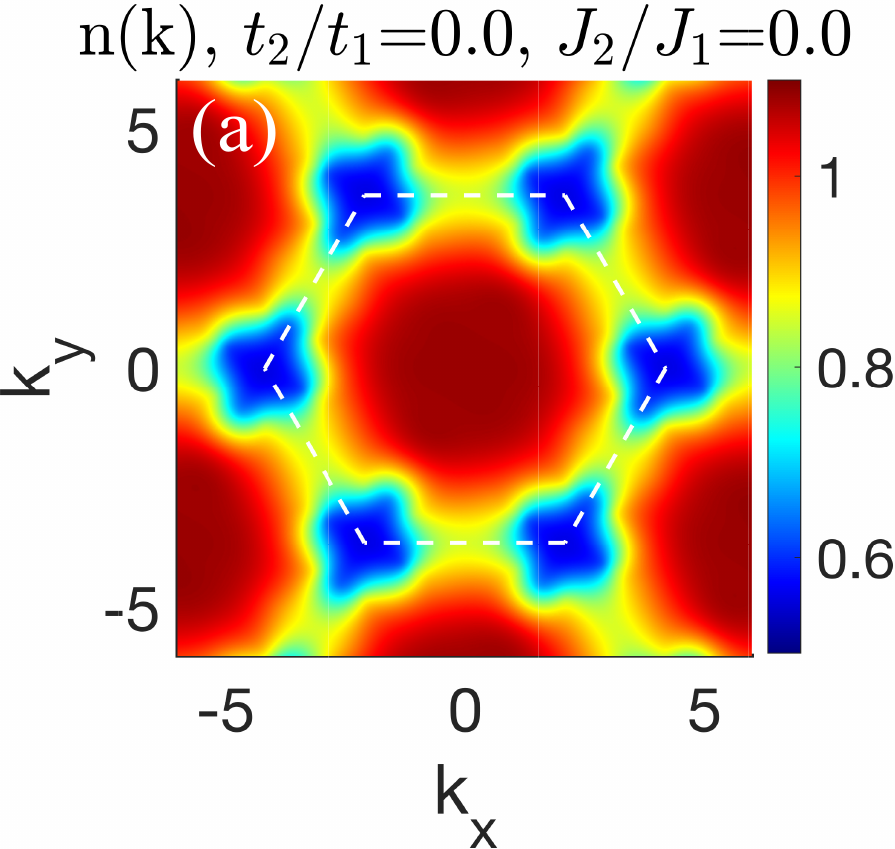}
\includegraphics[width=0.325\linewidth]{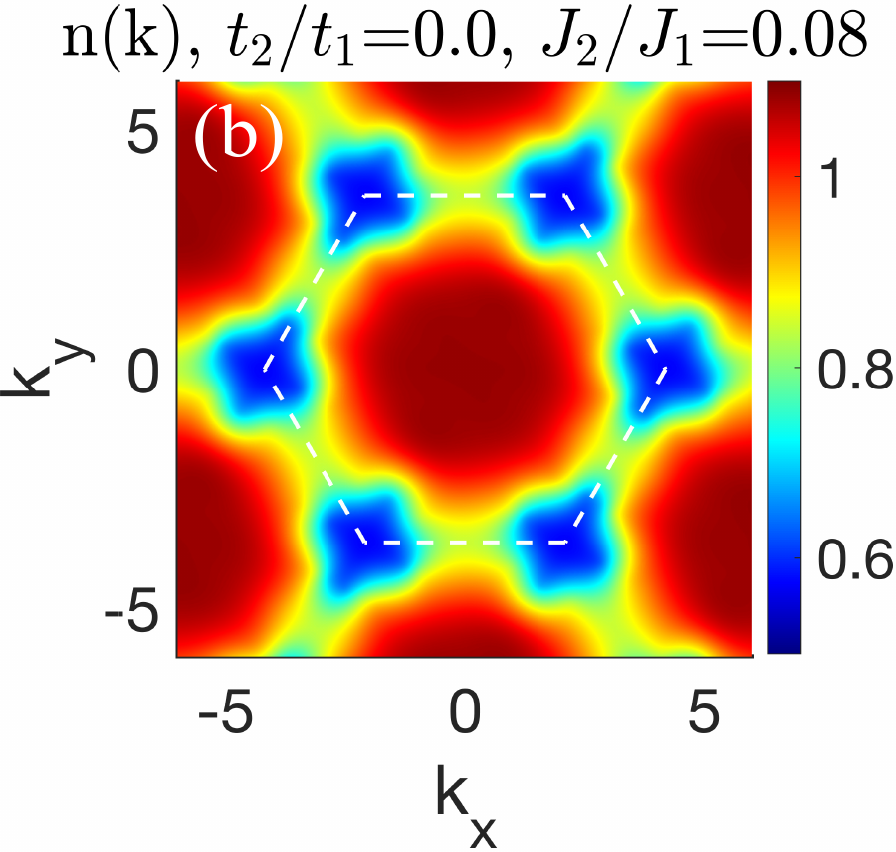}
\includegraphics[width=0.325\linewidth]{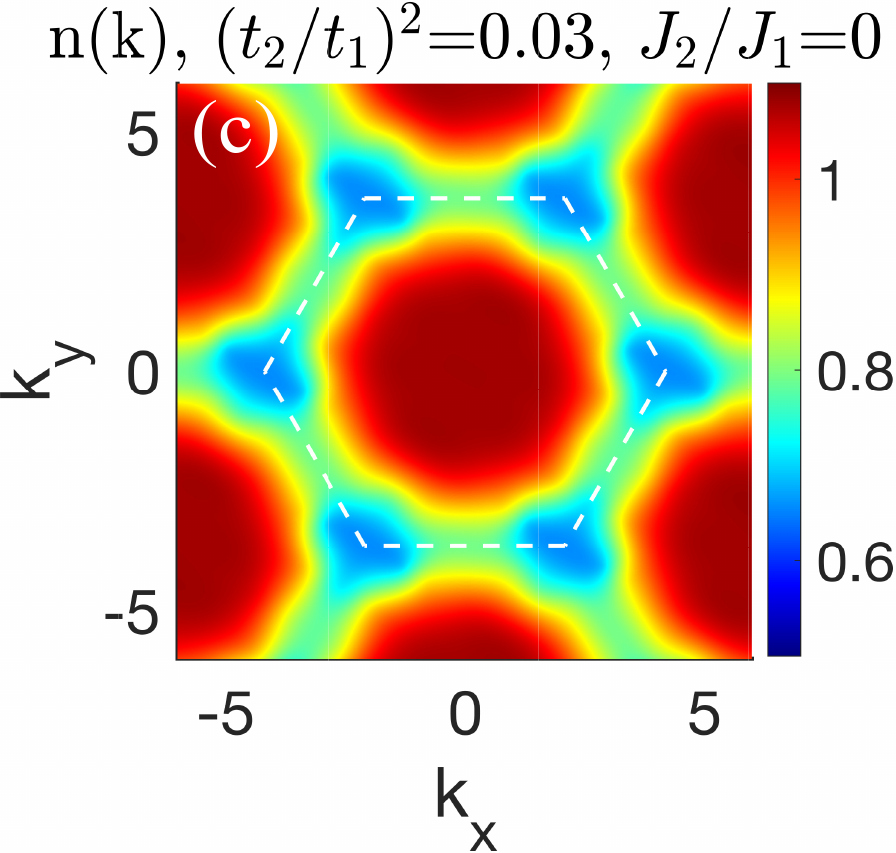}
\includegraphics[width=0.325\linewidth]{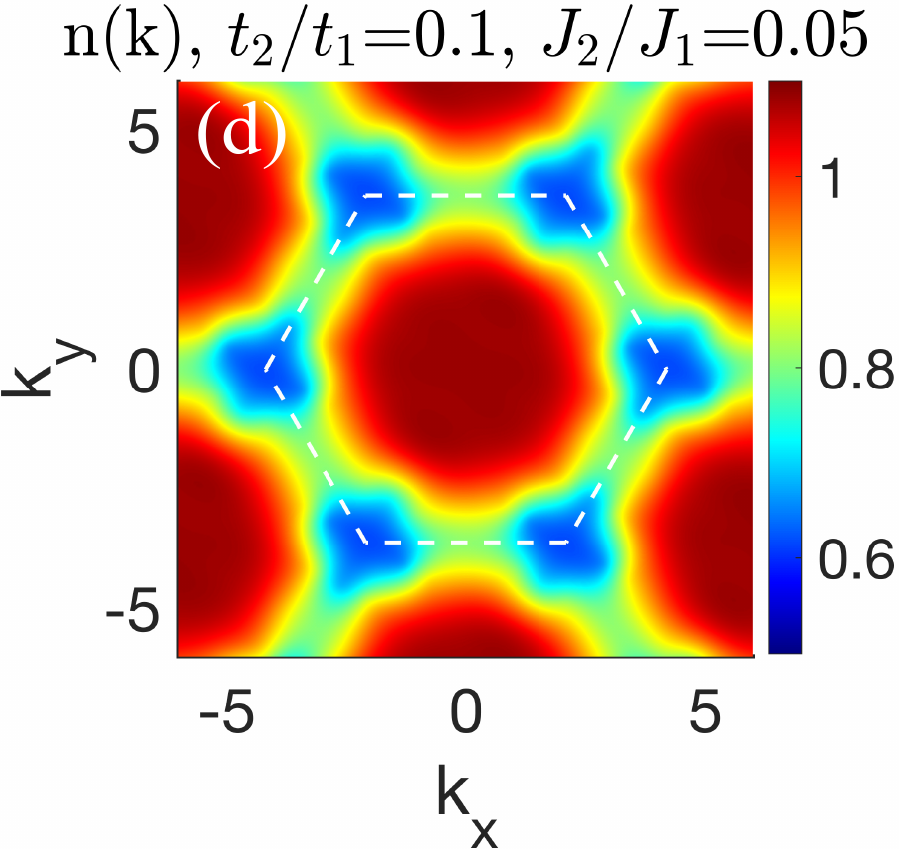}
\includegraphics[width=0.325\linewidth]{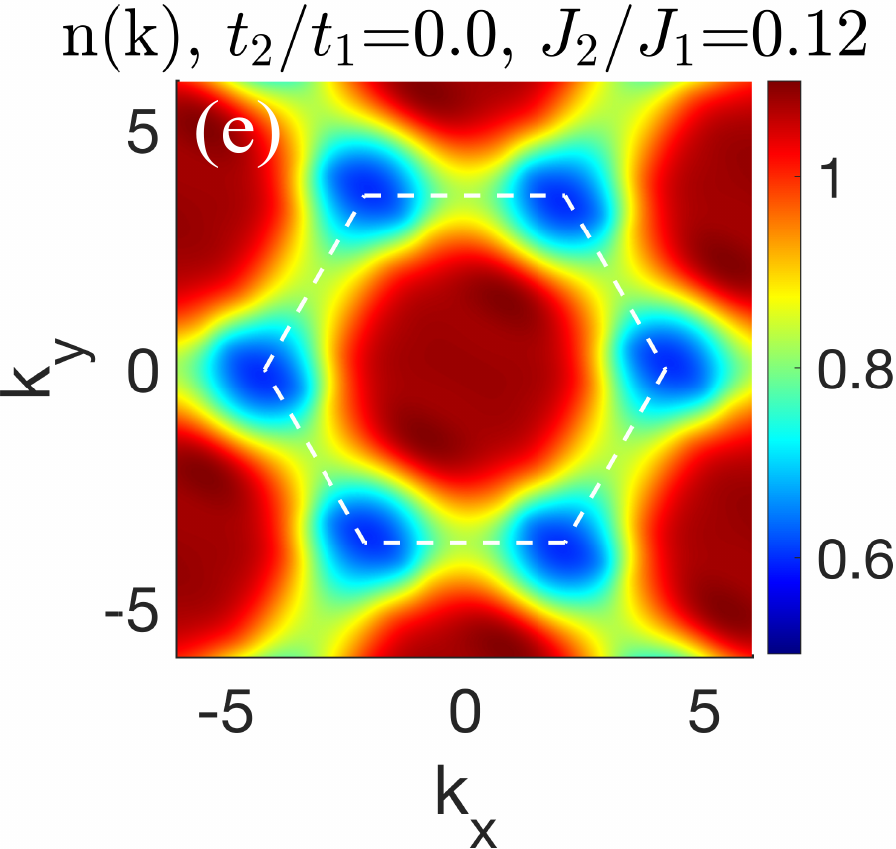}
\includegraphics[width=0.325\linewidth]{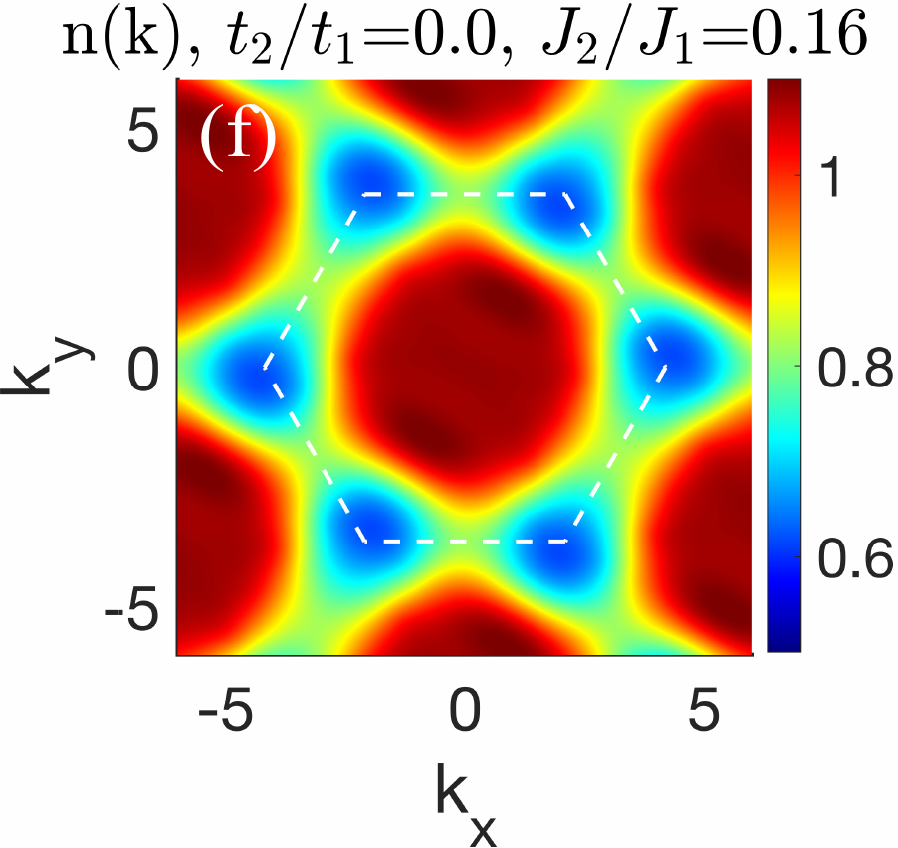}
\includegraphics[width=0.325\linewidth]{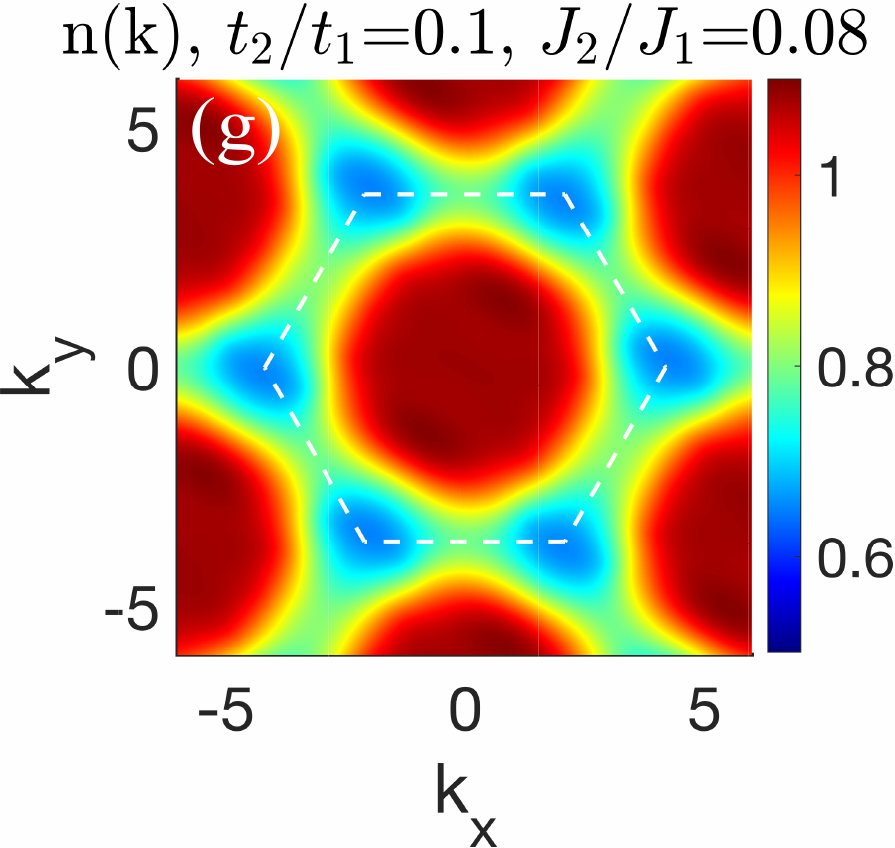}
\includegraphics[width=0.325\linewidth]{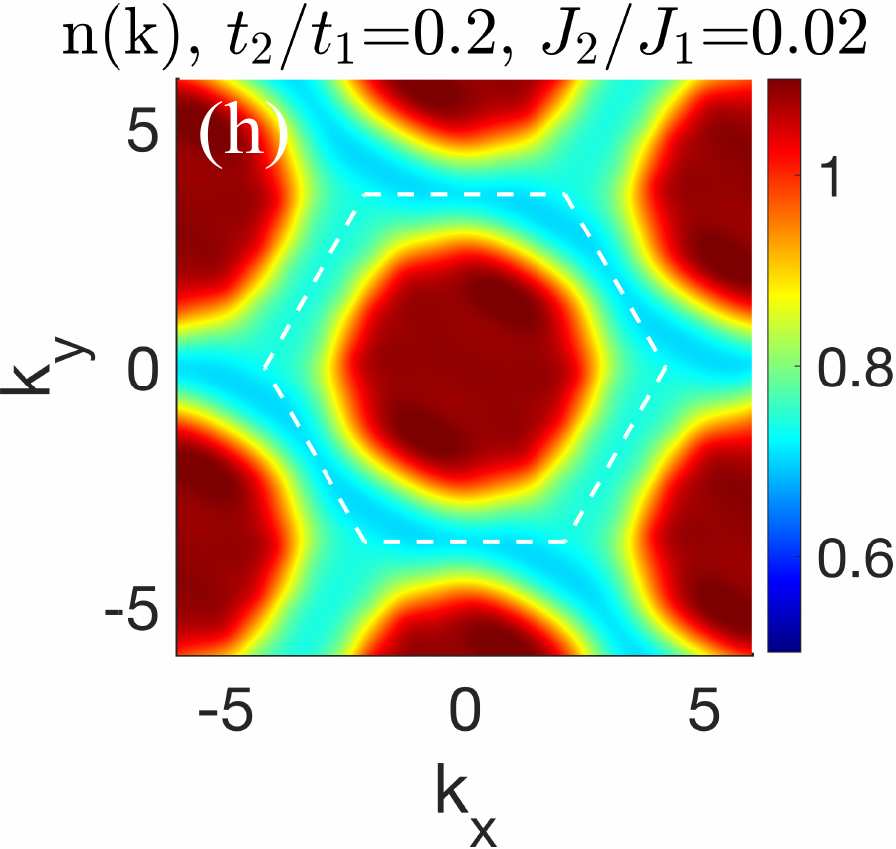}
\includegraphics[width=0.325\linewidth]{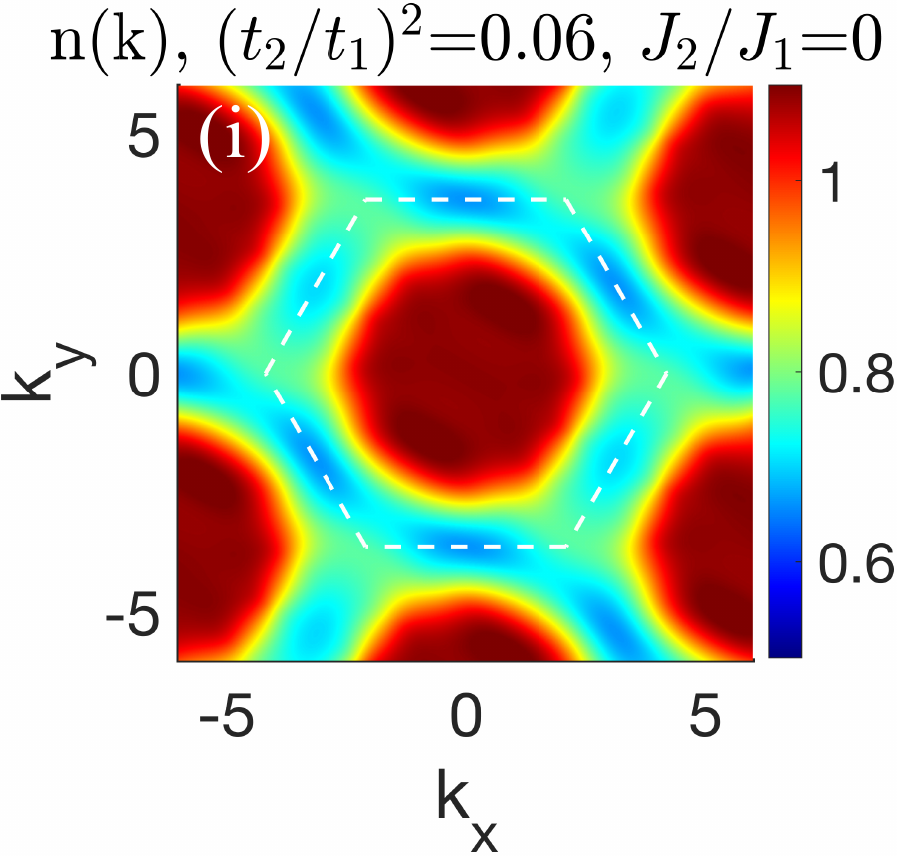}
\caption{Electron densities in the momentum space $n({\bf k})$ of different couplings at $\delta = 1/12$. $n(\bf k)$ is calculated by taking the Fourier transformation for the single-particle correlations of the middle $24\times 6$ sites ($L_y = 6$). The dashed white hexagon denotes the Brillouin zone. The parameter points in (a)-(d) locate in the CDW/SDWF phase. (e)-(i) belong to the TSC phase. The $M=10000$ data are shown here, which converges well with bond dimension.}
\label{FigS_nk}
\end{figure}

In Fig.~\ref{FigS_nk}, we show the electron occupation number in the momentum space $n({\mathbf{k}})$ of different couplings for $\delta = 1/12$ on the $L_y = 6$ cylinder.
$n({\mathbf{k}})$ is obtained by taking the Fourier transformation for the single-particle correlations of the middle $24\times 6$ sites on a long cylinder, namely $n({\mathbf{k}}) = \sum_{i,j,\sigma} \langle \hat{c}^{\dagger}_{i,\sigma} \hat{c}_{j,\sigma} \rangle e^{i {\mathbf{k}}\cdot ({\mathbf{r}}_i - {\mathbf{r}}_j)} / N_{m}$ ($N_m$ is the number of sites for computing the electron correlations).
In the CDW/SDWF phase, the electron density has a large electron pocket around the $\Gamma = (0,0)$ point and small hole pockets near the ${\bf K}$ points.
$n({\mathbf{k}})$ also shows an approximate $C_3$ rotational symmetry.
These features seem to be universal and independent of the tuning couplings in the CDW/SDWF phase, as shown in Figs.~\ref{FigS_nk}(a)-\ref{FigS_nk}(d).
The hole pockets at the ${\bf K}$ points suggest that the hole distribution may be related to the prominent SDWF.
In the $d+id$-wave TSC phase, tuning $J_2/J_1$ and $t_2/t_1$ seem to change $n({\mathbf{k}})$ differently.
With tuning $J_2/J_1$ for small $t_2/t_1$, the hole pockets still concentrate at the ${\bf K}$ points but $n({\mathbf{k}})$ shows an approximate $C_6$ rotational symmetry [Figs.~\ref{FigS_nk}(e)-\ref{FigS_nk}(g)]. 
On the other hand, the growing $t_2/t_1$ leads the hole pockets to extend along the boundaries of the Brillouin zone [Figs.~\ref{FigS_nk}(h) and \ref{FigS_nk}(i)]. These observations illustrate the common and distinct hole dynamics in different quantum phases.
In the mean-field theories, the change of the Chern number is usually associated with the change of the Fermi surface topology.
Our results indicate that the pairing gap function in the momentum space may change its shape with tuning couplings, but the gap remains opened so there is no change of the Chern number.

\section{Emergent $d$-wave superconductivity from the CDW/SDWF with pseudogap-like behaviors by increasing doping level and system width}
\label{grand_canonical}

\begin{figure}
\centering
\includegraphics[width=0.85\linewidth]{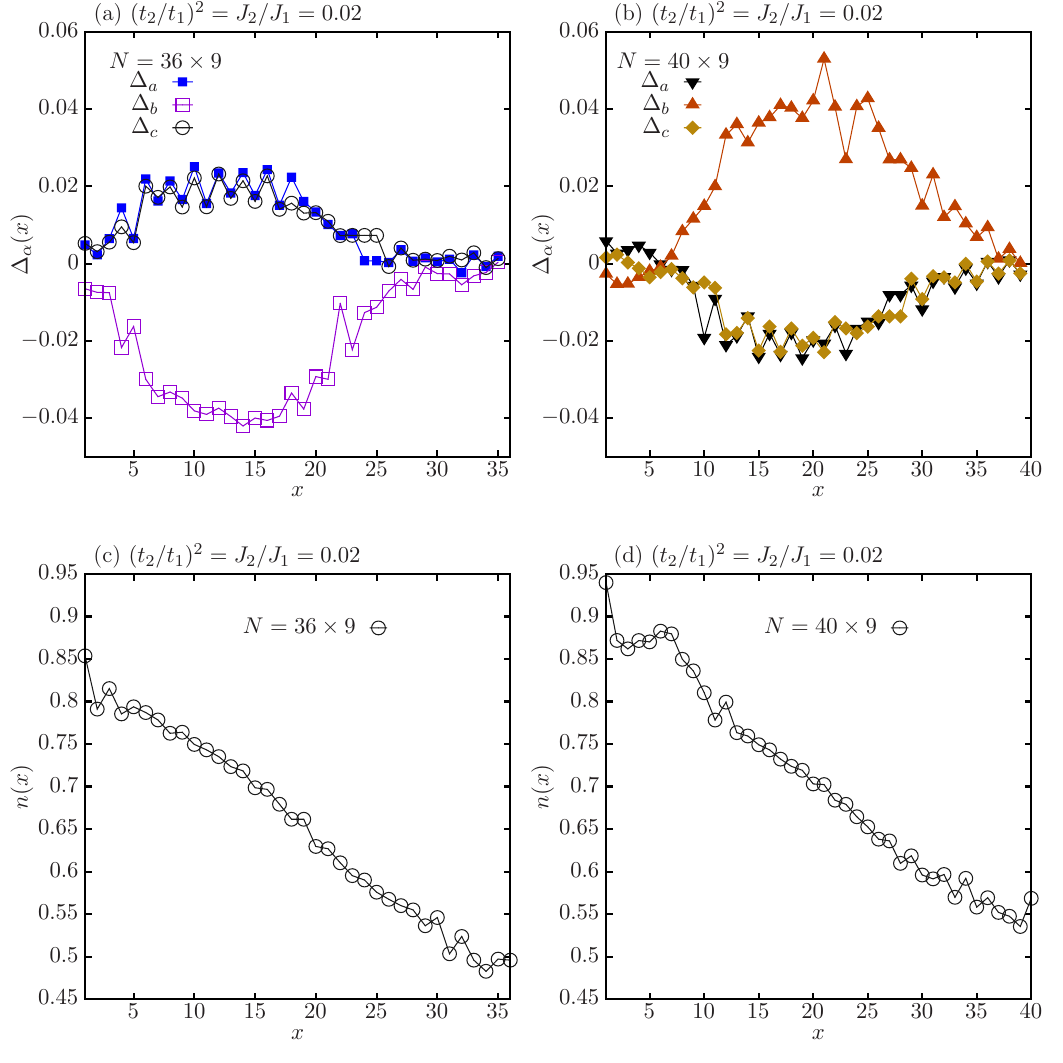}
\caption{The SC order and electron density at $(t_{2}/t_{1})^2 = J_{2}/J_1 = 0.02$ on the (a)/(c) $N = 36\times 9$ cylinder and the (b)/(d) $N = 40\times 9$ cylinder. Results are obtained with $M=8000$ in the grand canonical ensemble.}
\label{FigS_SC_order_density}
\end{figure}

In support of the Fig. 4(e) of the main text, we compare the SC orders $\Delta_{\alpha}(x)$ and the electron density $n(x)$ on different column $x$ with tuning the chemical potential. As shown in Figs.~\ref{FigS_SC_order_density} (a) and \ref{FigS_SC_order_density}(c), the SC order has sudden increase when electron density is below $0.8$, which corresponds to the doping level of 20\%. Similar results on a different $L_{x}$ can be seen by comparing Figs.~\ref{FigS_SC_order_density} (b) and \ref{FigS_SC_order_density}(d), confirming that the CDW/SDWF with PGL phase has a tendency to evolve into $d$-wave SC by increasing doping level and cylinder width $L_{y}$.

\section{Numerical data and program code availability}
\label{data_code}

Results of our study are presented within the article and its Supplementary. The digital data and the codes implementing the calculations are available on \href{https://github.com/hyx15903/TSC_Tri_t1t2J1J2}{GitHub}.

\end{document}